\documentclass[12pt]{article}

\usepackage{latexsym}

\usepackage[font=small,labelfont=bf]{caption}

\usepackage{graphics}
\usepackage{graphicx}
\usepackage{epstopdf}
\usepackage{amssymb}
\usepackage{tabularx}
\usepackage{caption}
\usepackage{subcaption}
\usepackage{hyperref}
\usepackage{tabularx}
\usepackage{gensymb}
\usepackage{mathtools,amsmath,amssymb,mathrsfs,color,esint}
\usepackage{array}
\usepackage{makecell}
\usepackage{float}
\usepackage{dcolumn}
\begin{document}


\title{Shadows of rotating traversable wormholes surrounded by plasma}

\author{
Tsanimir Angelov$^{1}$\footnote{E-mail: \texttt{tsangelov@phys.uni-sofia.bg}}, \,
Rasim  Bekir$^{1}$\footnote{E-mail: \texttt{rbekir@uni-sofia.bg}}, \,
Galin Gyulchev$^{1}$\footnote{E-mail: \texttt{gyulchev@phys.uni-sofia.bg}},  \, \\
Petya Nedkova$^{1}$\footnote{E-mail: \texttt{pnedkova@phys.uni-sofia.bg}}, \,
Stoytcho Yazadjiev$^{1,2}$\footnote{E-mail: \texttt{yazad@phys.uni-sofia.bg}}\\ \\
 {\footnotesize${}^{1}$ Faculty of Physics, Sofia University ``St. Kliment Ohridski'',}\\
  {\footnotesize 5 James Bourchier Boulevard, Sofia~1164, Bulgaria } \\
    {\footnotesize${}^{2}$ Institute of Mathematics and Informatics,}\\
{\footnotesize Bulgarian Academy of Sciences, Academic G. Bonchev 8, } {\footnotesize Sofia 1113, Bulgaria}}
\date{}
\maketitle

\begin{abstract}
We study the influence of the plasma environment on the shadows of stationary axisymmetric wormholes. We consider a sample of several wormhole solutions and plasma distributions for which  the Hamilton-Jacobi equation for the light rays is separable. This allows us to derive analytical expressions for the shadow boundary and examine in detail the behavior of the photon regions as the plasma frequency varies. We observe that plasma profiles which depend only on radial coordinate lead to common evolution of the photon region which does not depend on the wormhole metric and is consistent also with the Kerr black hole. In contrast, for plasma profiles with angular dependence the evolution of the photon region is specific for every spacetime. This creates opportunities for distinguishing wormholes observationally. We further investigate the formation of forbidden regions in the plasma medium where light cannot propagate. They lead to the formation of plasma frequency ranges where the shadow is no longer observable and we show that this phenomenon is characteristic for all the configurations in our sample. We obtain the critical frequencies for which the shadow vanishes and demonstrate that for all the wormholes they are lower than the critical frequencies for the Kerr black hole in the same environment. This implies that there exist plasma frequency ranges in which the Kerr black hole casts a shadow but wormholes do not, creating a strong observational signature for discriminating between compact objects. In the frequency ranges where both black hole and wormhole shadows exist the wormhole shadows are consistently smaller than those for the Kerr black hole. As the plasma frequency grows the discrepancy progresses showing that plasma medium facilitates the experimental detection of wormholes. Finally we consider aberrational effects on the wormhole shadows. They further increase the deviation from black holes making wormholes more easily detectable.
\end{abstract}

\section{Introduction}

General relativity predicts that gravity can affect light rays. This was experimentally tested more than a century ago with the first observation of gravitational lensing of light by Eddington and his team. They observed a small change in the position of background stars during the 1919 solar eclipse. The theory predicts that the angle of deflection of the light rays can be arbitrarily large near compact objects with sufficiently strong gravitational field such as black holes, wormholes and gravastars. This may cause some photon trajectories to be trapped by  the compact object and as an consequence a distant observer would see a dark area in this direction of their sky. This phenomenon is known as the compact object's shadow. Thanks to the recent advances of the observational capacities the Event Horizon Telescope collaboration has obtained the first images of the shadows of the compact objects at the center of M87* as well as at the center of our own galaxy Sgr A* \cite{EHT1}-\cite{EHT3}. The shadow is closely related to the metric of the spacetime and encodes information about the nature of the compact object which is causing it. Currently the observational data is consistent with the expectations for the Kerr black hole. However, it is not restrictive and admits other interpretations such as black holes in the modified theories of gravity or exotic compact objects without an event horizon \cite{Moffat:2015}-\cite{Deliyski:2025}. The near-future improvement of the Event Horizon Telescope may lead to sufficient resolution to break to some extent the degeneracy between the different spacetimes and  provide more precise information of the nature of the compact objects at the center of M87* and Sgr A*.

Some of the possible alternatives of supermassive black holes are wormholes which represent bridges that connect two universes or two regions of the same universe. They have been studied within general relativity for the first time  by Flamm \cite{Flamm:1916} and subsequently by Einstein and Rosen \cite{EinsteinRosen}. These early examples are not traversable i.e. a human being would not be able to cross though the tunnel.  The traversability conditions were studied by Morris and Thorne \cite{Morris:1988} where they derived the general geometry of a static spherically symmetric traversible wormhole. Their work was subsequently generalized by Teo for the stationary axially symmetric  case \cite{Teo:1998}. The matter and light propagation in the spacetime of rotating traversable wormholes were investigated in a series of works including their shadows \cite{Nedkova:2013}-\cite{Kunz:2023}, accretion disks in their vicinity \cite{Shaikh:2019a}-\cite{Vincent:2020}, quasi-periodic oscillations \cite{Deligianni:2021}-\cite{Deligianni:2021a} and linear polarization of the emission from the accretion disk \cite{Nedkova:2023}

The analytical investigation of the shadow usually considers that the compact object is illuminated by a uniform distribution of light sources and the light propagates in vacuum \cite{Perlick:2022a}. However, in astrophysical conditions the compact object is surrounded by an accretion disk and the presence of plasma would affect the photon propagation. In most wavelengths the effect is negligible but in the radio range the influence of the medium may be strong enough to lead to observational consequences \cite{Rogers:2015}-\cite{Mao:2014}. 

The effect of plasma on the gravitational lensing has been studied analytically in various astrophysical scenarios. The first works considered small gravitational  and plasma deflection when both sources were decoupled and linearized gravitational theory could be applied \cite{Muhleman:1966}-\cite{Bliokh:1989}. The strong gravitational lensing around compact objects in plasma environment was first studied by Perlick \cite{Perlick:2000} where the deflection angle for the Schwarzschild black hole in time-independent plasma distribution was calculated without any approximations. Lensing effects produced by the influence of the plasma in static spherically symmetric spacetime  were further investigated by Bisnovatyi-Kogan and Tsupko in \cite{Tsupko:2009}-\cite{Tsupko:2010} (for a review see \cite{Tsupko:2015}). These studies were extended for rotating black holes calculating the deflection angle in the equatorial plane in plasma environment around the Kerr black \cite{Perlick:2000}, and off-equatorial gravitational lensing in the small deflection limit \cite{Morozova:2013}. The influence of the plasma on the formation of multiple images was investigated in \cite{Tsupko:2013}-\cite{Liu:2016}, while its effect on the circular photon orbits was studies in \cite{Perlick:2024}-\cite{Perlick:2025}.

Plasma effects on the formation of the black hole shadow was initially studied for spherically symmetric black holes in \cite{Perlick:2015} demonstrating the influence of the dispersive medium on its angular size. The shadow of the Kerr black hole surrounded by a plasma medium was derived analytically in \cite{Perlick:2017} for plasma distributions which allow for the separation of variables in the Hamilton-Jacobi equation. The method was further extended for a general stationary and axisymmetric spacetime with separable geodesic equations which includes also some traversable wormhole geometries \cite{Perlick:2022}. This formalism was subsequently applied for a variety of compact objects in the modified theories of gravity \cite{Badia:2021}-\cite{Briozzo:2023}.

The purpose of this work is to consider the influence of the plasma distribution on the shadow for a  range of wormhole spacetimes investigating its deformation compared to black holes or a vacuum environment. In addition, we study aberrational effects which need to be taken into account if we do not assume a static observer.  A lot of research has been conducted which considers a static observer. However, this is an improbable scenario given that the compact objects which cause the shadows are spread out across the entire Universe. Their velocities relative to us lead to further deformations of the shadow silhouette and modify the location of the shadow in the observer's sky.  Previous research on this topic includes investigations of the shadow of the Kerr black holes in vacuum
\cite{Grenzebach:2015}, \cite{Penrose}, and an extension of these studies to black holes in the modified theories in gravity surrounded by a plasma medium \cite{Briozzo:2023}.

The paper is organized as follows. In chapter 2 we begin with a brief description of the wormhole spacetimes which we consider. In chapter 3 we  separate the variables in the Hamilton-Jacobi equation for a class of traversable rotating wormholes. This helps us to uncover the structure of the photon region which contains the spherical photon orbits and obtain the shadow boundary. In chapter 4 we present our results for the photon regions for several types of wormhole spacetimes and plasma distributions investigating also their evolution as the plasma density increases. We further obtain the shadow boundaries and make comparison between the wormhole shadows and the shadow of the Kerr black hole for  various plasma distributions. Chapter 5 focuses on aberrational effects on the shadow boundary and size caused by the motion of the observer. Chapter 6 contains our conclusions. We work in geometrical units assuming that $\hbar=G=c=1$.

\section{Traversable wormholes}

We consider stationary and axisymmetric wormholes which are traversable in the sense that they allow travelers to cross through the wormhole throat and reach another asymptotically flat spacetime.  This requires that the metric does not contain singularities or horizons and we can define suitable coordinates such that the metric functions are everywhere regular and non-zero. The general form of such geometries was derived by Teo  \cite{Teo:1998} who generalized the previous results of Morris and Thorne \cite{Morris:1988} for static spherically symmetric wormholes in order to include rotation.

The stationary and axially symmetric wormhole metric takes the following form

\begin{equation}\label {WH_metric}
d s^2=-N^2 dt^2+\left(1-\frac{b}{r}\right)^{-1} dr^2+r^2 K^2\left[d\theta^2+\sin ^2 \theta(d\phi-\omega dt)^2\right],
\end{equation}
where $r, \theta$ and $\phi$ are spherical coordinates and the metric functions depend only on  $r$ and $\theta$. For asymptotically flat spacetime the metric functions should possess the following behavior at infinity

\begin{eqnarray}\label{(6.66)}
&&N=1-\frac{m}{r}+\mathrm{O}\left(\frac{1}{r^2}\right), \quad \,\,\,  \frac{b}{r}=\mathrm{O}\left(\frac{1}{r}\right), \quad \,\,\, K=1+\mathrm{O}\left(\frac{1}{r}\right), \nonumber \\[2mm]
&&\omega=\frac{2 J}{r^3}+\mathrm{O}\left(\frac{1}{r^4}\right), 
\end{eqnarray}
where $M$ denotes the ADM mass of the wormhole and $J$ is its angular momentum. In order to avoid singularities the redshift function $N$ should be  everywhere finite and non-zero.  The function $b(r,\theta)$  determines the shape of the wormhole and needs to obey $\partial_\theta b(r, \theta)=0$ and $\partial_r b(r, \theta)<1$ at the throat in order to ensure the characteristic  flare-out geometry. It possesses a coordinate singularity at the wormhole throat $b=r_0$ which can be avoided by using the global coordinate $\ell$ defined as

\begin{equation}
d \ell= \pm \frac{dr}{K \sqrt{1-b/r}}.
\end{equation}
It takes the range $\ell \in (-\infty, +\infty)$ as $\pm\infty$ correspond to the two asymptotical ends of the asymptotically flat regions connected by the wormhole throat while the throat is located at $\ell =0$.  The metric function $K(r,\theta)$ needs to be regular everywhere and determines the proper radial distance.

We will consider three classes of wormholes belonging to the Teo geometry which allow separation of variables in the Hamilton-Jacobi equation for the null geodesics. The first two classes possess metric functions which depend only on the radial coordinate given by the expressions

\begin{equation}\label{WH1_metric}
b(r)=r_0,  \quad \,\,\, \omega(r) = \frac{2J}{r{^3}} \quad\,\,\, K(r)=1.
\end{equation}
where $r_0$ is a constant. The redshift function takes the form  $N=e^{-r_0/r}$ for the first class of wormholes which we denote by WH1 while we have  $N=e^{-r_0/r-r_0^2/r^2}$ for the second class denoted by WH2.

The third metric which we consider (WH3) possesses the form

\begin{equation}\label{WH2_metric}
d s^2=\Omega(r, \theta)\left(-d t^2+\left(1-\frac{r_0^2}{r^2}\right)^{-1}d r^2+r^2[d \theta^2+\sin^2{\theta} (d\phi-\omega dt)^2 ]\right),
\end{equation}
where we have
\begin{equation}
    \Omega(r, \theta)=1+\frac{(4 a \cos \theta)^2}{r_0^3 r}.
\end{equation}
This type of wormhole geometries was  previously studied in \cite{Perlick:2022} in the context of shadows in plasma environments and here we further expand their results.

\section{Light propagation in plasma in the spacetime of traversable wormholes}

In this section we derive analytical expressions for the wormhole shadow boundary taking into account the influence of the plasma medium on the proton trajectories. We consider in our calculations the wormhole spacetimes given by  Eq. ($\ref{WH1_metric}$) which we denoted by WH1 and WH2, while the shadow boundary for the WH3 spacetime given by Eq. ($\ref{WH2_metric}$) was previously obtained in \cite{Perlick:2022} in a similar way.

Light propagation in non-magnetized pressureless plasma is described by the following Hamiltonian \cite{Perlick:2017}

\begin{equation}
H(x,p) = \frac{1}{2}\left(g{^{\mu\nu}}p{_\mu}p{_\nu} +\omega{_p}(x){^2}\right),
\end{equation}
in terms of the spacetime coordinates $x$ and canonical momenta $p$ while $\omega{_p}(x)$ is the plasma electron frequency. It is proportional to the electron density $N{_e}(x)$

\begin{equation}
\omega{_p}(x)= \frac{4\pi e{^2}}{m{_e}}N{_e}(x),
\end{equation}
\noindent
where $e$ and $m{_e}$ are the electron charge and mass. The photon trajectories are solutions to the Hamilton's equations. Alternatively, we can consider the Hamilton-Jacobi equation for the particle propagation

\begin{equation}\label{10}
H\left(x{^\mu},\frac{\partial S}{\partial x{^\mu}}\right) = -\frac{\partial S}{\partial \lambda},
\end{equation}
where $S(x)$ is the Jacobi action and $\lambda$ is an affine parameter along the trajectories. We are interested in spacetimes in which we can separate the variables in the Hamilton-Jacobi equation. We can take advantage of the constants of motion $p_t=\omega_0$ and $p_{\phi}= L$ for the photon trajectories in a stationary axisymmetric spacetime, where $\omega_0$ is  the plasma frequency measured by an observer on a constant $t$ line at infinity, while $L$ is the specific angular momentum. Then we can consider the separation ansatz for the Jacobi action

\begin{equation}
 S = \frac{\mu{^2}}{2} \lambda - \omega{_0}t + L{_z}\phi + S{_r}(r) + S{_\theta}(\theta),
\end{equation}
where $\mu$ is the particle mass and the functions $S{_r}(r)$ and $S{_\theta}(\theta)$ depend only on the specified coordinates. Using wormhole metric given by Eq. ($\ref{WH_metric}$) the Hamilton-Jacobi equation takes the form
\begin{eqnarray}
&& \left(\frac{\partial S}{\partial \theta}\right)^2 - \frac{r{^2}K{^2}}{N{^2}}(\omega L{_z}-\omega{_0}){^2} +r{^2}K{^2}(1-b/r)\left(\frac{\partial S}{\partial r}\right)^2  \\[5pt] \nonumber
&&+ \frac{L{_z}{^2}}{\sin{\theta}{^2}}  +\omega{_p}{^2}r{^2}K{^2} +r{^2}K{^2}\mu{^2} =0.
\end{eqnarray}
We can separate the variables in this equation provided that we assume that the plasma frequency is described by the following ansatz

\begin{equation}\label{plasma}
\omega{_p}{^2}(r,\theta) = \frac{f{_r}(r)+f{_\theta}(\theta)}{r{^2}K{^2}}.
\end{equation}
This leads to the equations

\begin{eqnarray}\label{separ1} 
&& \left(\frac{\partial S}{\partial \theta}\right)^2 = Q -  \frac{L^2}{\sin{^2}{\theta}} - f{_\theta}(\theta), \\[2mm]
&& r{^2}K{^2}(1-b/r)\left(\frac{\partial S}{\partial r}\right)^2 = \frac{r{^2}K{^2}}{N{^2}}(\omega{_0} -\omega L{_z}){^2} - f{_r}(r) - r{^2}K{^2}\mu{^2}  - Q, \nonumber
\end{eqnarray}
where $Q$ is a separation constant. Denoting by $T(\theta)$ and $R(r)$ the expressions on the right-hand side of the equations

\begin{eqnarray}\label{separ2}
&&\Theta(\theta) = Q - \frac{L{_z}{^2}}{\sin{^2}{\theta}} - f{_\theta}(\theta), \\[2mm]
&&R(r) =  (\omega{_0} - \omega L{_z}){^2} -(N{^2}\mu{^2} + \frac{N{^2}}{r{^2}K{^2}}(f{_r}(r) + Q)), \nonumber
\end{eqnarray}
the Jacobi action obtains the form

\begin{eqnarray} 
S(\lambda,x{^\mu},\alpha) = \frac{\mu{^2}}{2} \lambda - \omega{_0}t + L{_z}\phi + \int^{r} \frac{\sqrt{R}}{\sqrt{N{^2(1-b/r)}}} \,dr + \int^{\theta} \sqrt{\Theta} \,d\theta.
\end{eqnarray}

Using the property that partial derivatives of the Jacobi action with respect to
the integrals of motion are constant it is possible to obtain the equations of motion.  Since we are interested in photon propagation, we should set
subsequently the mass of the particle $\mu$ to zero. Thus we obtain the following system of equations

\begin{eqnarray}\label{geodesic_eq}
&&\frac{N}{\sqrt{(1-b/r)}} \frac{d r} {d\lambda} =  \sqrt{R}, \quad \,\,\, r{^2}K{^2}\frac{d \theta} {d \lambda} =  \sqrt{\Theta} \\[2mm] \nonumber
&&N{^2} \frac{d \phi} {d \lambda} = \omega(\omega{_0}-\omega L{_z}) + \frac{N{^2}L{_z}}{r{^2}K{^2} \sin{^2}{\theta}}, \\[2mm] \nonumber
&&N{^2} \frac{dt}{d\lambda} = \omega{_0}-\omega L{_z}.
\end{eqnarray}

The plasma frequency $\omega_0$ is a scale parameter in these equations and it can be eliminated by rescaling the affine parameter as $\lambda \rightarrow \omega_0\lambda$ and introducing the impact parameters

\begin{eqnarray}
 \xi = \frac{L{_z}}{\omega{_0}}, \quad \,\,\, \eta = \frac{Q}{\omega{_0}{^2}}.
\end{eqnarray}
\noindent
Then the functions $R(r)$ and $\Theta(\theta)$  take the form

\begin{eqnarray}\label{29}
&& R(r) =  (1- \omega \xi){^2} - \frac{N{^2}}{r{^2}K{^2}}(\overline{f}{_r}(r) + \eta), \\[2mm] \nonumber
&& \Theta(\theta) = \eta - \frac{\xi{^2}}{\sin{\theta}{^2}} - \overline{f}{_\theta}(\theta), 
\end{eqnarray}
where 
\begin{eqnarray}
    \overline{f}{_\theta}(\theta) =  \frac{f{_\theta}(\theta)}{\omega{_0}{^2}}, \quad \,\, \overline{f}{_r}(r) =  \frac{f{_r}(r)}{\omega{_0}{^2}}.
\end{eqnarray}

\subsection{Wormhole shadow}

Light rays around wormholes can be divided into two families depending on their trajectory - light rays that scatter away to infinity, and light rays that are trapped by the wormhole and never reaching a distant observer. The second type of trajectories form dark directions in the observer's sky which correspond to the wormhole shadow. The two families are separated by unstable spherical photon orbits which determine the shadow boundary. Using Eqs. ($\ref{geodesic_eq}$) we can obtain the shadow boundary in the impact parameter space. We consider the radial equation which can be written in the form of an energy-like equation
\begin{equation}
\left(\frac{dr}{d\lambda}\right)^2 + V{_{eff}} = 1, \quad \, V{_{eff}} = 1 -\frac{1}{N{^2}}\left(1-\frac{b}{r}\right)R(r),
\end{equation}
by introducing the effective potential $V_{eff}$. The unstable spherical orbits correspond to the maxima of the effective potential determined by

\begin{equation}\label{35}
V{_{eff}} = 1, \quad \, \frac{d V{_{eff}}}{d r} = 0, \quad \, \frac{d{^2} V{_{eff}}}{d r{^2}} \leq 0.
\end{equation}
This set of equations gives a relation between the impact parameters $\xi$ and $\eta$, which defines a curve in parameter space corresponding to the shadow boundary. In the case of wormholes which are symmetric with respect to the wormhole throat we may have two families of unstable spherical photon orbits which contribute to the shadow boundary \cite{Nedkova:2018}. The first family is located outside the wormhole throat and it is obtained in a similar way as for black holes. We consider the equations

\begin{equation}\label{WH_shadow}
 R(r) = 0, \quad \, \frac{dR}{dr} = 0, \quad \, \frac{d{^2}R}{dr{^2}} \geq 0, 
\end{equation}
\noindent
which lead to the following expressions for the impact parameters $\xi$ and $\eta$

\begin{eqnarray}
\xi{_{1,2}} = \frac{-(\omega ' - 2 \Sigma \omega) \pm \sqrt{(\omega ' - 2 \Sigma \omega){^2} - 4(\Sigma \omega{^2} -\omega' \omega)(\frac{N{^2}}{2r{^2}K{^2}}\overline{f}'{_r}(r) + \Sigma)}}{2(\Sigma \omega{^2} -\omega' \omega)}, \nonumber 
\end{eqnarray}
\begin{eqnarray}\label{WH1_shadow}
&&\eta = \frac{r{^2}K{^2}}{N{^2}}(1-\omega \xi){^2} - \overline{f}{_r}(r), \nonumber \\[2mm]
&&\Sigma = \frac{1}{2}\frac{d}{dr}\ln\left(\frac{N{^2}}{r{^2}K{^2}}\right),
\end{eqnarray}
parameterized by orbit's radial location. Here we denote by prime the derivative with respect to the radial coordinate $r$.
\noindent

The second family of unstable spherical photon orbits result from maxima of the effective potential located at the wormhole throat $r = r{_0}$. The set of equations which determine this family are

\begin{equation}
 1 - \frac{b(r)}{r} = 0, \quad \, R(r) = 0, \quad \, \frac{dR}{dr} \geq 0,
\end{equation}
\noindent
giving an implicit relation between the impact parameters $\xi$ and $\eta$

\begin{eqnarray}\label{WH2_shadow}
r^2K^2(\omega\xi -1)^2 - N^2(\eta-\overline{f}_r(r))_{|_{r=r_0}} = 0.
\end{eqnarray}

Thus, the region in the impact parameter space which corresponds to the shadow boundary of the wormhole classes WH1 and WH2  is obtained by combining the solutions of Eqs. ($\ref{WH1_shadow}$) and ($\ref{WH2_shadow}$).

In order to obtain the shadow boundary for the wormhole family  WH3 we take advantage of the results obtained in \cite{Perlick:2022}. In this work the variables in the Hamilton-Jacobi equations were separated and an expression for the radial function $R(r)$ was derived. In the notations of Eqs.\eqref{separ1} and \eqref{separ2} it takes the form

\begin{equation}
    R(r) =  (1- \omega \xi){^2} - \frac{1}{r{^2}}(\overline{f}{_r}(r) + \eta).
\end{equation}
Using this expression the shadow boundary is obtained as solution to the conditions given by Eqn. ($\ref{WH_shadow}$). We should note that in order to obtain valid solutions for the trajectories building up the shadow rim the angular equation ($\ref{geodesic_eq}$) should be well-defined. This imposes additional constraints on the impact parameters which we will discuss in the following section.

In our subsequent analysis we will compare the influence of the plasma distribution on the shadow boundary for the different classes of wormhole as well for the Kerr black hole. For the purpose we normalize the solutions by their mass, or equivalently we set $m =1$. Since the deformation of the shadow is most pronounced at high rotation rate we perform our analysis for spin parameter $a = 0.999m$. This normalization determines the throat location for the WH1 and WH2 classes of solutions as $r_0=m =1$. For the WH3 solution the Komar mass is given by the expression

\begin{equation}
    m=\frac{8}{3}\frac{a^2}{r_0^3},
\end{equation}
which  defines the value of the radial coordinate on the throat in this case.

\subsection{Photon region and forbidden region}

The photon region provides a useful representation for the ranges of the radial and angular coordinates $r$ and $\theta$ on the circular photon orbits. Using the equations $R(r, \xi, \eta)=0$ and $R^{'}(r, \xi, \eta)=0$, which define the circular orbits,  we obtain expressions for the impact parameters $\xi=\xi(r)$, $\eta = \eta(r)$ in terms of the circular orbit's radial location. Plugging in these expressions in the condition that the angular  equation ($\ref{geodesic_eq}$)  is well-defined, i.e. $T(\theta)>0$, we obtain the locus of the circular photon orbits in the $(r,\theta)$ parameter space. Usually the photon  region is visualized more conveniently in the Cartesian coordinates $(x,z)$ defined as

\begin{equation}\label{photon region_coord}
    r^2=x^2+z^2, \qquad \sin {\theta}= \frac{z}{r}.
\end{equation}
The photon region for the WH1 and WH2 classes of solutions is obtained by considering the inequality

\begin{equation}\label{WH1_photon region}
    T(\theta) = \eta - \frac{\xi^2}{\sin \theta ^2}-\overline{f}_\theta(\theta)\geq 0,
\end{equation}
and substituting the relations for the impact parameters obtained in Eqs. $\ref{WH1_shadow}$.For the WH3 case we use a similar inequality derived previously in \cite{Perlick:2022}

\begin{equation}\label{WH3_photon region}
    \left(3 r \pm \sqrt{9 r^2-4 r\overline{f}_r^{\prime}(r)}\right)^2-16 \left(\overline{f}_r(r)+\overline{f}_{\theta}(\theta)\right)\geq \frac{r^4}{4 a^2 \sin ^2 \vartheta}\left(r \mp \sqrt{9 r^2-4 r \overline{f}_r^{\prime}(r)}\right)^2 .
\end{equation}
In our studies we will consider the photon region formed only by the unstable circular photon orbits, i.e. we will further impose the condition $R^{''}\geq 0$.

Due to the dispersive properties of the plasma medium light propagation in it is possible only for certain ratios of the photon frequency and plasma electron frequency. The index of refraction can be expressed as \cite{Perlick:2017}

\begin{equation}
    n^2(x, \omega_{ph}) =  1 - \frac{\omega_p(x)^2}{\omega_{ph}(x)^2},
\end{equation}
where $\omega_{ph}(x)$ is the photon frequency with respect to the plasma, and $\omega_p(x)$ is the plasma electron frequency. Light can propagate in such medium only  when the refractive index is well defined, i.e. it is satisfied that

\begin{equation}
    \omega_p(x)\geq\omega_{ph}(x).
\end{equation}
For any stationary and axisymmetric metric in Boyer-Lindquist-type coordinates $(t,r,\theta, \phi)$ this condition can be also expressed in the form \cite{Perlick:2017}

\begin{equation}
    \omega^2_0 \geq - g_{tt}\,\omega^2_p(r,\theta),
\end{equation}
in terms of the metric function $g_{tt}$ and the photon frequency at infinity $\omega_0$. Given a specific plasma distribution the last inequality defines  consider regions in the $(r,\theta)$ space which can be accessed by  the photon trajectories. The complementary regions where light propagation is prohibited are called forbidden regions. The existence of forbidden regions is a fundamental property of light propagation in a dispersive medium which causes some distinctive properties of the compact object's shadows. Therefore, in the following analysis we visualize them together with the photon regions to provide more complete understanding for the shadow formation.

\subsection{Celestial coordinates}
To construct the shadow we introduce the following zero angular momentum observer (ZAMO) tetrad
 
 \begin{align}
  e_0&=\left.\frac{1}{N} \left(\partial_t+\omega\partial_\phi\right)\right|_{(r_o,\theta_o)},
  \label{eq:tw1}
  \\
  e_1&=\left.\frac{1}{rK}\partial_\theta\right|_{(r_o,\theta_o)},
  \label{eq:tw2}
  \\
  e_2&= \left.\frac{1}{rK\sin\theta}
  \partial_\phi\right|_{(r_o,\theta_o)},
  \label{eq:tw3}
  \\
  e_3&=-\left.\left(1-\frac{b}{r}\right)^{1/2}\partial_r\right|_{(r_o,\theta_o)}.
  \label{eq:tw4}
\end{align}

The tetrad is orthonormal and at observer's position $(r{_o},\theta{_o})$. For each light ray $\tau{\lambda}$  it is possible to write the tangent vector as
\begin{equation}\label{lambdadot}
  \dot{\tau}=\dot{r}\partial_r+\dot{\theta}\partial_\theta+\dot{\phi}\partial_\phi+\dot{t}\partial_t \, .
\end{equation}
\noindent
The overdot signifies a derivative with respect to the affine parameter $\lambda$. The tangent vector at the observer's location can be represented alternatively as

\begin{equation}\label{lambda_obs}
  \dot{\lambda}=\left.-\alpha e_0+\beta(\sin\Theta\cos\Phi e_1+\sin\Theta\sin\Phi e_2+\cos\Theta e_3)\right|_{(r_o,\theta_o)},
\end{equation}
where the factors $\alpha$ and $\beta$ are positive. The factors $\alpha$ and $\beta$ are related by the equation $g(\dot{\lambda}, \dot{\lambda}) = - \omega_{pl}^2$ which leads to the condition

\begin{equation}
  \alpha^2-\beta^2=\left.\omega_{pl}^2\right|_{(r_o,\theta_o)}.
\end{equation}
The factor $\alpha$ can be determined for each light ray by the following relation

\begin{gather}\label{alphaeq1}
  \alpha=g(\dot{\lambda},e_0) = \frac{1}{N}\left(g{_{tt}}\dot{t}+g{_{t\phi}}(\omega \dot{t}+\dot{\phi})+g{^{\phi\phi}}\omega \dot{\phi}\right)=\left. \frac{1}{N}(\omega L{_z}-\omega{_0})\right|_{(r_o,\theta_o)}.
\end{gather}

The celestial coordinates for the observer are $\Phi$ and $\Theta$, which correspond to the colatitude and the azimuthal angle respectively. By comparing the equations eq.\eqref{lambdadot} and eq.\eqref{lambda_obs}, we can parameterize the shadow by the celestial coordinates of the observers in the following way

\begin{equation*}\label{celes:ThetaWh1}
  \sin{\Theta} =\left. \left(1+ \frac{\eta(r{_p}) + \overline{f}{_r} - \frac{r^2}{N^2}(1-\omega \xi(r{_p})){^2}}{\frac{r^2}{N^2}(1-\omega \xi(r{_p})){^2} - \overline{f}{_r} -\overline{f}{_\theta}}\right)\right|_{(r_o,\theta_o)},
\end{equation*}

\begin{equation*}\label{celes:PhiaWh1}
  \sin{\Phi} =\left. \frac{\xi(r{_p})}{\sin{\theta}\sqrt{\eta(r{_p}) -\overline{f}{_\theta}}}\right|_{(r_o,\theta_o)},
\end{equation*}
the above for WH 1 and WH 2 
and the following for WH 3:

\begin{gather}\label{alphaeq2}
  \alpha=g(\dot{\lambda},e_0) =\left. \frac{1}{\sqrt{\Omega}}(\omega L{_z}-\omega{_0})\right|_{(r_o,\theta_o)},
\end{gather}

\begin{equation*}\label{celes:ThetaWh3}
  \sin{\Theta} =\left. \left(1+ \frac{\eta(r{_p}) + \overline{f}{_r} - r^2(1-\omega\xi(r{_p})){^2}}{r^2(1-\omega\xi(r{_p})){^2} - \overline{f}{_r} -\overline{f}{_\theta}}\right)\right|_{(r_o,\theta_o)},
\end{equation*}

\begin{equation*}\label{celes:PhiaWh3}
  \sin{\Phi} =\left. \frac{\xi(r{_p})}{\sin{\theta}\sqrt{\eta(r{_p}) -\overline{f}{_\theta}}}\right|_{(r_o,\theta_o)},
\end{equation*}
where the parameter $r{_p}$ describes the location of the unstable spherical photon orbits. The upper and lower boundary of its domain are determined by the radius values of spherical light rays that have turning points
at$\theta =\theta_o$. 

To construct the shadow we use stereographic projections on the plane tangent to the celestial sphere at the pole $\Theta=0$ and we define in this plane the dimensionless Cartesian coordinates

\begin{gather}\label{eq:stereographic}
    X(r{_p})=-2\tan\left(\frac{\Theta(r{_p})}{2}\right)\sin(\Phi(r{_p})),\\
    Y(r{_p})=-2\tan\left(\frac{\Theta(r{_p})}{2}\right)\cos(\Phi(r{_p})).
\end{gather}

\subsection{Aberration of the shadow of wormhole}

The shape of shadow depends on observer's state of motion. In the previous section we explored the method for obtaining the shadow for an observer with a 4-velocity $e{_0}$. When the observer possesses a non-zero $3-$velocity vector, we expect both the shape and size of the shadow to change accordingly. Proceeding further, we denote the standard observer by $O$, while the new observer experiencing aberration is denoted by $O'$. The following derivation closely follows the works of Grenzebach \cite{Grenzebach:2015} and Briozzo et al \cite{Briozzo:2023}.

 If the observer $O'$, located at point $(r{_o},\theta{_o})$, moves with $3-$velocity $\mathbf{v}=(v{_1},v{_2},v{_3})$, $v < 1 = c$, with respect to the standard observer $O$, we have to modify the chosen tetrad, as follows:

  \begin{equation}
  \label{eq:newTetrad}
        \begin{split}
	& \widetilde{e}_{0} = \frac{v_{1}e_{1} + v_{2}e_{2} + v_{3}e_{3} + e_{0}}{
		\sqrt{1-v^{2}}}, \\[\smallskipamount]
		&\widetilde{e}_{1} = \frac{\bigl(1-v_{2}^{2}\bigr)e_{1} 
			+ v_{1}(v_{2}e_{2}+e_{0})}{ \sqrt{1-v_{2}^{2}} \; 
			\sqrt{1-v_{1}^{2}-v_{2}^{2}} }, \\[\smallskipamount]
		&\widetilde{e}_{2} = \frac{ e_{2} + v_{2}e_{0} }{ \sqrt{1-v_{2}^{2}} }, 
			\\[\smallskipamount]
		&\widetilde{e}_{3} = \frac{\bigl(1-v_{1}^{2}-v_{2}^{2}\bigr)e_{3} 
			+ v_{3}(v_{1}e_{1}+v_{2}e_{2}+e_{0})}{ \sqrt{1-v_{1}^{2}-v_{2}^{2}} \; 				\sqrt{1-v^{2}} }.
        \end{split}
    \end{equation}
 Similarly to the previous section, the tangent vector to any photon trajectory at the observer's position can be written in two different ways. First, in the coordinate basis
 \begin{equation}\label{lambdadot2}
  \dot{\tau}=\dot{r}\partial_r+\dot{\theta}\partial_\theta+\dot{\phi}\partial_\phi+\dot{t}\partial_t \, ,
\end{equation}
or, with the help of the modified tetrad,
\begin{equation}\label{lambda2}
  \dot{\tau}=\left.-\alpha \widetilde{e}_0+\beta(\sin\Theta\cos\Phi \widetilde{e}_1+\sin\Theta\sin\Phi \widetilde{e}_2+\cos\Theta \widetilde{e}_3)\right|_{(r_o,\theta_o)}.
\end{equation}
Here, we assume that $\Psi$ and $\Theta$ are the celestial coordinates of the observer $O'$. The factors $\alpha$ and $\beta$ have to be redefined to take into account the $4-$velocity of observer $O'$. Accordingly, we obtain the following expressions:
   \begin{equation}
        \begin{split}
        \alpha &=g(\dot{\tau},\Tilde{e}_0)=k_0^tp_t+k_0^rp_r+k_0^\theta p_\theta + k_0^\phi p_\phi, \\
        \beta  &=\sqrt{(k_0^tp_t+k_0^rp_r+k_0^\theta p_\theta + k_0^\phi p_\phi)^2-\omega_p^2},
        \end{split}
        \label{eq:alphabeta}
    \end{equation}
where $p_{\mu}$ is the photon $4-$momentum, and the components $k_i^\mu$ are defined through
 \begin{equation}
        \Tilde{e}_i=k_i^\mu\partial_\mu.
        \label{eq:Kelements}
    \end{equation}
The exact form of $k_i^\mu$ can be obtained substituting the standard tetrad into Eq. \eqref{eq:newTetrad}.
Proceeding further, by substituting Eq. \eqref{eq:Kelements} into Eq. \eqref{lambda2} and comparing the resulting expression with Eq.  \eqref{lambdadot2}, we arrive at the following system of equations for the  light rays: 
  \begin{equation}
        \label{eq.newgeodesics}
        \begin{split}
        &\dot{t} =-\alpha k_0^t+\beta k_1^t\sin\Theta\cos\Phi+\beta k_2^t\sin\Theta\sin\Phi+\beta k_3^t\cos\Theta, \\
        &\dot{r} =-\alpha k_0^r + \beta k_3^r \cos\Theta, \\
        &\dot{\theta} =-\alpha k_0^\theta + \beta k_1^\theta\sin\Theta\cos\Phi + \beta k_3^\theta\cos\Theta, \\
        &\dot{\phi} =-\alpha k_0^\phi + \beta k_1^\phi\sin\Theta\cos\Phi + \beta k_2^\phi\sin\Theta\sin\Phi
        + \beta k_3^\phi\cos\Theta.
        \end{split}
    \end{equation}

From this system, we derive the following expression for the celestial angles $\Phi$ and $\Theta$: 

\begin{subequations}
\begin{align}
	&\cos{\Theta} =\left. \frac{\dot{r} + \alpha k_0^r}{\beta k_3^r} \right|_{(r_o,\theta_o)},
	\label{aber:theta}
	\\
	&\sin{\Phi} =\left. \frac{\dot{\phi} + \alpha k_0^r
		- \frac{k_1^\phi}{k_1^\theta}(
			\dot{\theta} + \alpha k_0^\theta)
		- \beta(k_3^\phi-\frac{k_1^\phi}{k_1^\theta}k_3^\theta ) 
		\frac{\dot{r} + k_0^r}{\beta k_3^r}}{
		\beta k_{2}^\phi\sqrt{1- \bigl(\frac{\dot{r} + \alpha k_0^r}{\beta k_3^r}\bigr)^2} } \right|_{(r_o,\theta_o)},
	\label{aber:phi}
\end{align}
\end{subequations}
where the equations of motion should be substituted by their explicit form given in Eq. \eqref{eq.newgeodesics}. As previously, all the functions must be evaluated at observer's coordinates $(r_o,\theta_o)$, except for the impact factors $\eta$ and $\xi$, which must be evaluated at $(r_p,\theta{_p})$. In the limiting case where the $3-$velocity $\mathbf{v}$ vanishes, the expressions reduce to those corresponding to zero aberration, as found by Grenzebach \cite{Grenzebach:2015}. A similar conclusion was reached by Briozzo et al. \cite{Briozzo:2023}. Finally, we continue to use the functions $X(r_p)$ and $Y(r_p)$ in Eq. \eqref{eq:stereographic} for the stereographic projection.

\section{Wormhole shadow in a  plasma environment}

In this section we present our results on the influence of the plasma environment on the wormhole shadows. Our studies are further extended by considering aberrational effects demonstrating how a non-zero velocity of the observer would interplay with the shadow deformations due to the dispersive medium. In order to describe the plasma distribution we use several toy models proposed previously in \cite{Perlick:2017} which are consistent the separability condition ($\ref{plasma}$).  They should further satisfy the requirements

\begin{eqnarray}
    &&f_r(r)=Cr^k, \quad \,\,\, C \geq 0, \nonumber \\
    &&f_{\theta}(\theta)   \geq 0,
\end{eqnarray}
which ensure that the light propagation is well-defined. In particular we consider the following plasma distributions
\begin{align}
&    f_r=\omega_c^2\sqrt{m^3r}, \quad \, f_\theta=0,\label{sqrtrplasma}
\\[10pt]
&    f_r=4\omega_c^2m^2\sqrt{(m/r)}, \quad \, f_\theta=0,\label{Perlickplasma}
\\[10pt]
&    f_r=0,\quad \, f_\theta=\omega_c^2m^2(1+2\sin^2\theta).\label{thetaplasma}
\end{align}
in the notations of Eq. ($\ref{plasma}$) where $\omega_c$ is a constant with the dimension of a frequency. This allows us to compare the influence of the plasma distribution on the shadow of wormholes with previous black hole studies in similar conditions \cite{Perlick:2017}, \cite{Briozzo:2023}, and access the role of the spacetime geometry  on the dispersive effects also in horizonless spacetimes.

\subsection{Influence of the plasma environment on the wormhole shadows}

In this section we focus on studying plasma effects on the wormhole shadow for a static observed located at the asymptotic infinity at edge-on inclination. In particular we investigate all the possible combinations of the three wormhole spacetimes which we previously described with the plasma distributions given by Eqs. ($\ref{sqrtrplasma}$)-($\ref{thetaplasma}$) searching in this sample for common trends and distinctive signatures compared to black holes. Since the photon region and the forbidden region are geometrical properties which govern the shadow formation we examine them initially in each case and relate their morphology to the corresponding shadow properties. We observe that some of the combinations in our sample demonstrate similar behavior. Therefore, in this section we present only examples of qualitatively distinct phenomena while the rest of the cases are presented in the appendix for completeness.

In Fig. 1 we present the photon region and the forbidden region for the first configuration from our sample representing  a WH1 type of spacetime with a plasma distribution given by Eq. \eqref{sqrtrplasma}. The photon region (shown in orange) and the forbidden region (shown in grey) are investigated for various ratios of the plasma frequency $\omega_c$ and the photon frequency at infinity $\omega_0$. We observe that when the ratio between the two frequencies $\omega_c/\omega_0$ becomes sufficiency  large a forbidden region appear initially near the poles. As the plasma frequency further increases, the forbidden region expands toward the equatorial plane until it finally encompasses the full parameter space of the spherical photon orbits. The frequency ratio for which this is accomplished corresponds to a critical plasma frequency after which no shadow can be observed. The photon region for the Kerr black with the same plasma distribution possesses similar qualitative behavior when the plasma frequency is varied, however with  a different value of the critical frequency (see \cite{Perlick:2017}).
\begin{figure}[t!]
\centering
    \begin{tabular}{ cc}
    \includegraphics[width=6.05cm]{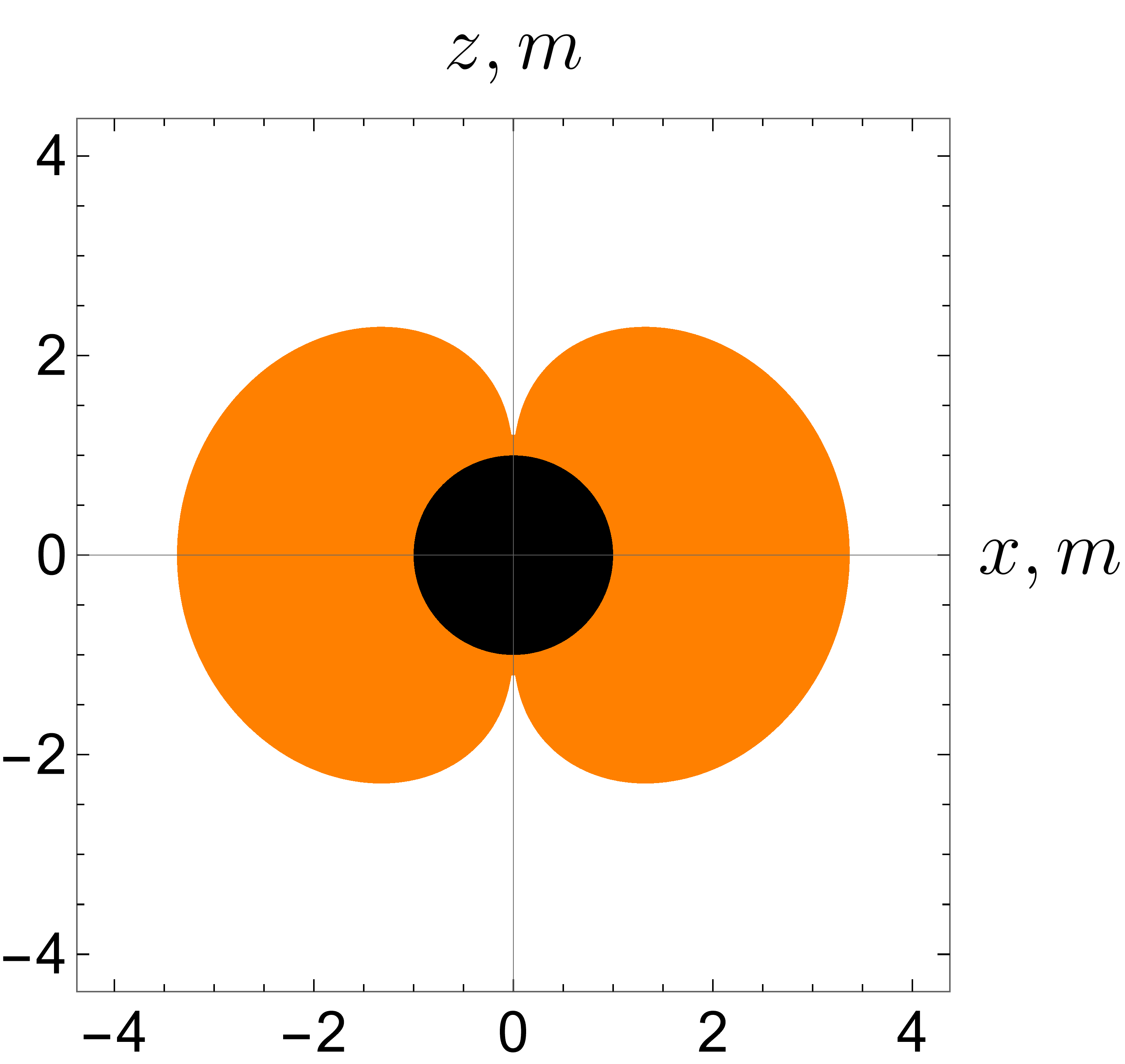}
    \includegraphics[width=6.05cm]{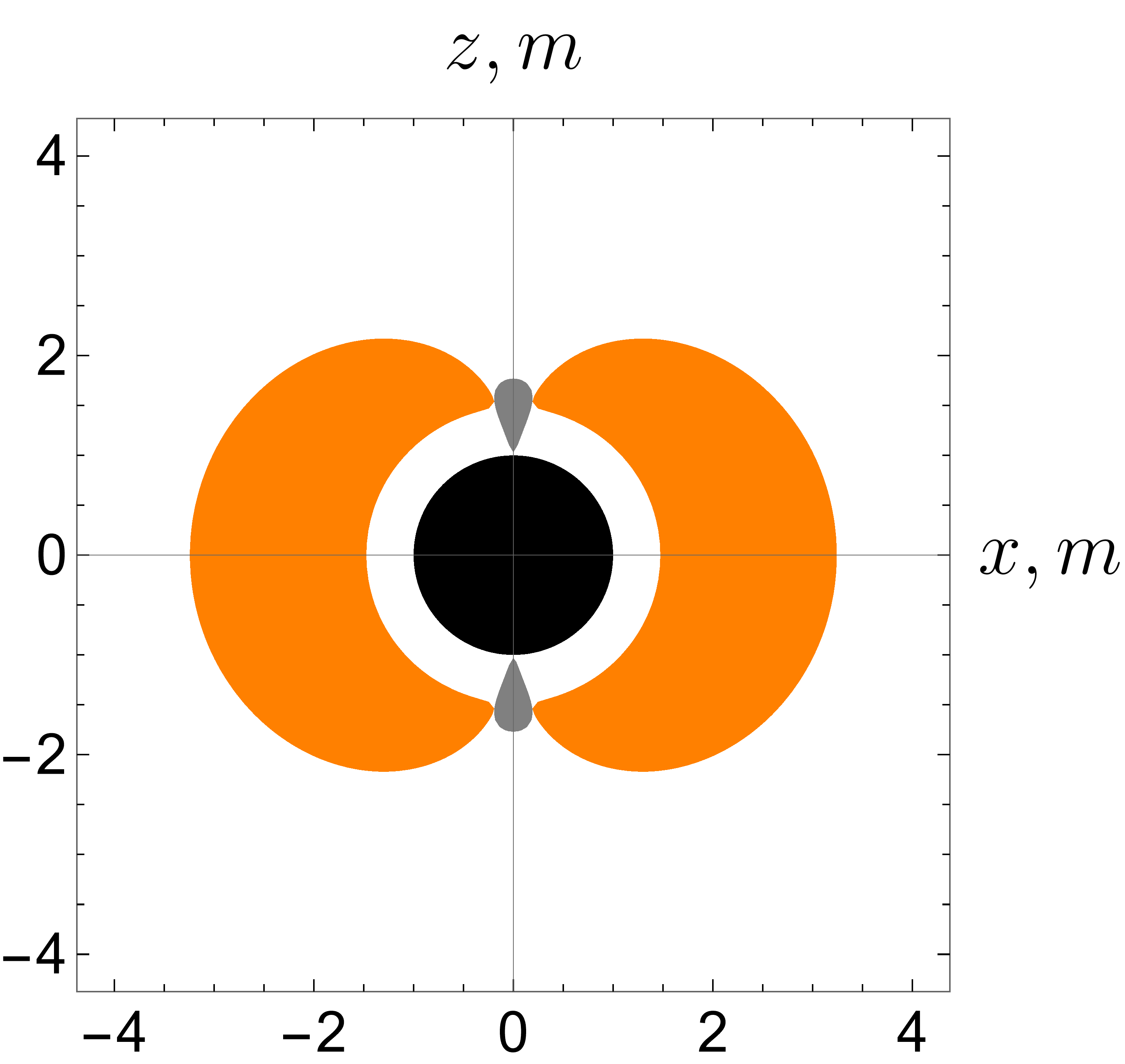} \\
    \includegraphics[width=6.05cm]{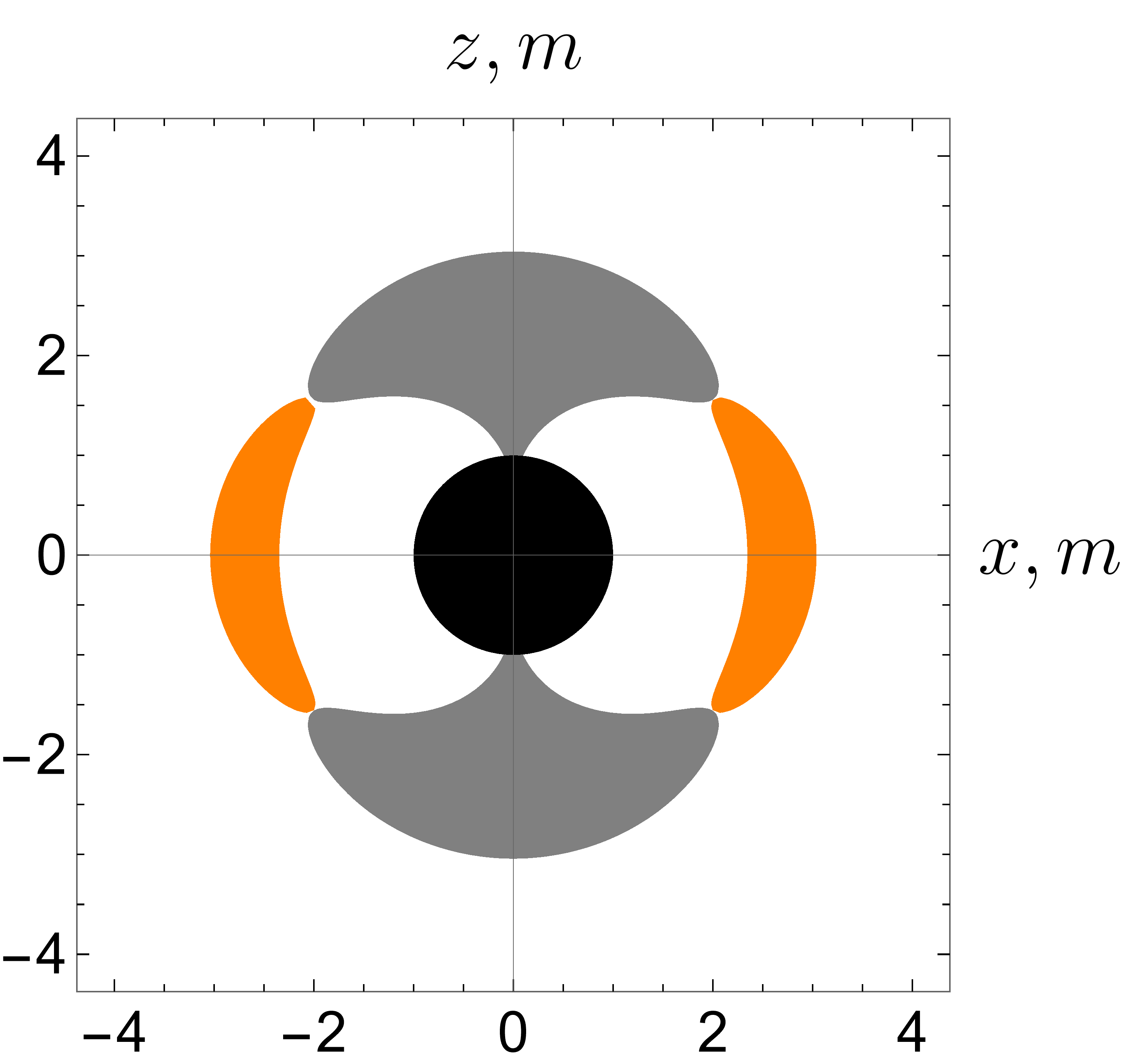}
    \includegraphics[width=6.05cm]{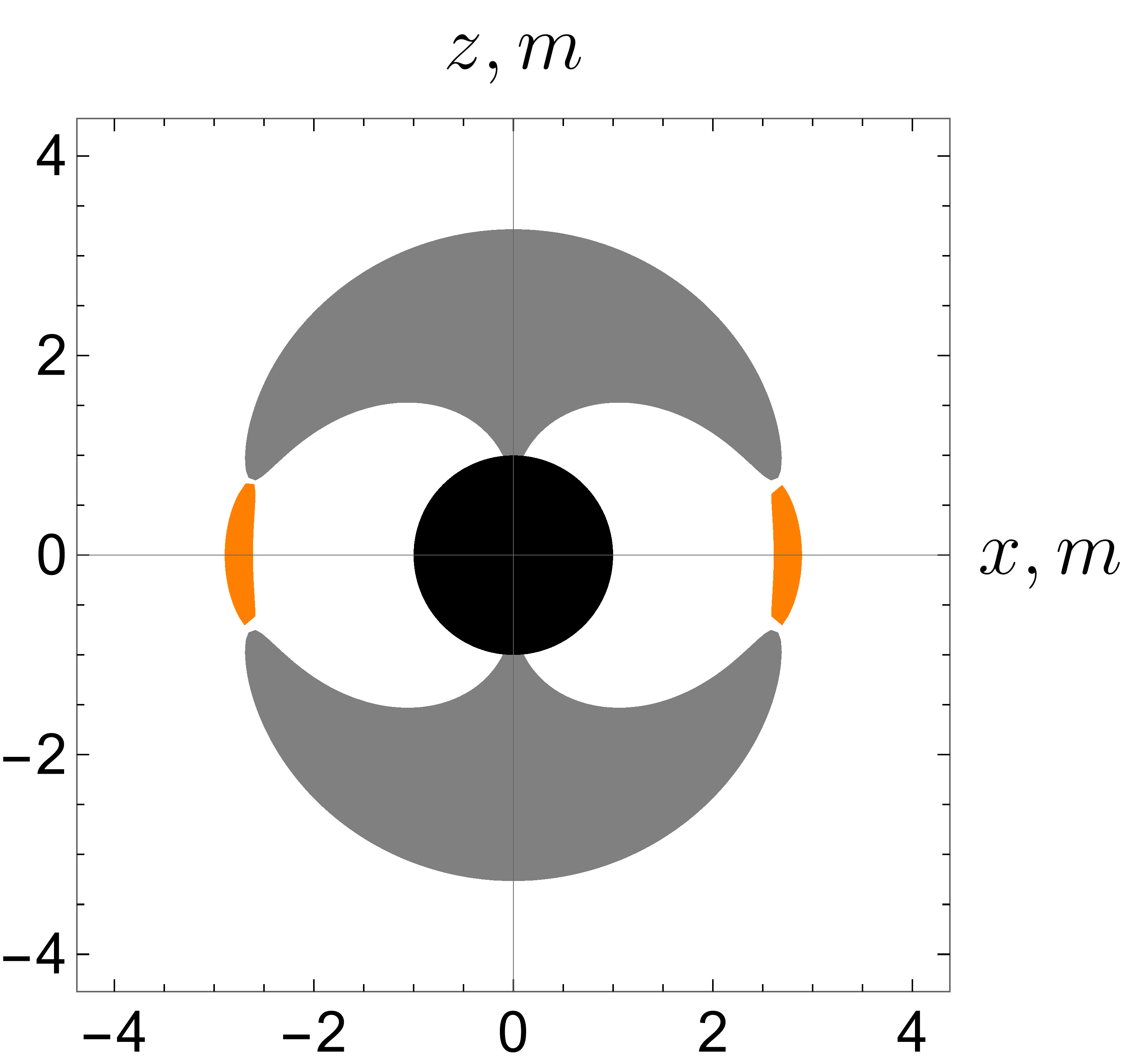}
\end{tabular}
    \caption{WH1 spacetime with plasma distribution $f_r = \omega{_c}{^2} \sqrt{m{^3}r}$, $f_{\theta} =0$, and spin parameter $a=0.999m$. The top-left panel corresponds to the frequency ratio $\omega{_c}/ \omega{_0} = 1$; the-top right to $\omega{_c}/ \omega{_0} = 2.7$, the bottom-left to $\omega{_c}/ \omega{_0} = 3.2$ and bottom-right has frequency $\omega{_c}/ \omega{_0} = 3.3$. The photon region is shown in orange, while the forbidden zone is indicated in grey. The forbidden region forms initially at the poles  and subsequently expands toward the equatorial plane as the plasma frequency increases.}
\end{figure}

\begin{figure}[H]
\centering
    \begin{tabular}{ cc}
\includegraphics[height=6.05cm]{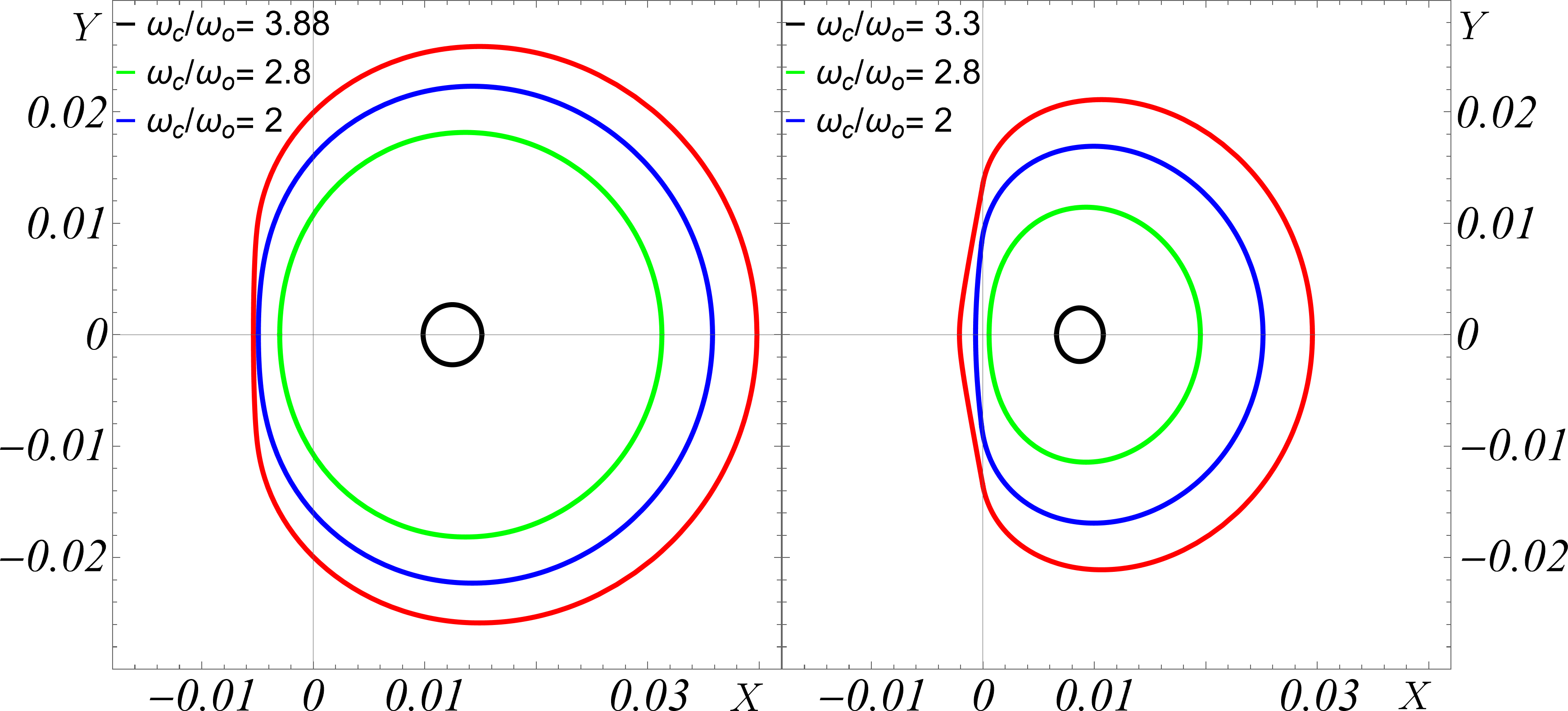}
         \end{tabular}
         \caption{Shadow of the Kerr black hole (left) compared to the shadow of WH1 spacetime (right) for plasma distribution $f_r = \omega{_c}{^2} \sqrt{m{^3}r}$, $f_{\theta}=0$, and spin parameter $a=0.999m$.} The observer is located at $r_O=200m$ and inclination angle $\theta_O=\pi/2$.
\end{figure}

\begin{figure}[t!]
\centering
    \begin{tabular}{ cc}
    \includegraphics[width=6.05cm]{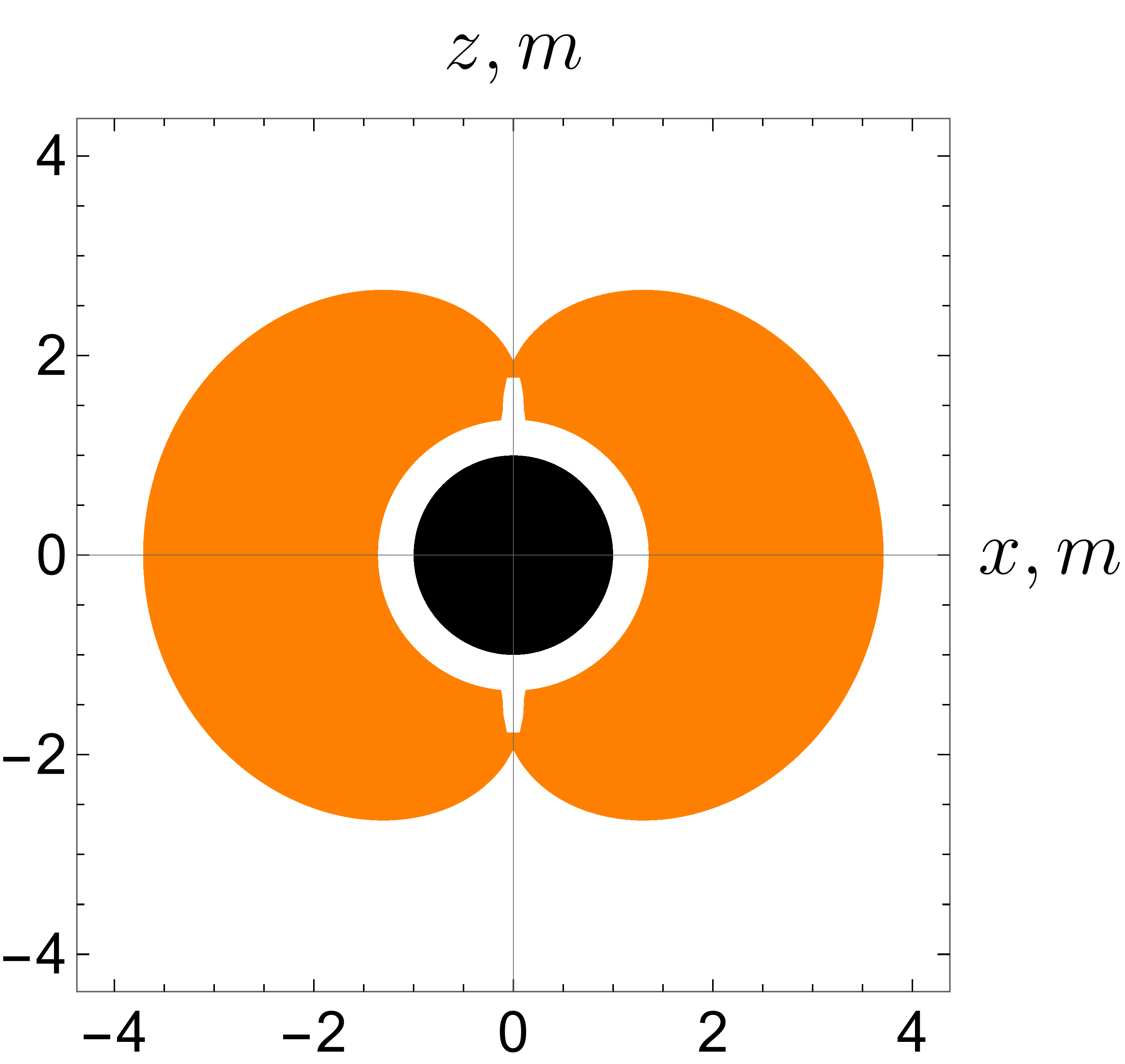}
    \includegraphics[width=6.05cm]{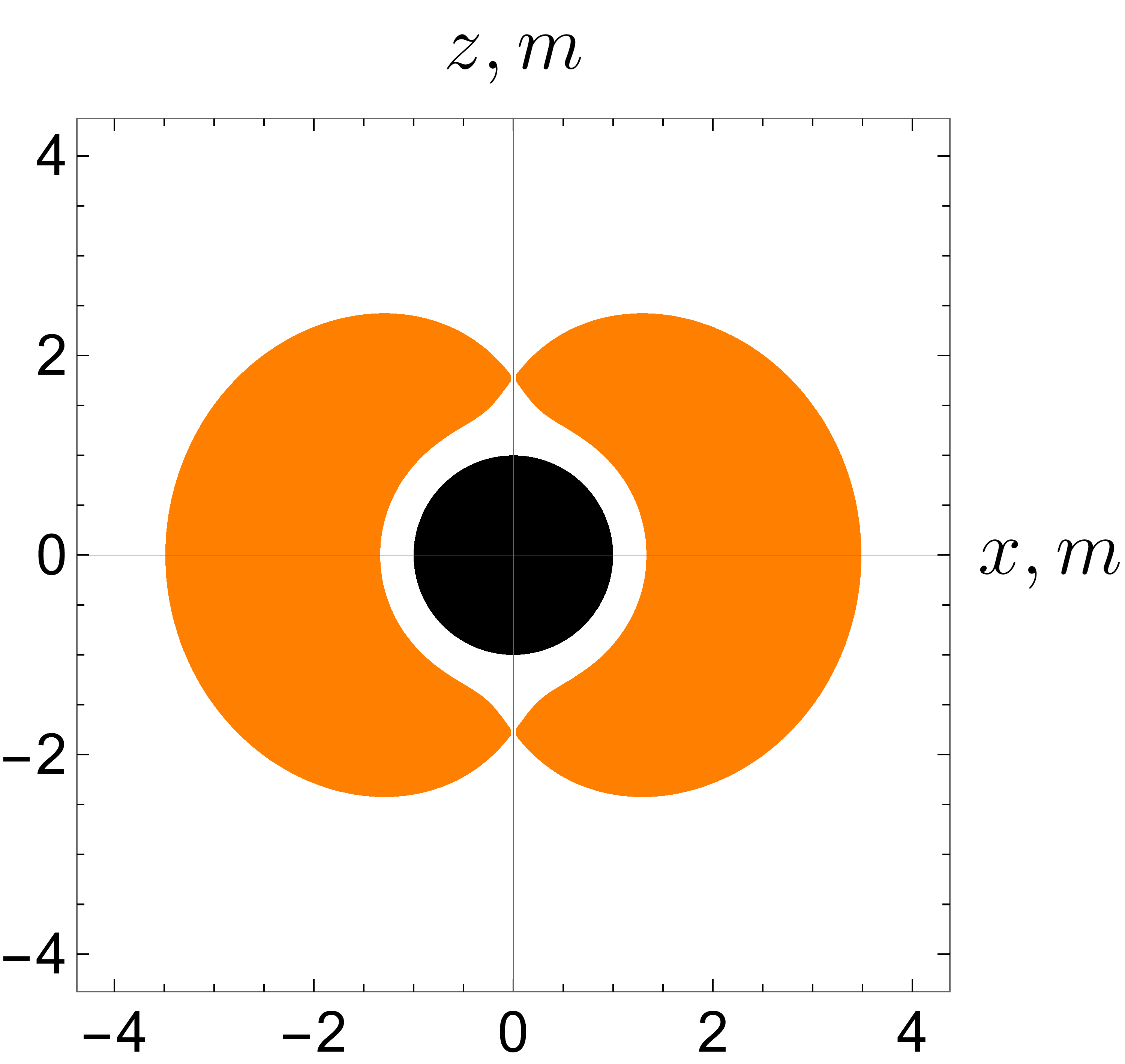} \\
    \includegraphics[width=6.05cm]{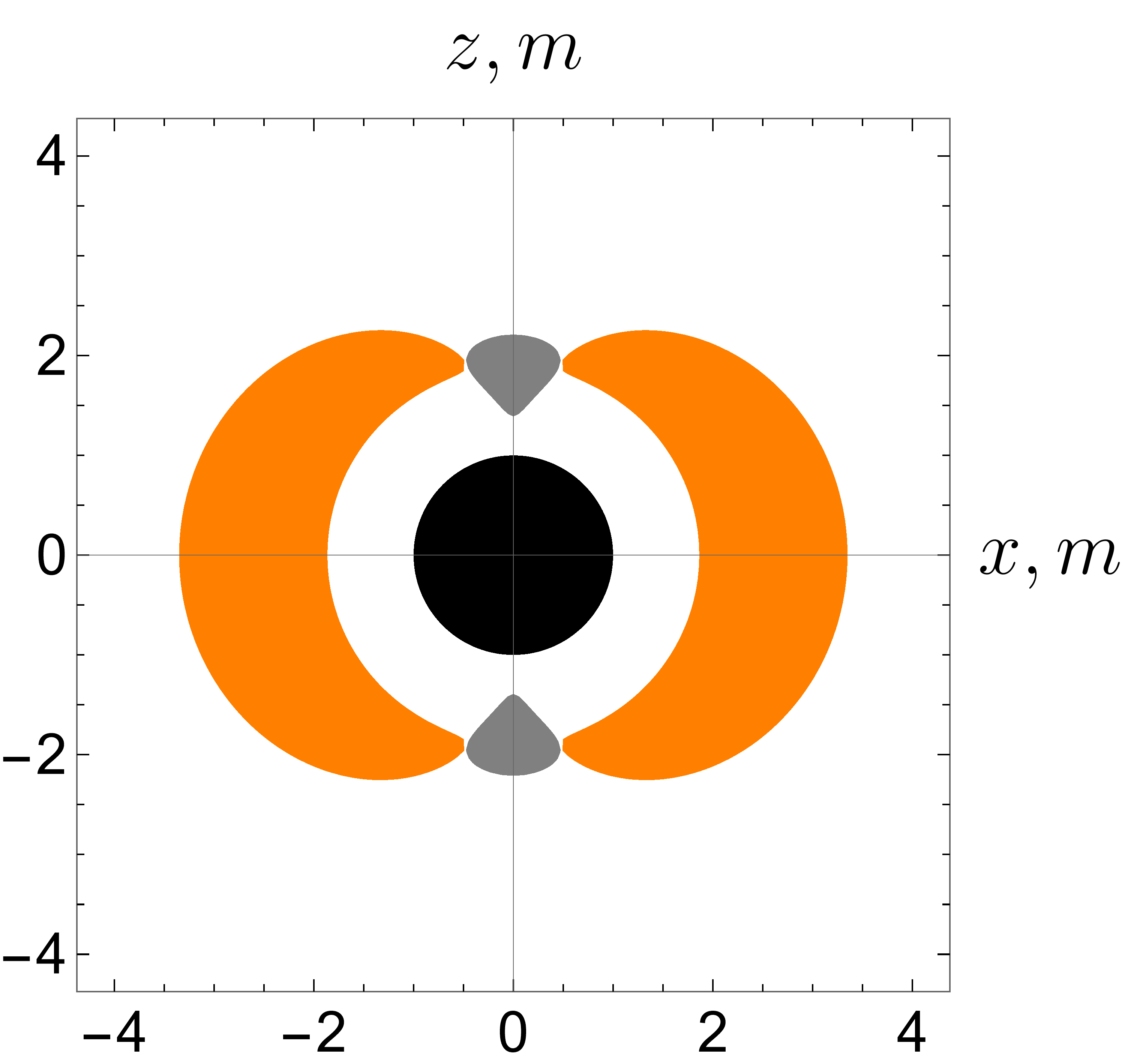}
    \includegraphics[width=6.05cm]{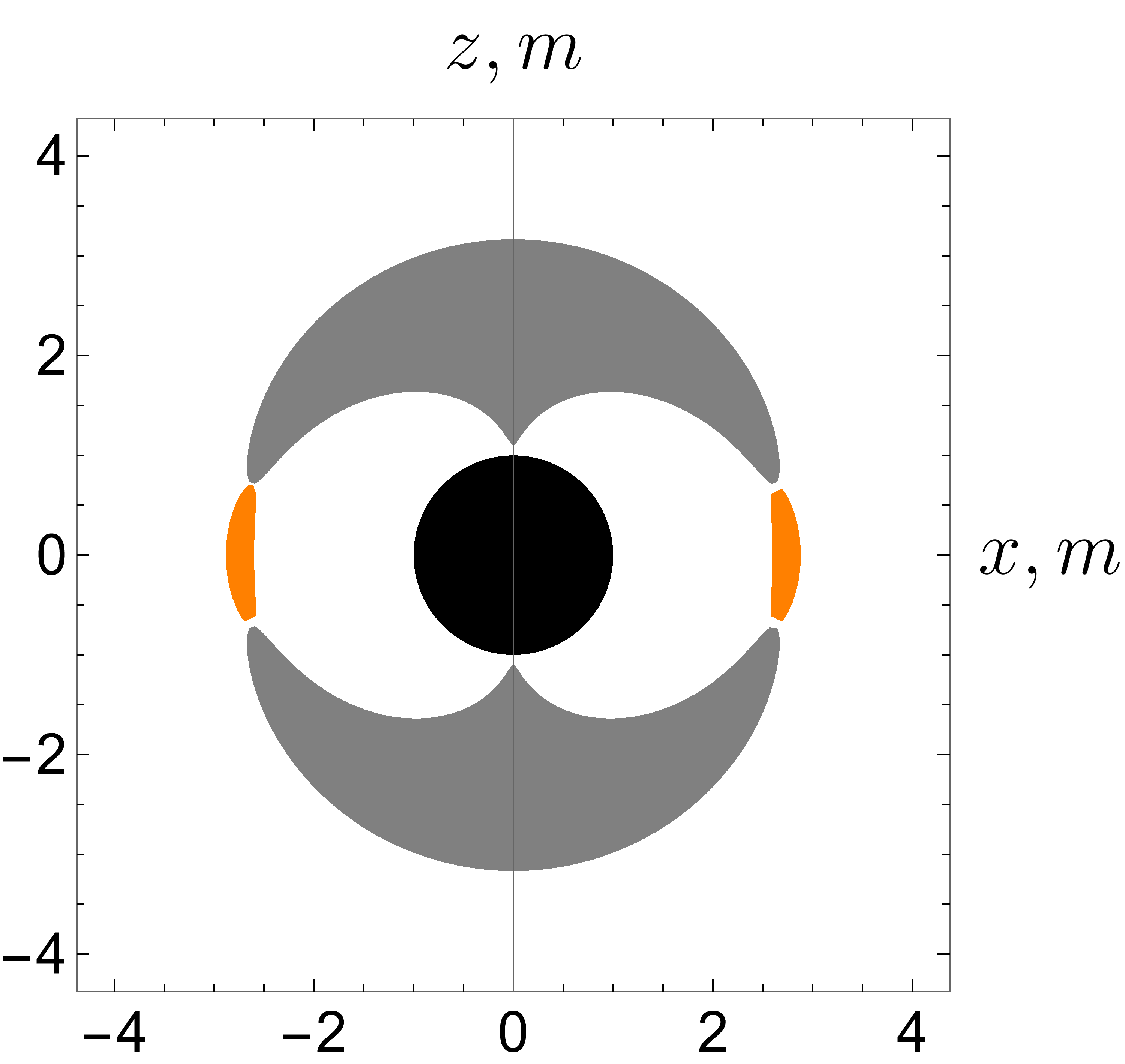} 
\end{tabular}
    \caption{WH2 spacetime with plasma distribution $f_r = 4\omega{_c}{^2}m{^2}\sqrt{m/r}$, $f_{\theta} =0$,  and spin parameters $a=0.999m$. The top-left corresponds to the frequency ratio $\omega{_c}/ \omega{_0} = 1$; the top-right to $\omega{_c}/ \omega{_0} = 2.2$, the bottom-left to $\omega{_c}/ \omega{_0} = 2.6$ and the bottom-right to $\omega{_c}/ \omega{_0} = 3.2$. The forbidden zone forms first near the poles and gradually expands toward the equatorial plane as the plasma frequency increases.}
\end{figure}

\begin{figure}[H]
\centering
    \begin{tabular}{ cc}
\includegraphics[height=6.05cm]{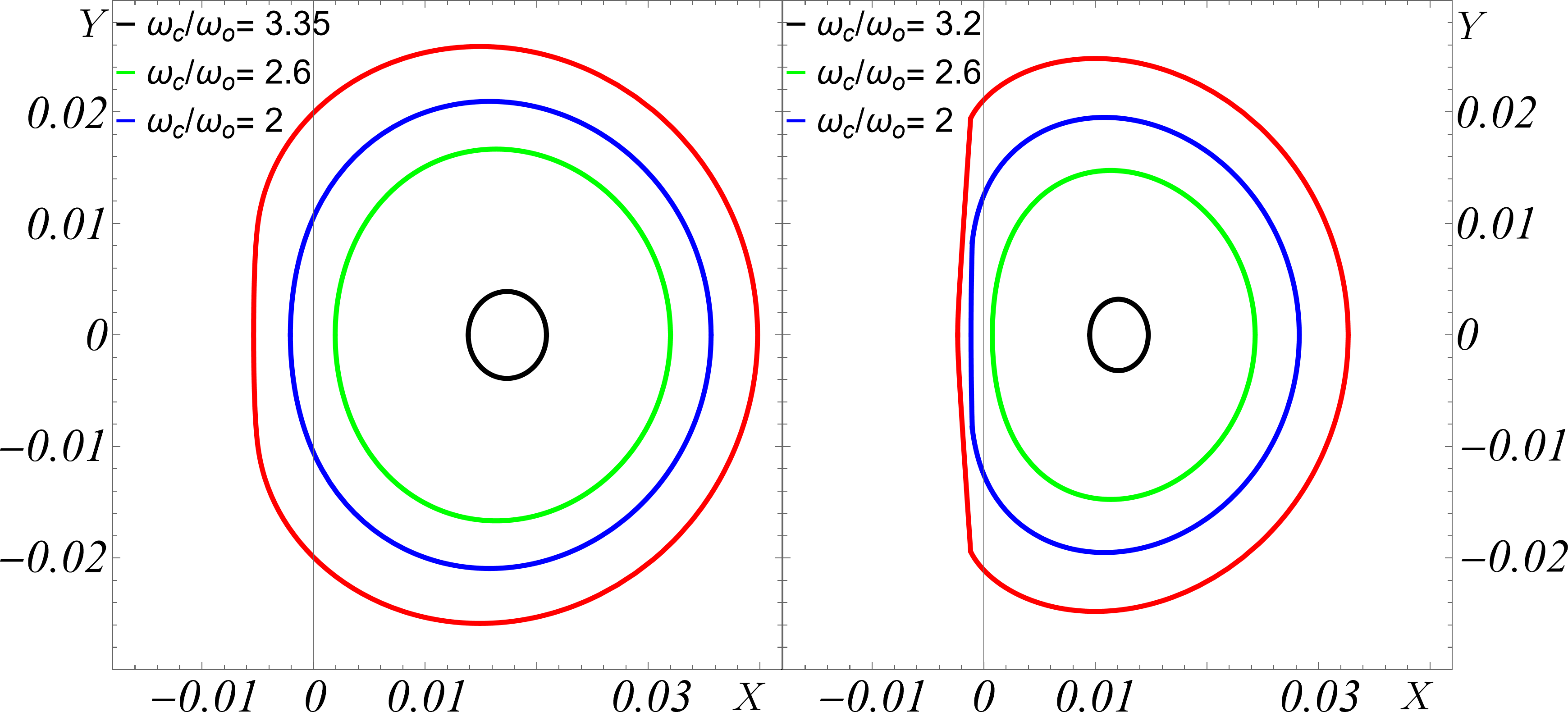}
         \end{tabular}         
    \caption{ Shadow of the Kerr black hole (left) compared to the shadow of the WH2 spacetime (right) for plasma distribution $f_r = 4\omega{_c}{^2}m{^2}\sqrt{m/r}$, $f_{\theta} =0$, and spin parameter $a=0.999m$.} The observer is located at $r_O=200m$ and inclination angle $\theta_O=\pi/2$.
\end{figure}

In Fig. 2 we present the wormhole shadow corresponding to this configuration in comparison with the shadow of the Kerr black hole with the same plasma distribution.  The wormhole shadow is consistently smaller than that of the Kerr black hole. As the frequency ratio $\omega{_c}/ \omega{_0}$ increases, the difference in size between the shadows of the two compact objects becomes more pronounced. The critical plasma frequency  beyond which the shadow ceases to be observable is $\omega{_c}^{crit}/ \omega{_0} \approx 3.32$ for the wormhole which is smaller than the critical plasma frequency $\omega{_c}^{crit}/ \omega{_0} \approx 3.9$ for the corresponding Kerr black hole. With increasing the plasma frequency $\omega{_c}/ \omega{_0}$, the influence of the inner family of unstable spherical orbits on the wormhole shadow diminishes, and finally the shadow of the wormhole is entirely determined by the outer family of unstable spherical orbits.   

In Fig. 3 we depict the photon region and the forbidden region for the configuration of a WH2 spacetime with the plasma model described by Eq. \eqref{Perlickplasma}. We see that the forbidden region emerges again near the poles and progressively extends toward the equatorial plane as the plasma frequency increases.  The corresponding shadow, displayed in Fig. 4, remains smaller than that of the Kerr black hole, with the size difference becoming increasingly evident for higher plasma frequencies. The critical plasma frequency for the wormhole when the shadow becomes no longer observable  is $\omega{_c}^{crit}/ \omega_0\approx 3.22$, which is lower than the corresponding Kerr value $\omega{_c}^{crit}/ \omega{_0} \approx 3.38$.

The next case  which we present is a WH1 type of spacetime  embedded in the plasma model described by Eq.~\eqref{thetaplasma}. The structure of the photon and forbidden regions is displayed in Fig.~5. The first forbidden zone forms along the equatorial plane, and as the plasma frequency increases further, a second, separate forbidden region develops near the poles at $\omega{_c}/ \omega{_0} \approx 2.88$. For higher values of the plasma frequency the two disjoint forbidden regions grow until they merge at some intermediate value of the polar angle at frequency ratio $\omega{_c}/ \omega{_0} \approx 3.05$. Forbidden regions which emerge initially at the equatorial plane are also observed for the Kerr and Kerr-Newman black holes with the same plasma distribution \cite{Perlick:2017}, \cite{Briozzo:2023}. In this case we should note that the critical plasma frequency for which the shadow can no longer be observed by an equatorial observer does not coincide with the plasma frequency for which the forbidden region encompasses the full parameter space for the spherical photon orbits. The critical frequency for an equatorial observer is determined by the value when an equatorial forbidden region appears. Beyond this value no shadow exists for an equatorial observer but an observer located at some intermediate inclination angle would still see a shadow. This phenomenon is visualized in further details in Figs. 10-11.

The wormhole shadow in this case is shown in Fig.~6. We see that it remains smaller than that of the Kerr black hole for all the plasma frequencies, and the difference in their apparent sizes becomes more noticeable with higher plasma frequencies. For the wormhole the critical plasma frequency is $\omega{_c}^{crit}/ \omega{_0} \approx 2.44$ which is lower than the value $\omega{_c}^{crit}/ \omega{_0} \approx 3$ obtained for the Kerr black hole.


\begin{figure}[t!]
\centering
    \begin{tabular}{ cc}
    \includegraphics[width=6.05cm]{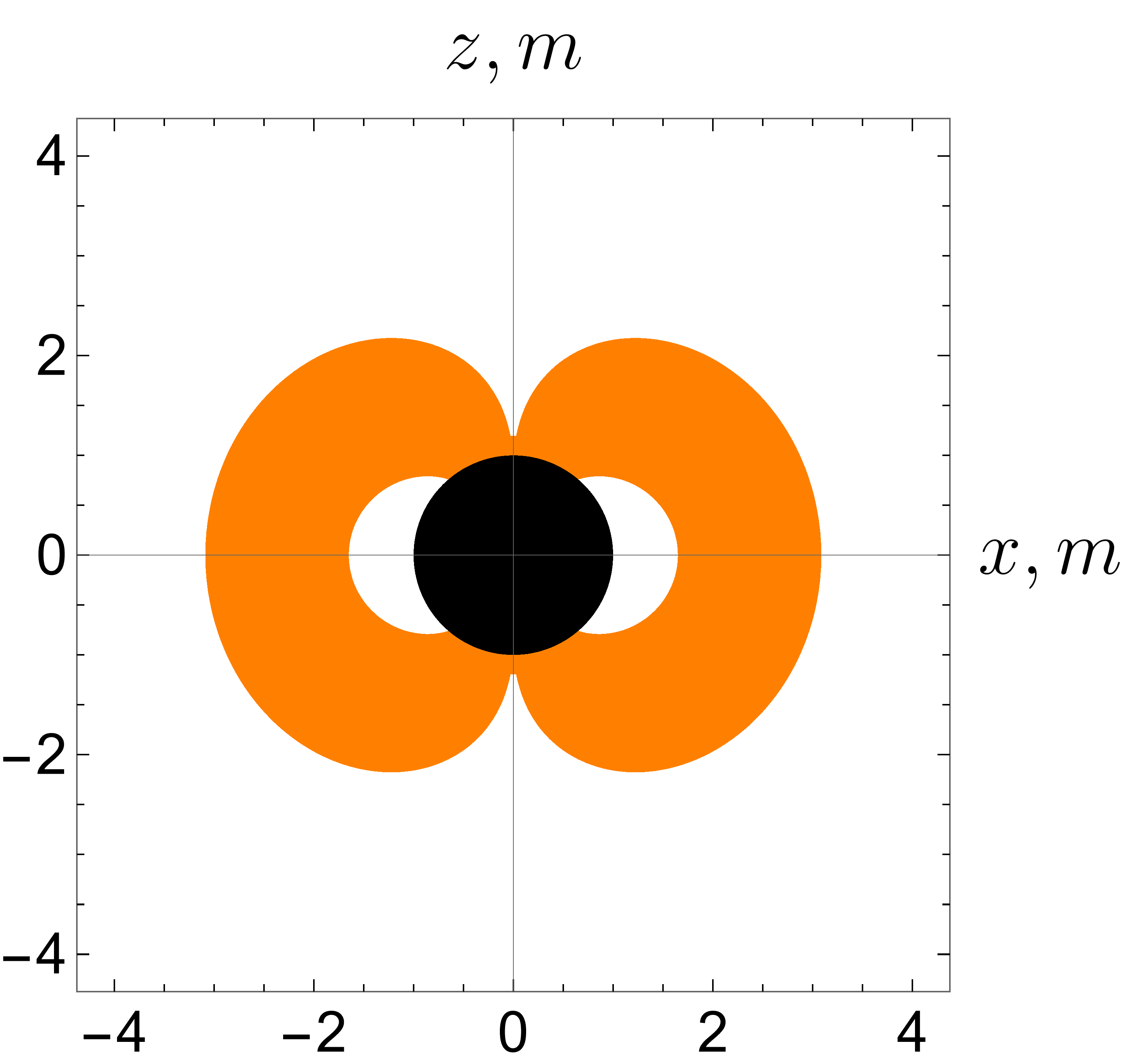}
    \includegraphics[width=6.05cm]{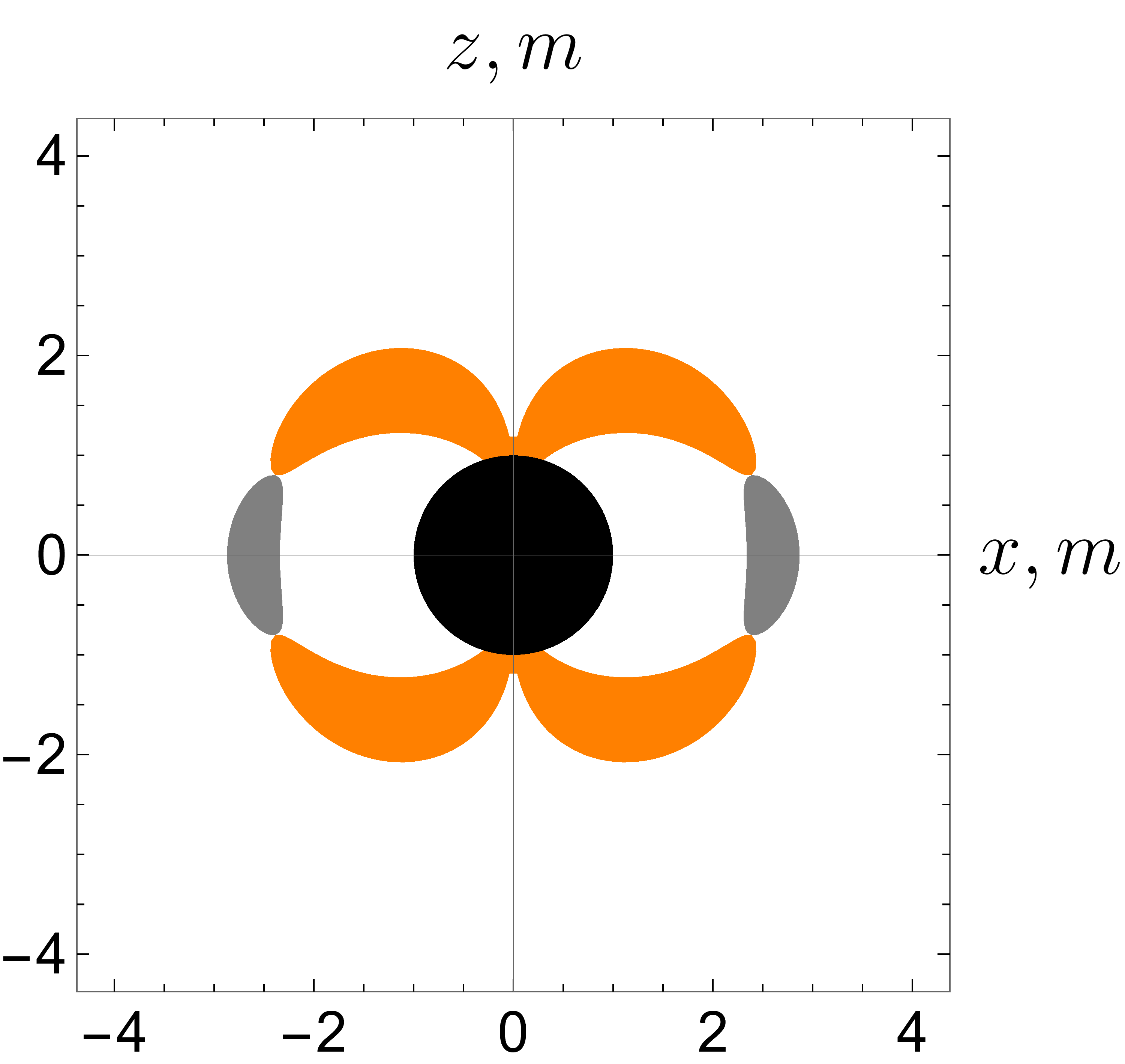} \\
     \includegraphics[width=6.05cm]{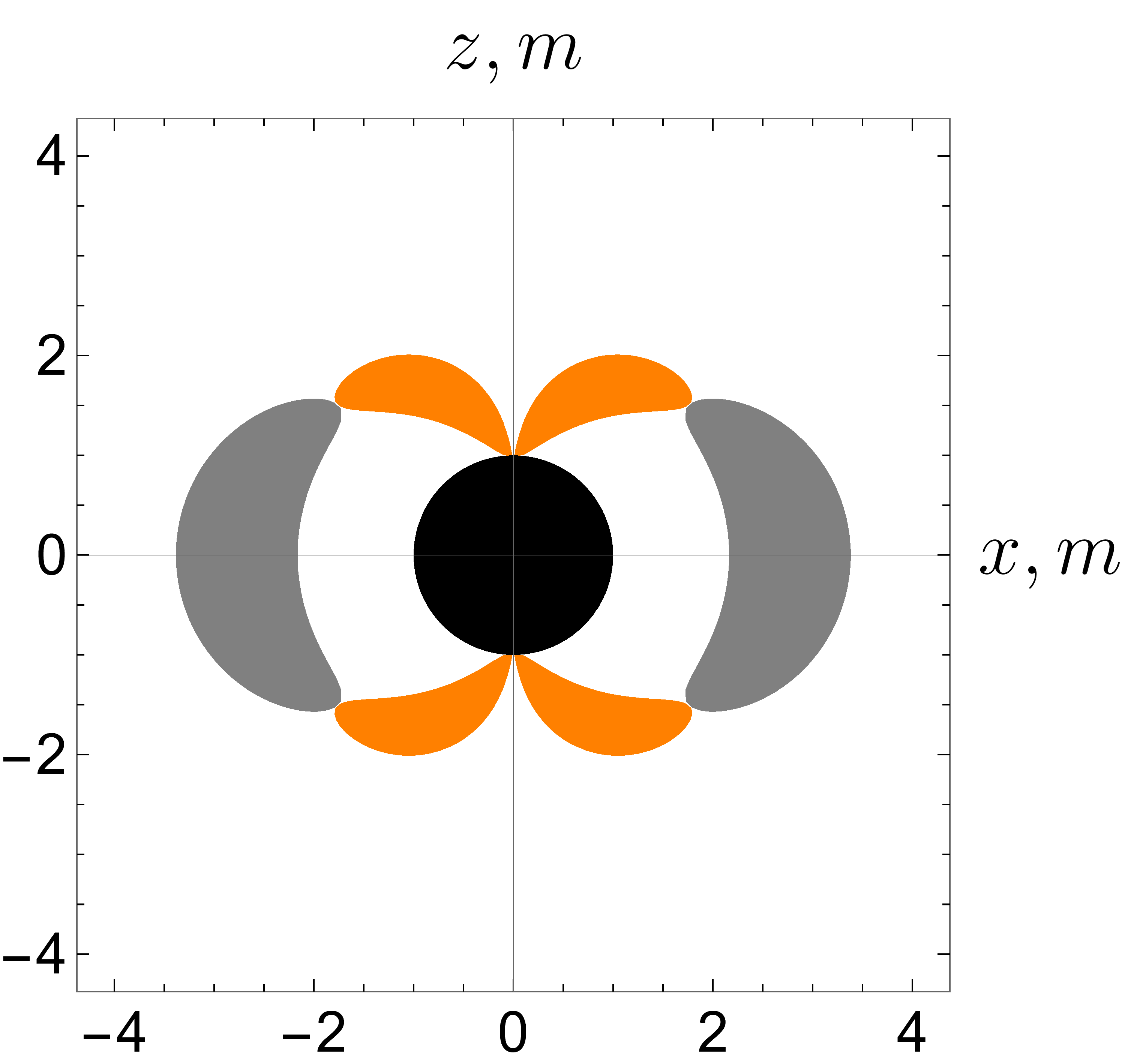}
     \includegraphics[width=6.05cm]{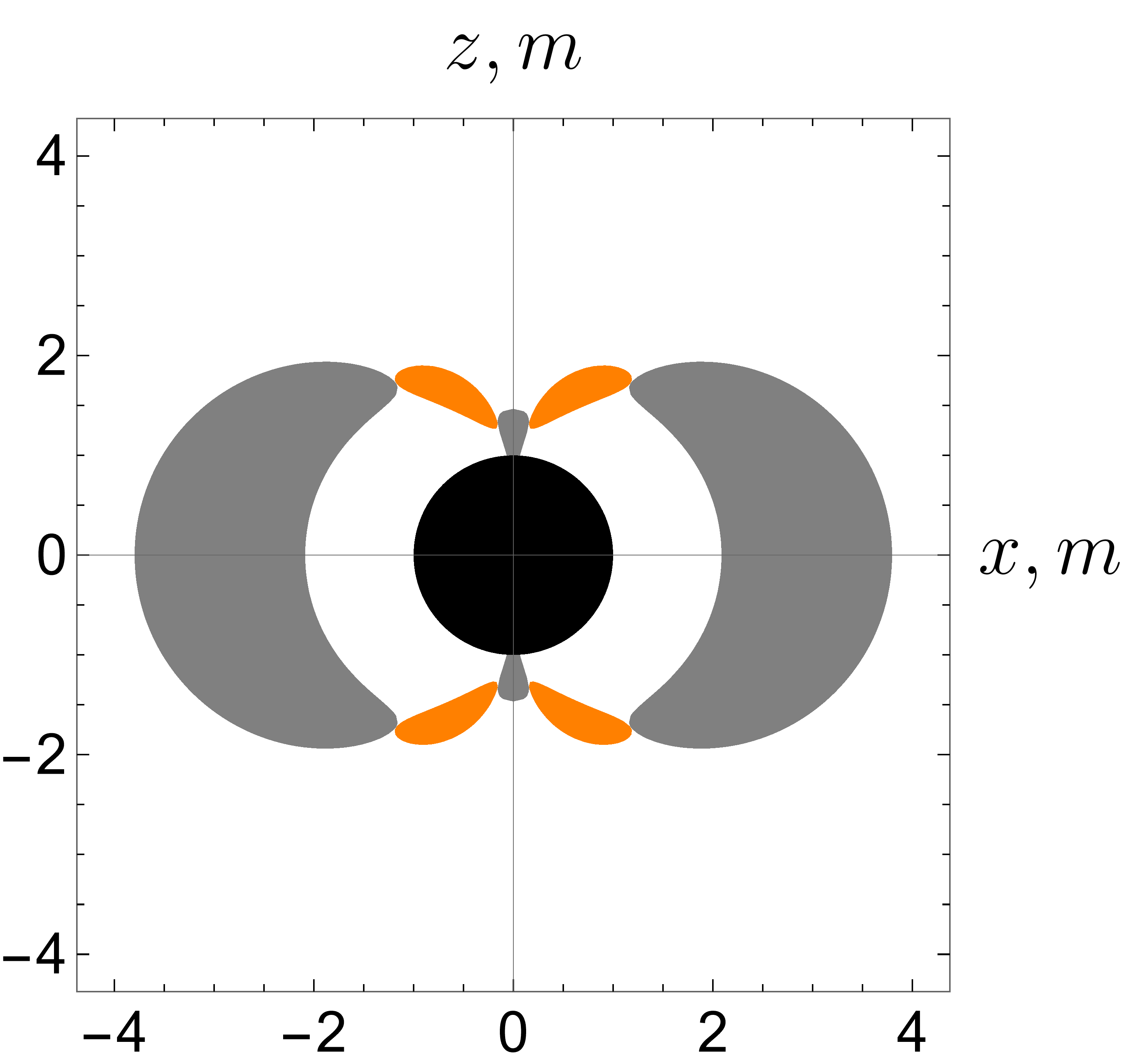}
     \end{tabular}
    \caption{WH1 spacetime with plasma distribution  $f_\theta = \omega{_c}{^2}m{^2}(1+2\sin{^2}\theta)$, $f_r =0$, and spin parameter $a=0.999m$.} The top-left figure corresponds to frequency ratio $\omega{_c}/ \omega{_0} = 2$; the top-right to $\omega{_c}/ \omega{_0} = 2.5$; the bottom-left to $\omega{_c}/ \omega{_0} = 2.7$; and the bottom right to $\omega{_c}/ \omega{_0} = 2.9$. The photon region is shown in orange, while the forbidden region is indicated in grey. The forbidden region first forms along the equatorial plane and expands towards the poles as the plasma frequency increases.
\end{figure}

\begin{figure}[H]
\centering
    \begin{tabular}{ cc}
\includegraphics[height=6.05cm]{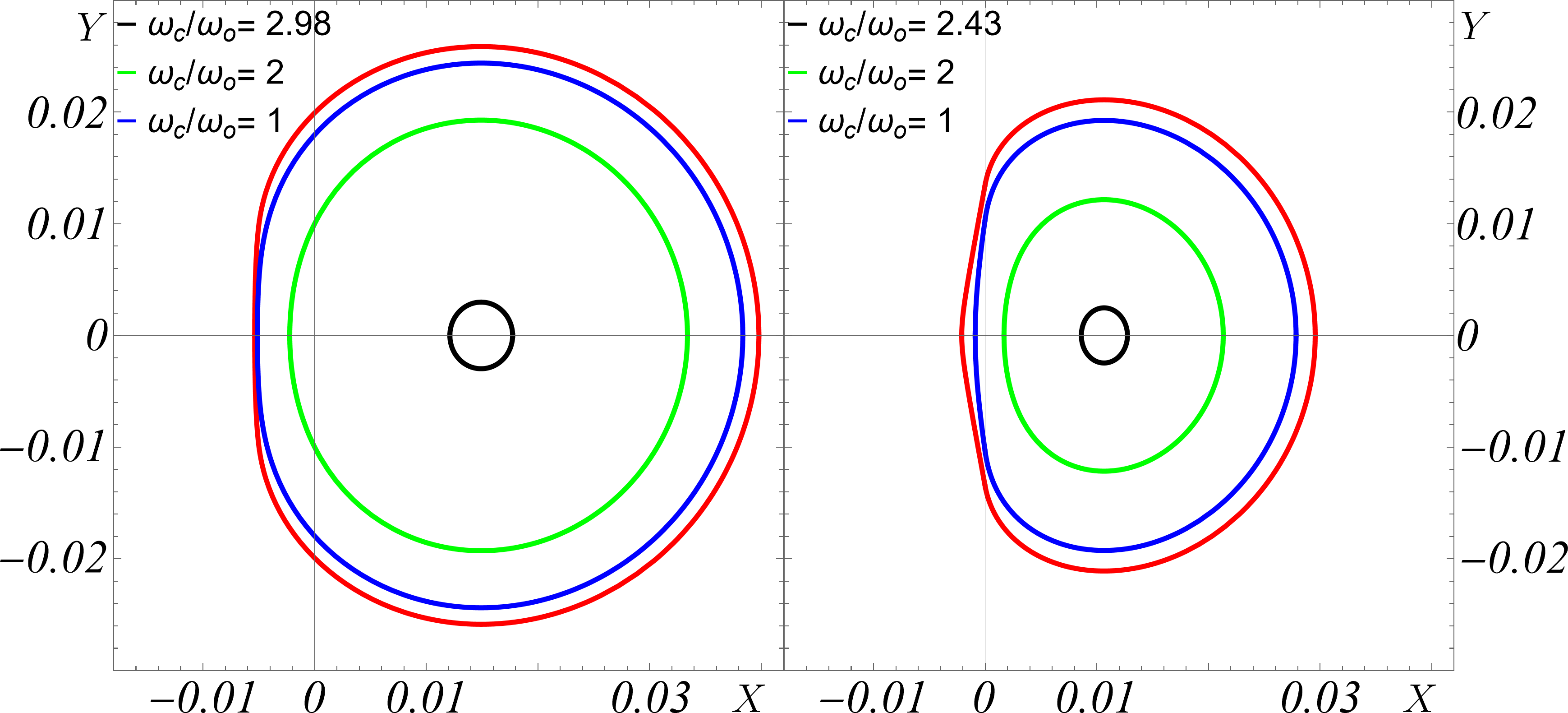}
         \end{tabular}         
    \caption{Shadow of the Kerr black hole (left) compared to the shadow of the WH1 spacetime (right) for plasma distribution $f_\theta = \omega{_c}{^2}m{^2}(1+2\sin{^2}\theta)$, $f_r = 0$, and spin parameter $a=0.999m$.  The observer is located at $r_O=200m$ and inclination angle $\theta_O=\pi/2$.}
\end{figure}

\begin{figure}[t!]
\centering
    \begin{tabular}{ cc}
    \includegraphics[width=6.05cm]{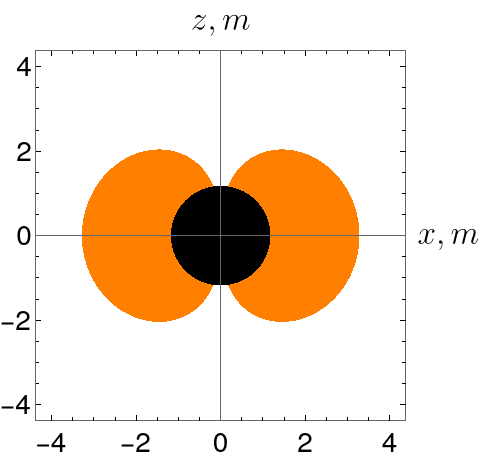}
    \includegraphics[width=6.05cm]{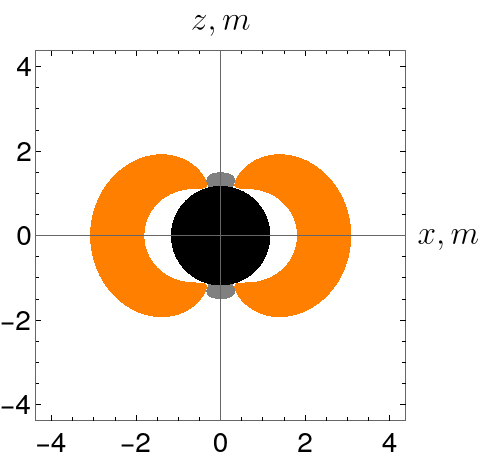} \\
     \includegraphics[width=6.05cm]{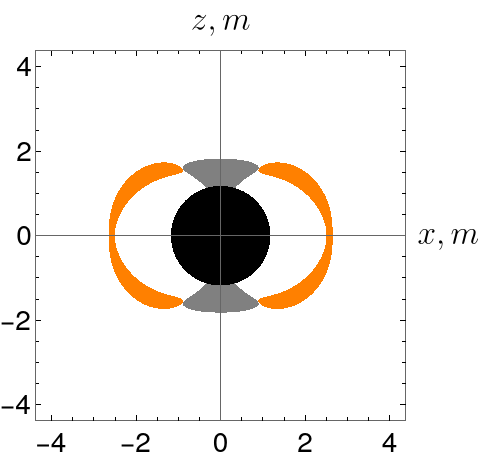} 
     \includegraphics[width=6.05cm]{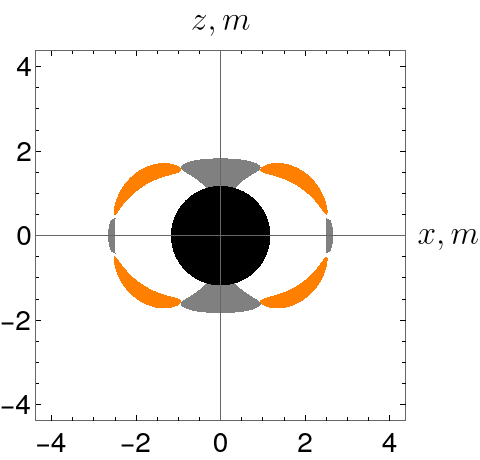}
     \end{tabular}
    \caption{WH3 spacetime with plasma distribution  $f_\theta = \omega{_c}{^2}m{^2}(1+2\sin{^2}\theta)$, $f_r=0$, and spin parameter $a=0.999m$.} The top-left figure corresponds to frequency $\omega{_c}/ \omega{_0} = 1$; the top-right figure to $\omega{_c}/ \omega{_0} = 1.5$, the bottom-left to $\omega{_c}/ \omega{_0} = 1.82$ and the bottom-right to $\omega{_c}/ \omega{_0} =1.83$. The photon region is shown in orange, while the forbidden region is indicated in grey. The forbidden region starts near the poles, while another forms at the equator, causing the shadow to vanish.
\end{figure}

\begin{figure}[H]
    \centering
    \includegraphics[height=6.05cm]{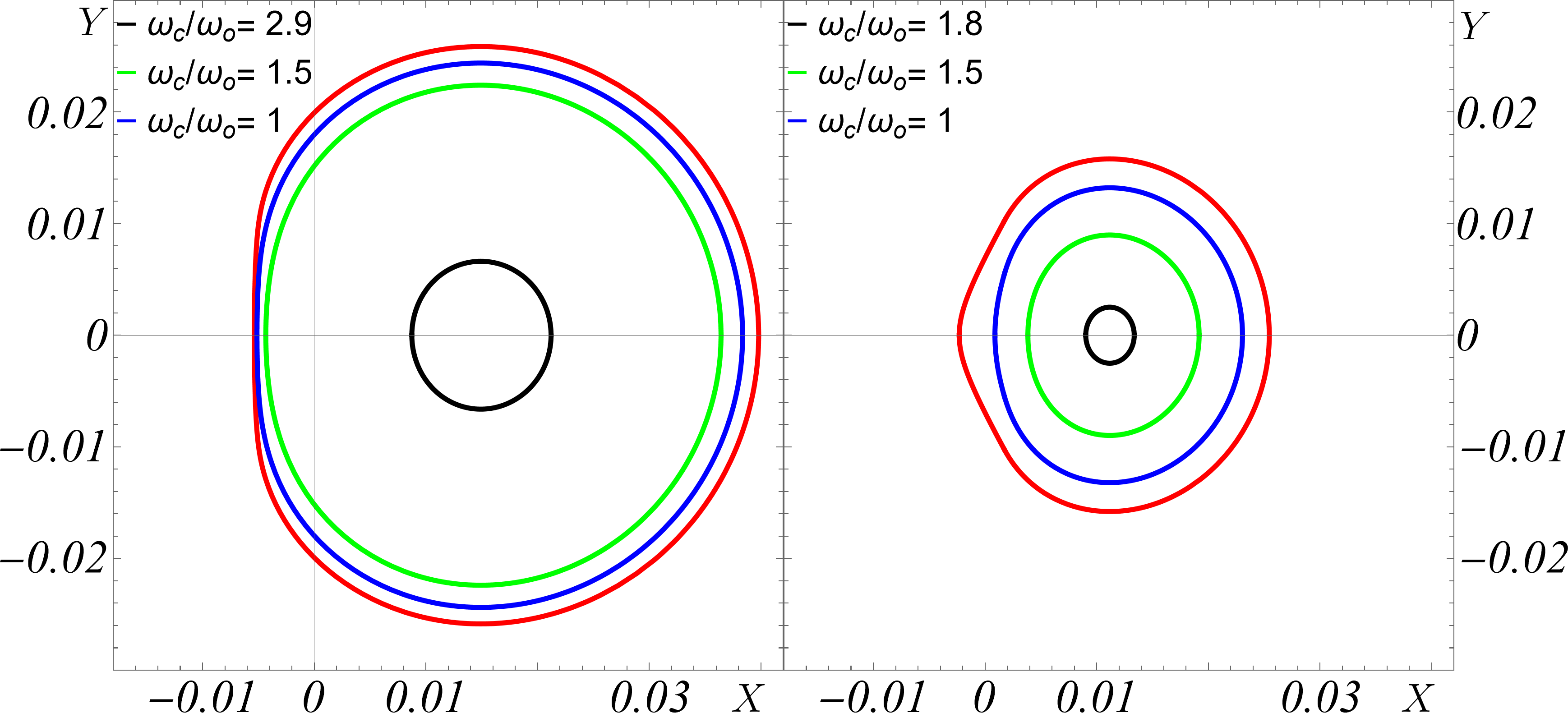}
    \caption{Shadow of the Kerr black hole (left) compared to the shadow of the WH3 spacetime (right) for plasma distribution $f_\theta = \omega{_c}{^2}m{^2}(1+2\sin{^2}\theta)$, $f_r= 0$, and spin parameter $a=0.999m$.  The observer is located at $r_O=200m$ and inclination angle $\theta_O=\pi/2$.}
    \label{fig:enter-label}
\end{figure}

The last configuration which we will present is a WH3 type of spacetime with plasma distribution given by Eq.~\eqref{thetaplasma}. In Fig. 7 we illustrate the photon and forbidden regions. In this case, the forbidden zone first emerges near the poles. However, with increasing plasma frequency, an additional forbidden region develops on the equatorial plane at $\omega{_c}/ \omega{_0} \approx 1.82$. The two photon regions grow for higher frequencies and finally merge at some intermediate polar angle. As in the previous case the shadow disappears for an equatorial observer for the critical plasma frequency when the equatorial forbidden region emerges. The shadow of the wormhole, depicted in Fig. 8, consistently remains smaller than that of the Kerr black hole, and the contrast in size becomes more pronounced as the plasma frequency grows. The critical plasma frequency for the wormhole is $\omega{_c}^{crit}/ \omega{_0} \approx 1.82$, which is lower than the corresponding value inherent to Kerr black hole. 

The remaining combinations of wormhole spacetimes and plasma distributions from our sample are presented in the appendix while list of the critical plasma frequencies is provided in Table 1. We observe the following trends as a result of our analysis. For the plasma distributions which depend only on the radial coordinate, i.e. given by Eqs. ($\ref{sqrtrplasma}$)-($\ref{Perlickplasma}$), the evolution of the photon region and the forbidden region depends primarily on the form of the plasma surrounding the wormhole rather than the spacetime metric itself. Consequently, we observe similar qualitative behavior for all the wormhole metrics which is further consistent with the behavior of the Kerr black hole and its generalizations in the modified theories of gravity studied in \cite{Perlick:2017}, \cite{Briozzo:2023}.

On the contrary, for the plasma distribution given by Eq. ($\ref{thetaplasma}$) which depends also on the polar angle $\theta$  we observe dependence on the spacetime metric. In Fig. 5 and Fig. 7 we demonstrated the evolution of the photon region for the WH1 and WH3 types of spacetimes while the case of WH2 metric is presented in Fig. 17 in the appendix. In this configuration the forbidden region initially appears on the equatorial plane and as we steadily increase the plasma frequency, it expands towards the poles. Approximately at the plasma frequency $\omega{_c}/ \omega{_0} \approx 4.24 $, the forbidden regions encompasses the wormhole, and the wormhole shadow can no longer be seen by any observers.  All the wormhole spacetimes possess qualitatively different evolution of the photon region with the variation of the plasma frequency despite the common model for the plasma distribution. This demonstrates further that the shadow formation in some wormhole spacetimes surrounded by appropriate plasma distributions may deviate qualitatively from the Kerr black hole allowing for observational signatures.

In Fig. 9 we present the shadows of all the wormhole configurations from our sample compared to the shadow of the Kerr black hole surrounded by the same plasma distributions. Based on the results of  our analysis we can make the following conclusions.Regardless of the plasma distribution the largest shadows correspond to the Kerr black hole followed by the shadows of the WH2, WH1 types of spacetimes, while the smallest shadows belong to the WH3 metric. 

\begin{figure}
    \centering
    \includegraphics[width=15.5cm]{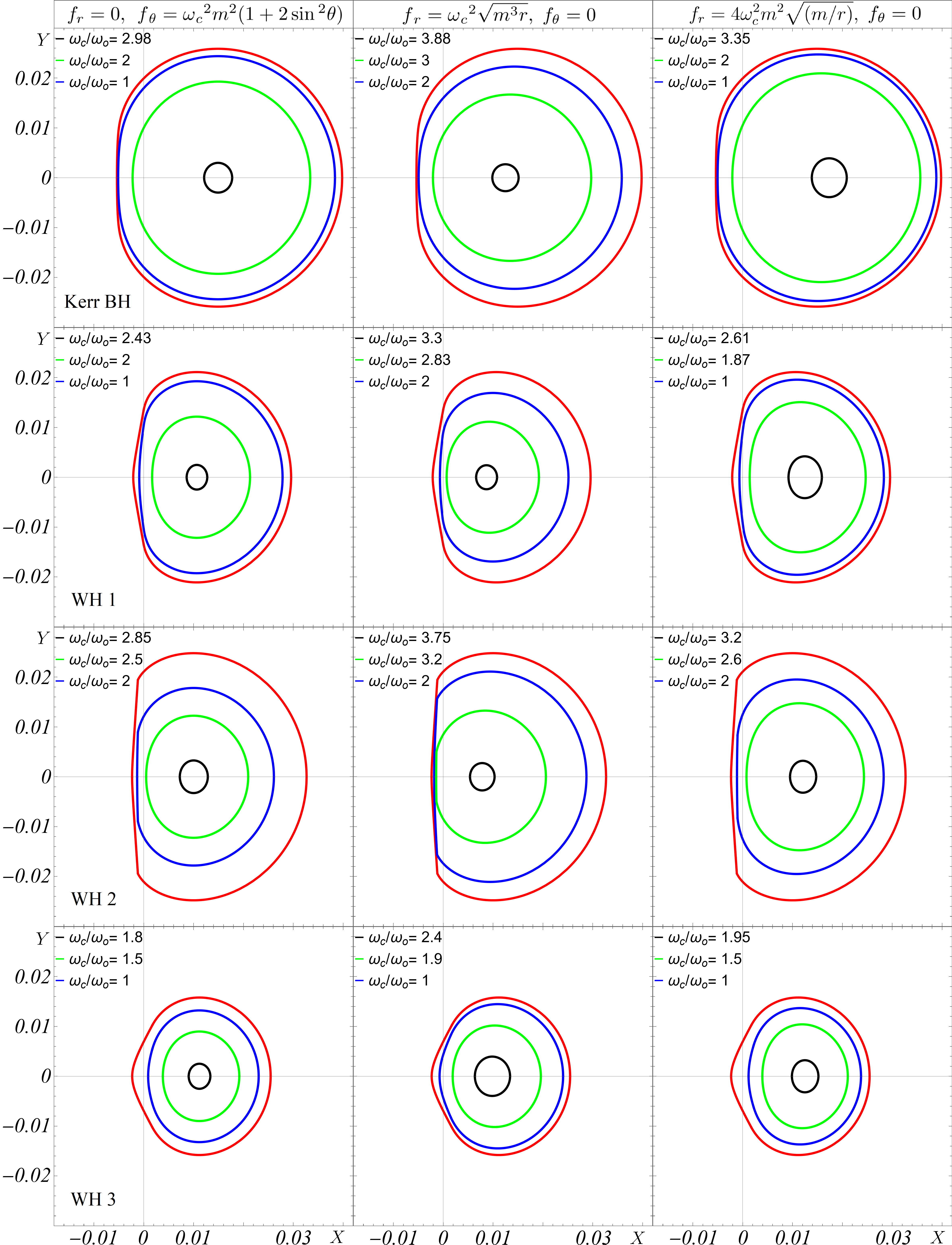}
    \caption{Comparison of the wormhole shadows from our sample and the shadow of the Kerr black hole for the different plasma distributions. In all the cases the observer is located at $r_O=200m$ and inclination angle  $\theta_O=\pi/2$, while the spin parameter is $a=0.999m$.} The red curves correspond to the vacuum case.
    \label{fig:enter-label}
\end{figure}

The range of the plasma frequencies of the matter surrounding the compact object for which a shadow exists for an equatorial observer depends both on the plasma distribution and the metric (see Table 1). However,  certain qualitative correlations can still be identified. The widest range of plasma frequencies  occurs for the plasma distribution given by Eq. \eqref{sqrtrplasma}, whereas the narrowest range is found for plasma distribution described by Eq. \eqref{thetaplasma} for all the analyzed metrics. 

Similar qualitative observations arise when the different metrics are  compared for a selected plasma distribution. The metrics can be ranked from the widest to the narrowest range of plasma frequencies   in the following order as the widest range corresponds to the Kerr black hole followed by the WH2 and WH1 spacetimes, while the smallest range occurs for the  WH3 metric. This is valid for for all the plasma distributions which we considered. 

\begin{figure}[t!]
    \centering
    \includegraphics[width=15.5cm]{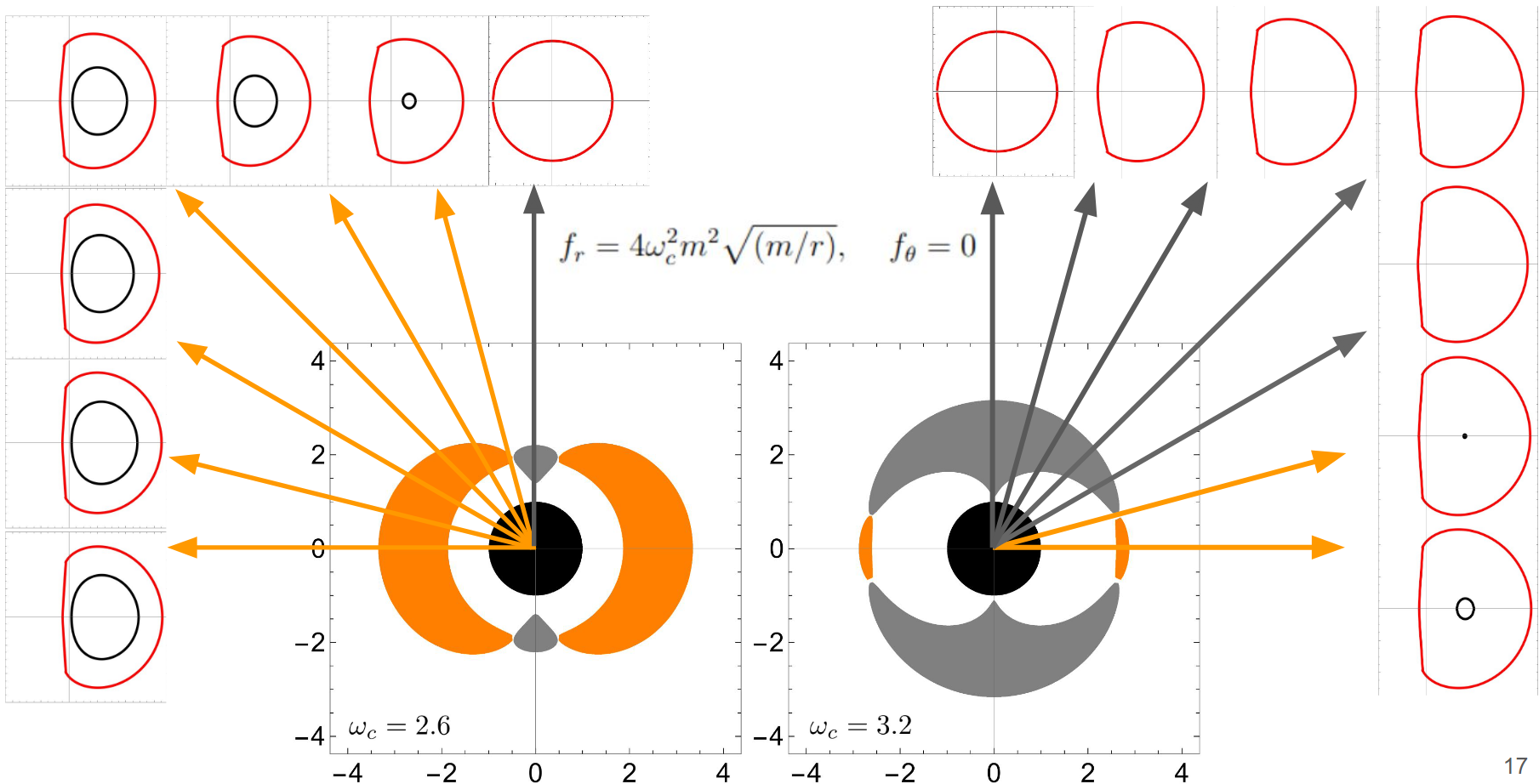}
    \caption{WH2 type of spacetime with plasma distribution $f_r=4\omega_c^2m^2\sqrt{(m/r)}$,$ f_\theta=0$, and spin parameter $a=0.999m$. We present the shadow boundary (in black) varying the inclination angle by $15^\circ$ and relate its deformation to the morphology of the photon region.} As the plasma frequency increases, observers at low inclinations can no longer see the shadow. The vacuum case is shown in red for comparison.
    \label{fig:enter-label}
\end{figure}

\begin{table}    \centering
    \begin{tabular}[t]{ |p{1.9cm} | p{4.5cm} |p{4.5cm}|p{4.5cm}|}
\hline
 & \hfil $f{_\theta}=\omega{_c}{^2}m{^2}(1+2\sin{^2}\theta)$  & \hfil $f{_r}=\omega{_c}{^2} \sqrt{m{^3}r}$ & \hfil $f{_r}=4\omega{_c}{^2} m{^2}\sqrt{m/r}$ \\[1.2ex]
 \hline
\hfil Kerr BH & \hfil 3 & \hfil3.9 & \hfil 3.38 \\
\hfil WH 1    & \hfil 2.44 & \hfil 3.32 & \hfil 2.66\\
\hfil WH 2    & \hfil 2.87 & \hfil 3.77 & \hfil 3.22 \\
\hfil WH 3    & \hfil 1.82 & \hfil 2.47 & \hfil 1.99\\
\hline
    \end{tabular}
     \caption{The critical frequency  $\omega{_c}^{crit}/\omega{_0}$ for the configurations of wormhole spacetimes and plasma distributions in our sample The critical frequency corresponds to the upper limit of the plasma frequency for which the shadow of the object remains visible for an equatorial observer.}
    \label{tab:my_label}
\end{table}

With the increase of the plasma frequency, the initial difference in size between the shadow of the Kerr BH and the three types of wormhole becomes larger. This may be due to the shorter range of the plasma frequency for the wormholes compared to that of the Kerr black hole.  

\begin{figure}[t!]
    \centering
    \includegraphics[width=15.5cm]{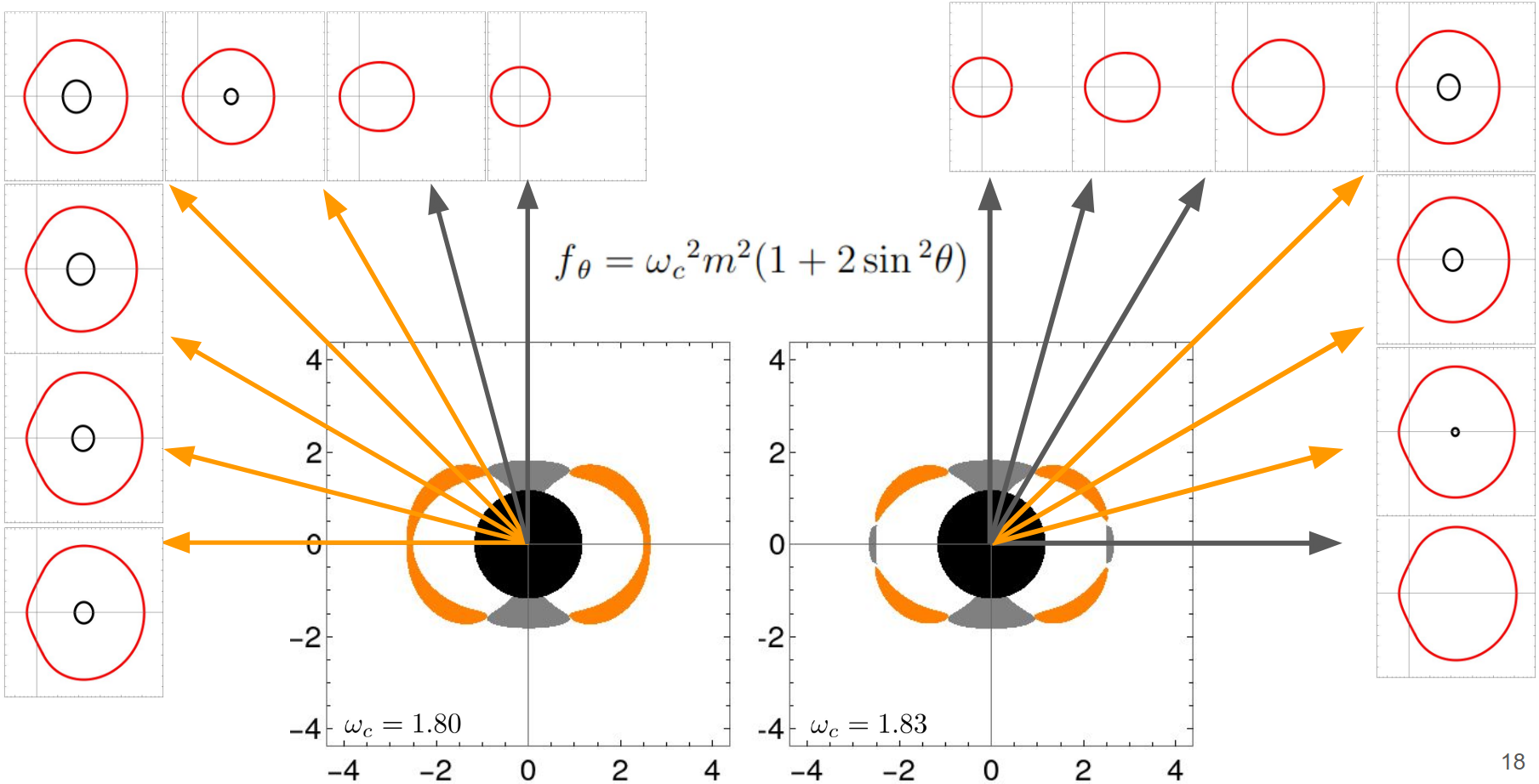}
    \caption{WH3 type of spacetime with plasma distribution  $f_\theta=\omega_c^2m^2(1+2\sin^2\theta)$, $f_r=0$, and spin parameter $a=0.999m$. We present the shadow boundary (in black) varying the inclination angle by $15^\circ$ and relate its deformation to the morphology of the photon region.} As the plasma frequency increases, observers at low inclinations can no longer see the shadow. Above the critical plasma frequency, the shadow ceases to exist also for equatorial observers, while those at intermediate inclinations can still observe it. The vacuum case is shown in red for comparison.
    \label{fig:enter-label} 
\end{figure}

The inner family of unstable photon orbits exhibits a qualitatively similar evolution in for all the wormholes. Initially, the inner family contributes primarily to the left side of the wormhole shadow, and the outer family of unstable photon orbits contributes to the right side of the shadow. As the plasma frequency increases, the inner family's contribution to the shadow diminishes, and ultimately the wormhole shadow becomes entirely determined by the outer family of unstable photon orbits. 

For completeness we will briefly focus on the shadows for an observer located outside the equatorial plane. In Figs. 10-11 we study the deformation of the shadow as the inclination angle varies for two wormhole configurations from our sample. We visualize the shadows by increasing the inclination angle with a step of $15^\circ$  and relate the modifications in the shadow boundary and size to the morphology of the photon region at the corresponding inclination. We observe that when the inclination decreases the shadows become more circular. This is  expected since the compact objects are axially symmetric, and viewed from the rotational axes their shadows should also be symmetric. On the other hand when plasma is present, the the observer's inclination determines whether they will be able to observe a shadow. As presented in Figs.10-11, observers who have a forbidden region lying between them and the wormholes do not see a shadow. This is denoted by grey arrows in the figures. If a photon region exist between the observer and the wormhole, the arrows are marked in orange and a shadow is observable. 

\begin{figure}[t!]
    \centering
    \includegraphics[width=15.5cm]{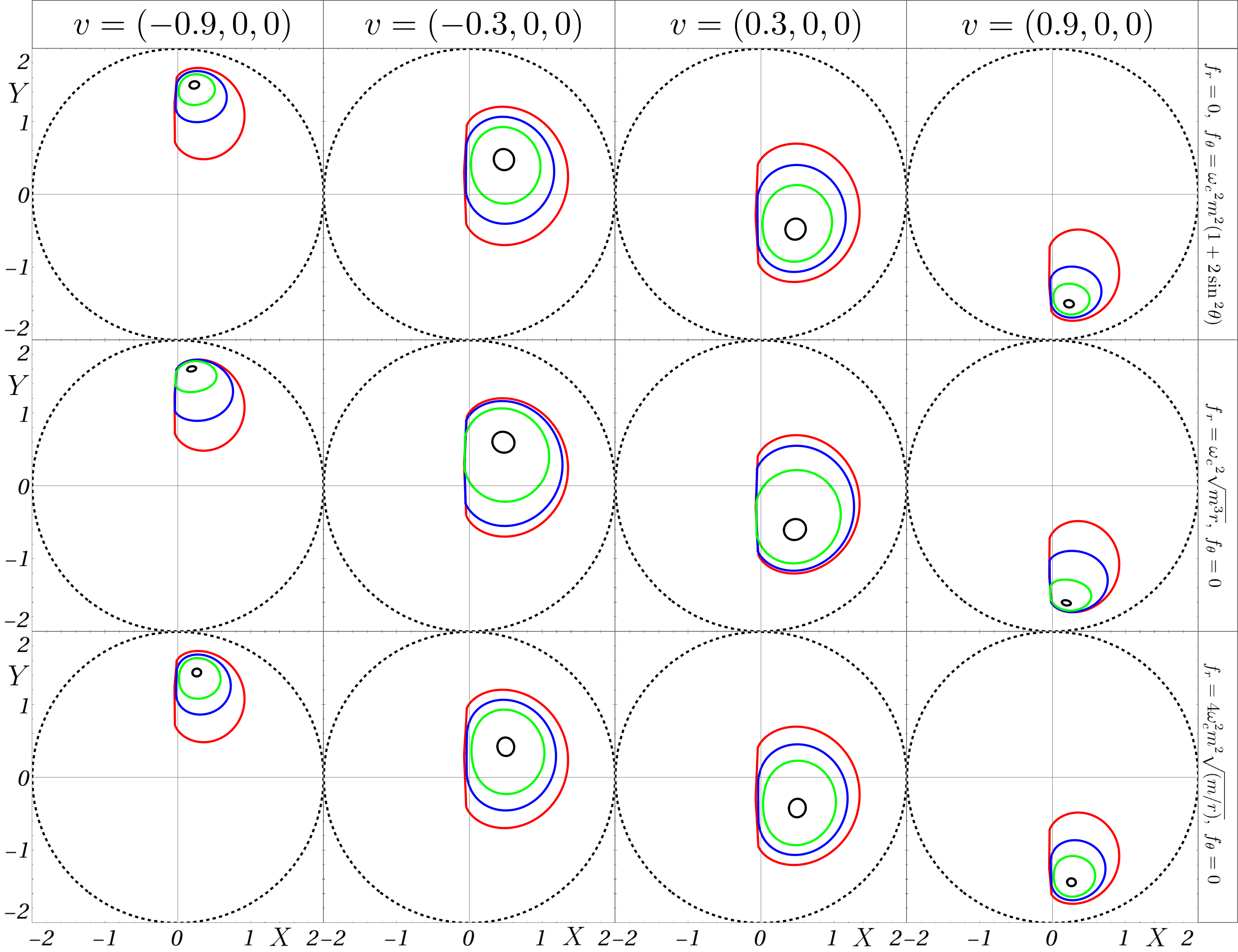}
    \caption{Shadows of WH 2$-$type wormholes for a moving observer located at $r_O=5m$, $\theta_O=\pi/2$, and $a=0.999m$. Each column corresponds to a different observer velocity, while each row represents a distinct plasma distribution: the first row $f_r=0, f_\theta=\omega_c^2m^2(1+2\sin^2\theta)$; the second row $f_r=\omega_c^2\sqrt{m^3r}, f_\theta=0$; the third row $f_r=4\omega_c^2m^2\sqrt{(m/r)},f_\theta=0$. The red contours display the vacuum case. For the first row: $\omega_c/\omega_0=2$ (blue), $\omega_c/\omega_0=2.5$ (green), and $\omega_c/\omega_0=2.85$ (black). For the second row: $\omega_c/\omega_0=2.2$ (blue), $\omega_c/\omega_0=3.2$ (green), and $\omega_c/\omega_0=3.75$ (black). For the third row: $\omega_c/\omega_0=2$ (blue), $\omega_c/\omega_0=2.6$ (green), and $\omega_c/\omega_0=3.2$ (black).}
    \label{fig:enter-label}
\end{figure}

In Fig. 10, WH2 type of spacetime with plasma distribution  $f_r=4\omega_c^2m^2\sqrt{(m/r)}$, $f_\theta=0$, and spin parameter  $a=0.999m$ is displayed. As we already discussed the plasma frequency determines a critical inclination above which the observers cannot observe the shadow. In this case, the equatorial observer is the one who can detect the shadow at the highest plasma frequency. In Fig. 11, we see a different situation with WH3 type of spacetime  and plasma distribution  and $f_\theta=\omega_c^2m^2(1+2\sin^2\theta)$, $f_r=0$, and spin parameter $a=0.999m$. In this example that the shadow remains visible for some observers at intermediate inclinations  even for higher plasma frequencies than the critical plasma frequency for an equatorial observer. 

\begin{figure}[t!]
    \centering
    \includegraphics[width=15.5cm]{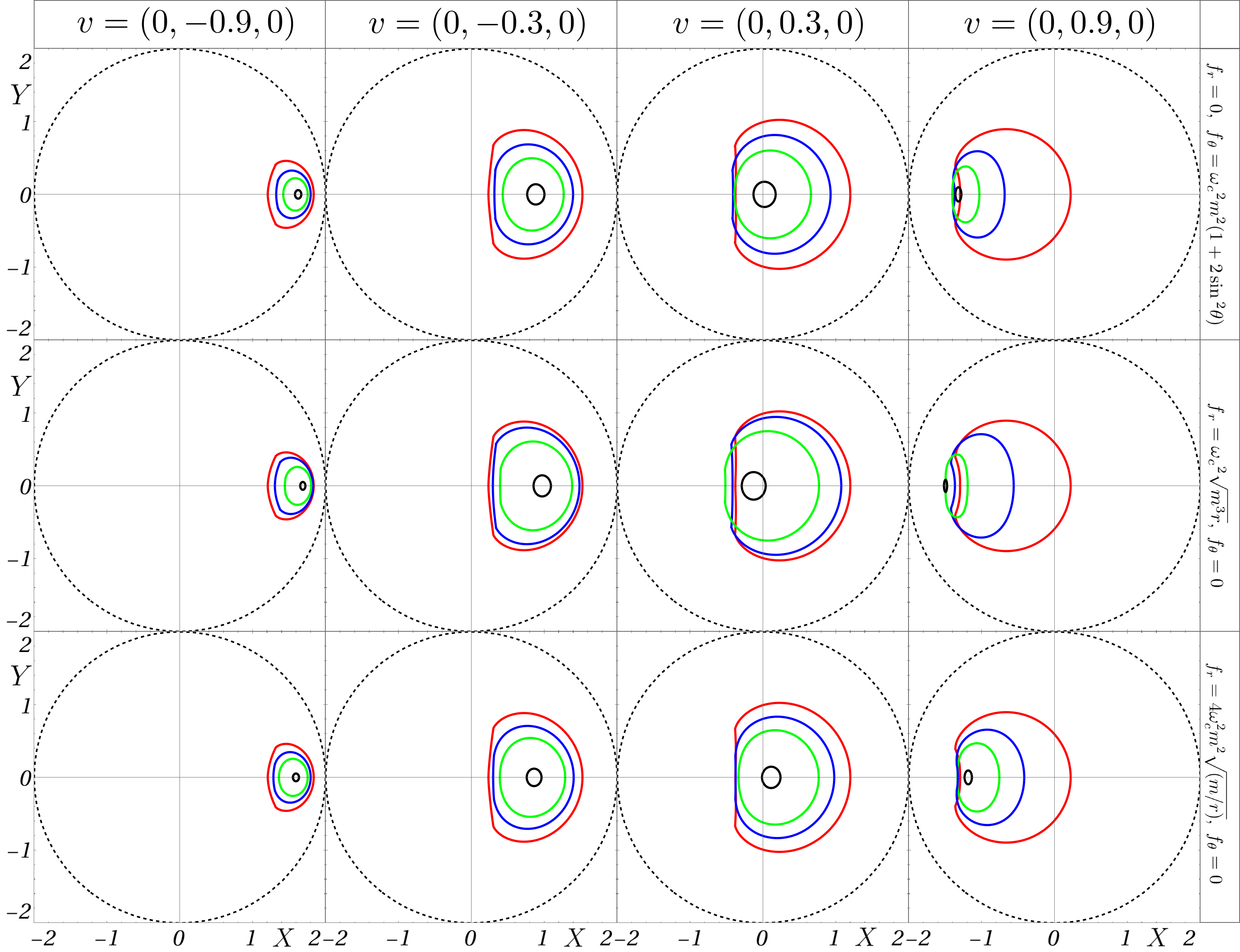}
    \caption{Shadows of WH 2$-$type wormholes for a moving observer located at $r_O=5m$, $\theta_O=\pi/2$, and $a=0.999m$. Each column corresponds to a different observer velocity, while each row represents a distinct plasma distribution: the first row $f_r=0, f_\theta=\omega_c^2m^2(1+2\sin^2\theta)$; the second row $f_r=\omega_c^2\sqrt{m^3r}, f_\theta=0$; and the third row $f_r=4\omega_c^2m^2\sqrt{(m/r)},f_\theta=0$. The red contours display the vacuum case. For the first row: $\omega_c/\omega_0=2$ (blue), $\omega_c/\omega_0=2.5$ (green), and $\omega_c/\omega_0=2.85$ (black). For the second row: $\omega_c/\omega_0=2.2$ (blue), $\omega_c/\omega_0=3.2$ (green), and $\omega_c/\omega_0=3.75$ (black). For the third row: $\omega_c/\omega_0=2$ (blue), $\omega_c/\omega_0=2.6$ (green), and $\omega_c/\omega_0=3.2$ (black).}
    \label{fig:enter-label}
\end{figure}

\subsection{Aberrational effects on the wormhole shadows}

For the cases with aberrations, we have chosen two observational distances. First, the observer is located at $r_O=5m$, and then we switch to $r_O=200m$. The deviations at $r=5m$ are more pronounced and more interesting to explore, however such a situation is unlikely in realistic astrophysical conditions where the observer is located at asymptotic infinity. Choosing observer's location  at $r=200m$ ensures that they are outside the strongly curved region of the spacetime and closer to the asymptotically flat regime. 

\begin{figure}[t!]
    \centering
    \includegraphics[width=15.5cm]{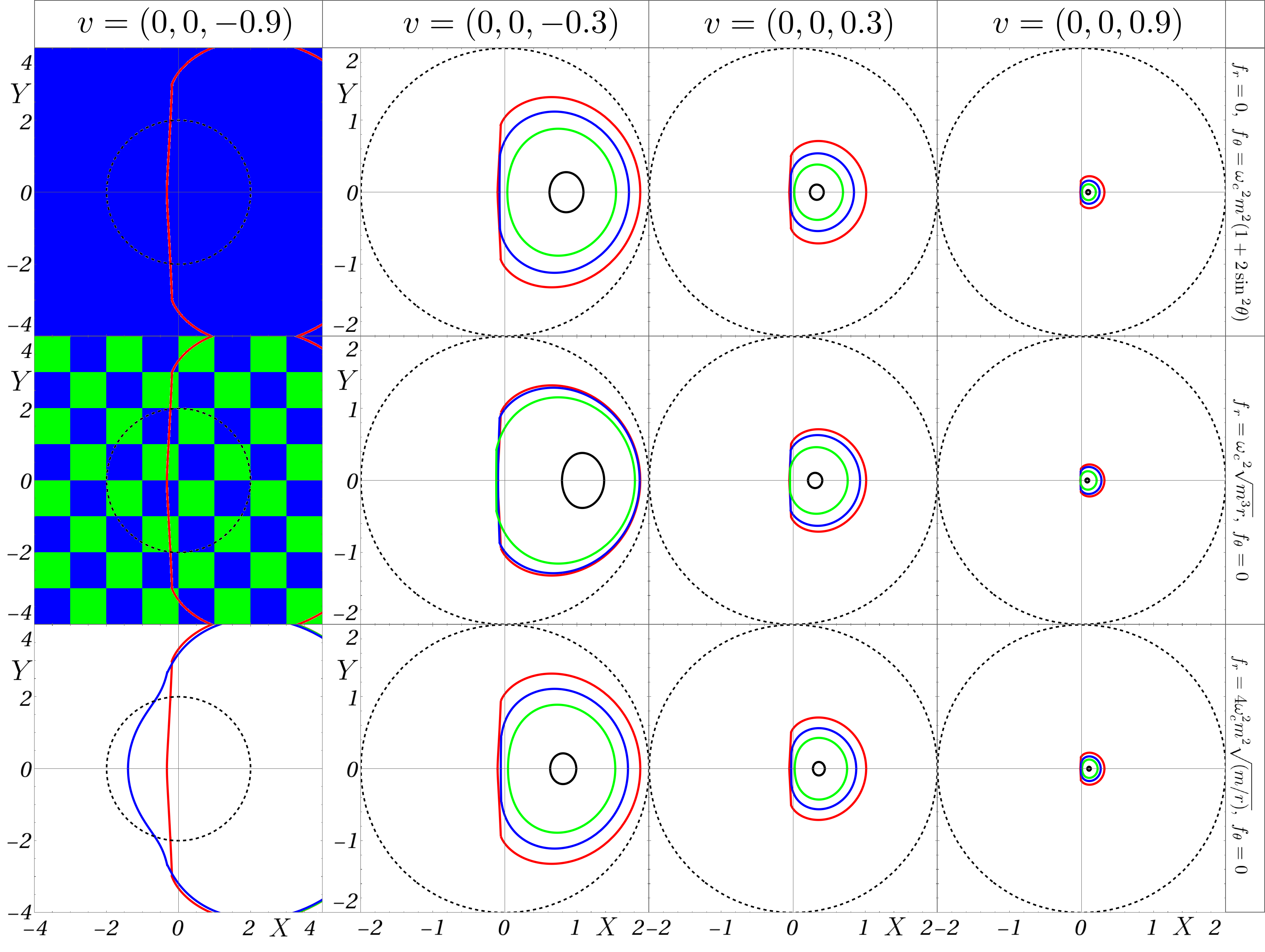}
    \caption{Shadows of WH 2$-$type wormholes for a moving observer located at $r_O=5m$, $\theta_O=\pi/2$, and $a=0.999m$. Each column corresponds to a different observer velocity, while each row represents a distinct plasma distribution: the first row $f_r=0, f_\theta=\omega_c^2m^2(1+2\sin^2\theta)$; the second row $f_r=\omega_c^2\sqrt{m^3r}, f_\theta=0$; and the third row $f_r=4\omega_c^2m^2\sqrt{(m/r)},f_\theta=0$. The red contours indicate the vacuum case. First row: $\omega_c/\omega_0=2$ (blue), $\omega_c/\omega_0=2.5$ (green), and $\omega_c/\omega_0=2.85$ (black). In the top-left panel, the blue background indicates that the corresponding contour covers more than half of the observer’s sky, extending from $X=-14.5$ to $X=6.5$. Additionally, the green contour lies outside the field of view to the right ($X=7.1$ to $X=99.4$), and the black one is also outside the field view ($X=8.8$ to $X=12.2$). Second row: $\omega_c/\omega_0=2.2$ (blue), $\omega_c/\omega_0=3.2$ (green), and $\omega_c/\omega_0=3.75$ (black). In the leftmost panel, the blue and green backgrounds indicate that both contours occupy more than half of the observer’s sky spanning $X=-8.4$ to $X=6.9$ for the blue case and $X=-51$ to $X=8.5$ for the green one. The black case lies outside of the field of view to the right ($X=14.5$ to $X=23.7$). Third row: $\omega_c/\omega_0=2$ (blue), $\omega_c/\omega_0=2.6$ (green), and $\omega_c/\omega_0=3.2$ (black). The green and black contours lie outside the field of view to the right, spanning $X=6$ to $X=24$ and $X=6.3$ to $X=7$, respectively.}
    \label{fig:enter-label}
\end{figure}

While exploring the wormhole's shadow from close proximity, we noticed significant variations in both size and shape, including considerable overlapping of the shadows. When varying the first component of the 3-velocity we noticed that the higher the velocity, the greater the change in shape, as the shadows become vertically compressed. This effect is most pronounced for the plasma distribution given by Eq. \eqref{sqrtrplasma}, which also exhibits the strongest overlapping of the shadow contours, as shown in Fig. 12.

\begin{figure}
    \centering
    \includegraphics[width=15.5cm]{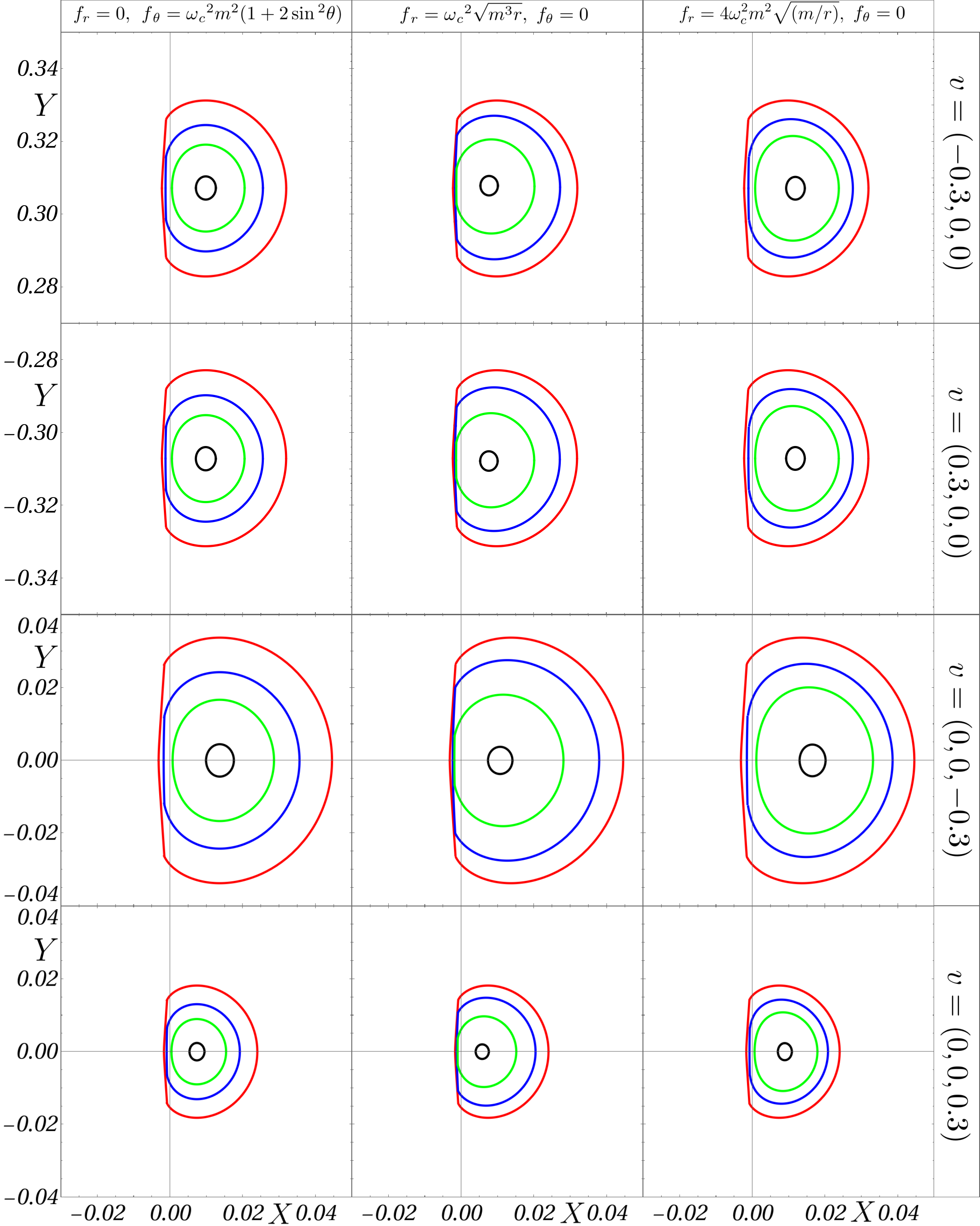}
    \caption{Shadows of WH 2$-$type wormholes for a moving observer located at $r_O=200m$, $\theta_O=\pi/2$, and $a=0.999m$. Each row corresponds to a different observer velocity, while each column represents a distinct plasma distribution: the first column $f_r=0, f_\theta=\omega_c^2m^2(1+2\sin^2\theta)$; the second $f_r=\omega_c^2\sqrt{m^3r}, f_\theta=0$; and third $f_r=4\omega_c^2m^2\sqrt{(m/r)},f_\theta=0$. The red contours correspond to the vacuum case. For the first column: $\omega_c/\omega_0=2$ blue, $\omega_c/\omega_0=2.5$ (green), and $\omega_c/\omega_0=2.85$ (black). For the second column: $\omega_c/\omega_0=2.2$ (blue), $\omega_c/\omega_0=3.2$ (green), and $\omega_c/\omega_0=3.75$ (black). For the third column: $\omega_c/\omega_0=2$ (blue), $\omega_c/\omega_0=2.6$ (green), and $\omega_c/\omega_0=3.2$ (black).}
    \label{fig:enter-label}
\end{figure}

\begin{figure}[t!]
    \centering
    \includegraphics[width=15.5cm]{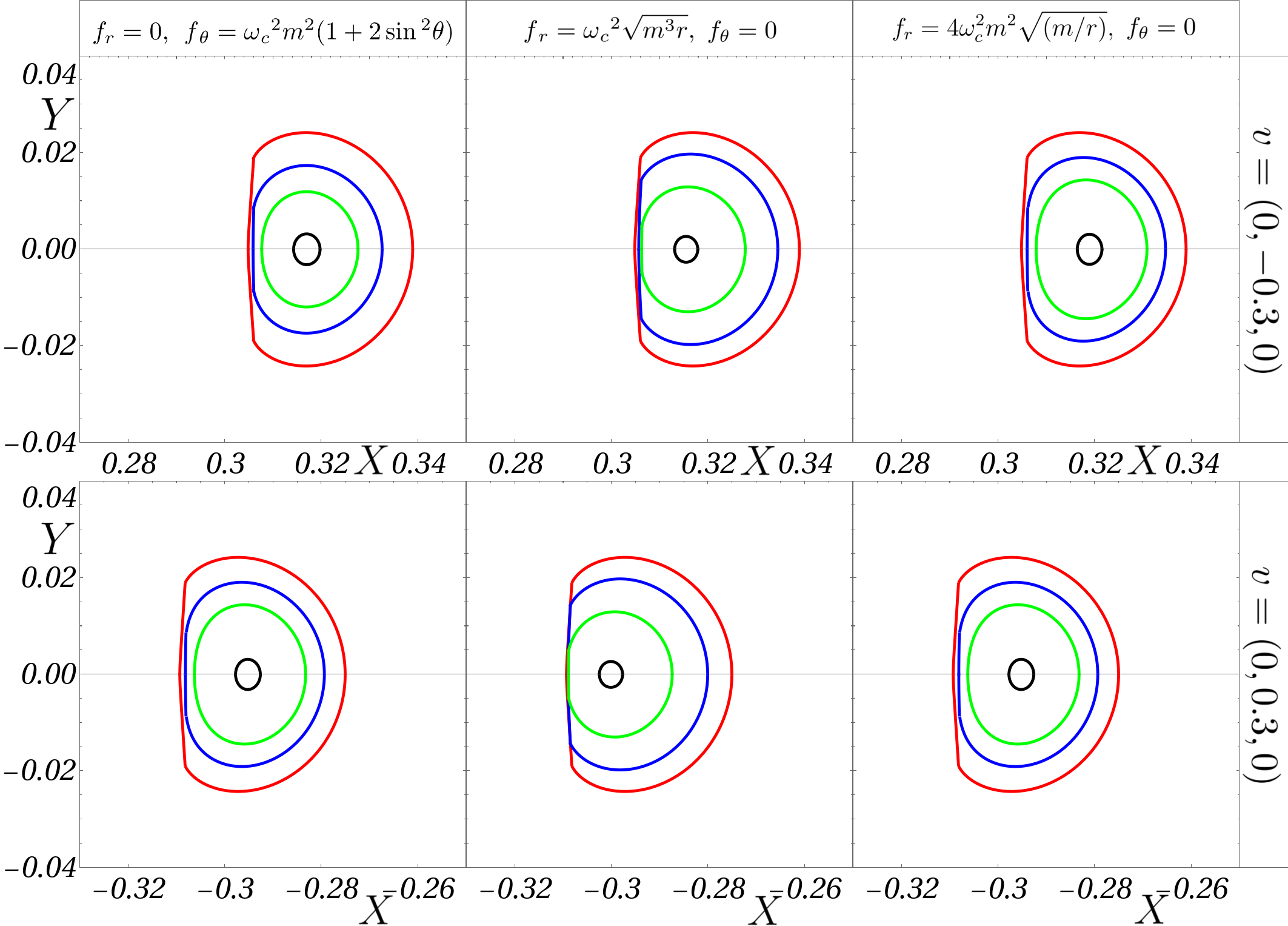}
    \caption{Shadows of WH 2$-$type wormholes for a moving observer located at $r_O=200m$, $\theta_O=\pi/2$, and $a=0.999m$. Each row corresponds to a different observer velocity, while each column represents a distinct plasma distribution: the first column $f_r=0, f_\theta=\omega_c^2m^2(1+2\sin^2\theta)$; the second $f_r=\omega_c^2\sqrt{m^3r}, f_\theta=0$; and the third $f_r=4\omega_c^2m^2\sqrt{(m/r)},f_\theta=0$. The red contours correspond to the vacuum case. For the first column: $\omega_c/\omega_0=2$ (blue), $\omega_c/\omega_0=2.5$ (green), and $\omega_c/\omega_0=2.85$ (black). For the second column $\omega_c/\omega_0=2.2$ (blue), $\omega_c/\omega_0=3.2$ (green), and $\omega_c/\omega_0=3.75$ (black). For the third column: $\omega_c/\omega_0=2$ (blue), $\omega_c/\omega_0=2.6$ (green), and $\omega_c/\omega_0=3.2$ (black).}
    \label{fig:enter-label}
\end{figure}

In Fig. 13, we have shown variations due to the second component of the 3-velocity, which also induces substantial changes in the shadows. As we can see in Fig. 9, the WH2 shadow in vacuum  displays a nearly flat left edge reminiscent of the Kerr black hole shadow. This feature disappears when $v_2$ aberrations are introduced. The left side becomes convex for negative velocities and concave for positive ones. This is a noticeable difference from the Kerr black hole case, shown in \cite{Grenzebach:2015}, \cite{Briozzo:2023} and provides another possible method to distinguish a WH2-type wormhole from a Kerr black hole. Another interesting feature is that the shadow contour in a plasma environment may not only overlap with the vacuum shadow but may lie entirely outside it. This effect is again most prominent for the plasma distribution given by Eq. \eqref{sqrtrplasma}. 

Fig. 14 shows aberrational effects on the shadows caused by variations in the third component of the 3-velocity. This produces significant changes in the sizes of the shadows for positive velocities and dramatic changes in size, shape and orientation of the shadows for negative velocities. We discovered that, depending on the velocity and the plasma distribution, shadows can cover more than half of the observer's celestial sphere or even lie entirely outside the observer's field of view. For $v_3= -0.9$ and for all of the plasma distributions, the black contours at high plasma frequency lie completely outside the graphs on the right. Furthermore, at lower plasma frequencies, the shadows can cover more than half of the celestial sphere, which is indicated by changes the background color corresponding to the regions fully occupied by the shadow.

In addition to placing the observer farther from the object (from $r=5m$ to $r=200m$), we restrict the observer's velocity to $30\%$ of the speed of light. Although this velocity is still high, it is more realistic than $90\%$ of the speed of light, and may leads to effect which are observable in the future missions. Figs. 15-16 show these cases for WH2. 

When varying the first and second components of the 3-velocity, we found that the shapes and sizes of the shadows remain nearly unchanged. However, it lead to modification of the \textit{direction} in which the observer would see the shadow. We determined that this shift in the celestial coordinates isindependent of the choice of the plasma distribution. For motion with $v=(\pm 0.3,0,0)$, the shadow shifts downwards or upwards, respectively, to $Y=\pm0.31$, corresponding to a $17.6$ displacement from the celestial pole, as shown in Fig. 13. Similarly, motion with, $v=( 0,\pm0.3,0)$, shifts the shadow left to right to $X=\pm0.31$, again corresponding to $17.6$, as illustrated in Fig. 14.

In the last case, varying the third component of the 3-velocity leaves the direction unchanged, but alters the size of the shadow, as shown in Fig. 13.

\section{Conclusion}

In this work we have studied the influence of the plasma environment on the shadow of stationary and axisymmetric traversable wormholes. For the purpose we have considered a sample of wormhole spacetimes and plasma distributions for which the Hamilton-Jacobi equation for the light propagation is separable. We have derived analytical expressions for the shadow boundary and investigated the morphology of the photon region which governs the shadow formation. In dispersive medium forbidden regions are formed where light cannot propagate and their structure depends on the plasma frequency. Due to this phenomenon the shadow of each compact object is relevant only for a particular plasma frequency range and beyond a certain critical frequency it cannot be observed.

We examined the behavior of these structure for the wormholes and plasma distributions in our sample for various plasma frequencies and compared to the Kerr black hole in the same environment. We observe the following trends. For plasma distributions which depend only on the radial coordinate the morphology of the photon region is determined by the plasma profile rather than the spacetime metric. Such distributions lead to similar behavior for all the wormholes in our sample which is also consistent with the Kerr black hole. For plasma distribution which depend also on the polar angle the structure of the photon region is determined by the spacetime metric and various configurations may occur for the same plasma profile. This creates qualitative distinctions between the different wormholes as well as from the Kerr black hole which may have observational consequences.

The shadows of all the configurations in our sample decrease when the plasma frequencies increases finally vanishing after a critical frequency value. This behavior is consistent with the Kerr black hole with quantitative differences. For all the wormhole spacetimes the decrease in the shadow size progresses faster with increasing the plasma frequencies than for the Kerr black hole. This creates higher discrepancies between the different compact which facilitate their detection by imaging experiments. Therefore, considering the plasma medium leads to more favorable results in accessing the necessary resolutions in the observations than purely vacuum environment. Moreover, the critical frequencies for all the wormhole spacetimes are lower than for the Kerr black hole leading to smaller ranges of the plasma frequencies when their shadows exists. This implies a strong observational signature since there exist plasma frequency ranges in which the Kerr black hole creates a shadow but wormholes do not. Finally we investigate the deformation of the shadows if we consider a moving observer. Aberrational effects further increase the discrepancies between the wormhole and black hole shadows and facilitate the experimental detection of wormholes.

\section*{Appendix}

In this appendix we explore the photon and forbidden regions for the metrics and plasma distribution combinations which were not discussed in the main text. This  completes our analysis, although the general features of the evolution remain similar to the main discussion. Fig. 17 shows the evolution for WH2 with plasma distribution \eqref{thetaplasma}, and we observe that the forbidden region appears on the equatorial plane and extends towards the poles. As for WH1 and WH3 an equatorial observer cannot see a shadow once the forbidden region develops.

Figs. 18-21 show the evolution of the photon regions and forbidden regions for WH1 with the plasma distribution given by Eq. \eqref{Perlickplasma} , WH2 with the distribution given by  \eqref{sqrtrplasma} and WH3 with the  distributions given by Eqs. \eqref{sqrtrplasma} and \eqref{Perlickplasma}. In all these cases, the forbidden regions first appear at the poles and extend toward the equatorial plane. Again, an equatorial observer will see a shadow as long as a photon region exists on the equatorial plane.

Figs. 22 shows the Kerr black holes in vacuum and in plasma environment as seen from the point of view of a moving observer for the third component of the 3-velocity. We noticed that the shadow is visible in front of the observer at higher plasma frequencies and higher velocities before shifting to the right, compared to WH2. This can be seen by comparing the green contours between Fig. 14 and Fig. 22. 

\begin{figure}[t!]
\centering
    \begin{tabular}{ cc}
    \includegraphics[width=4.8cm]{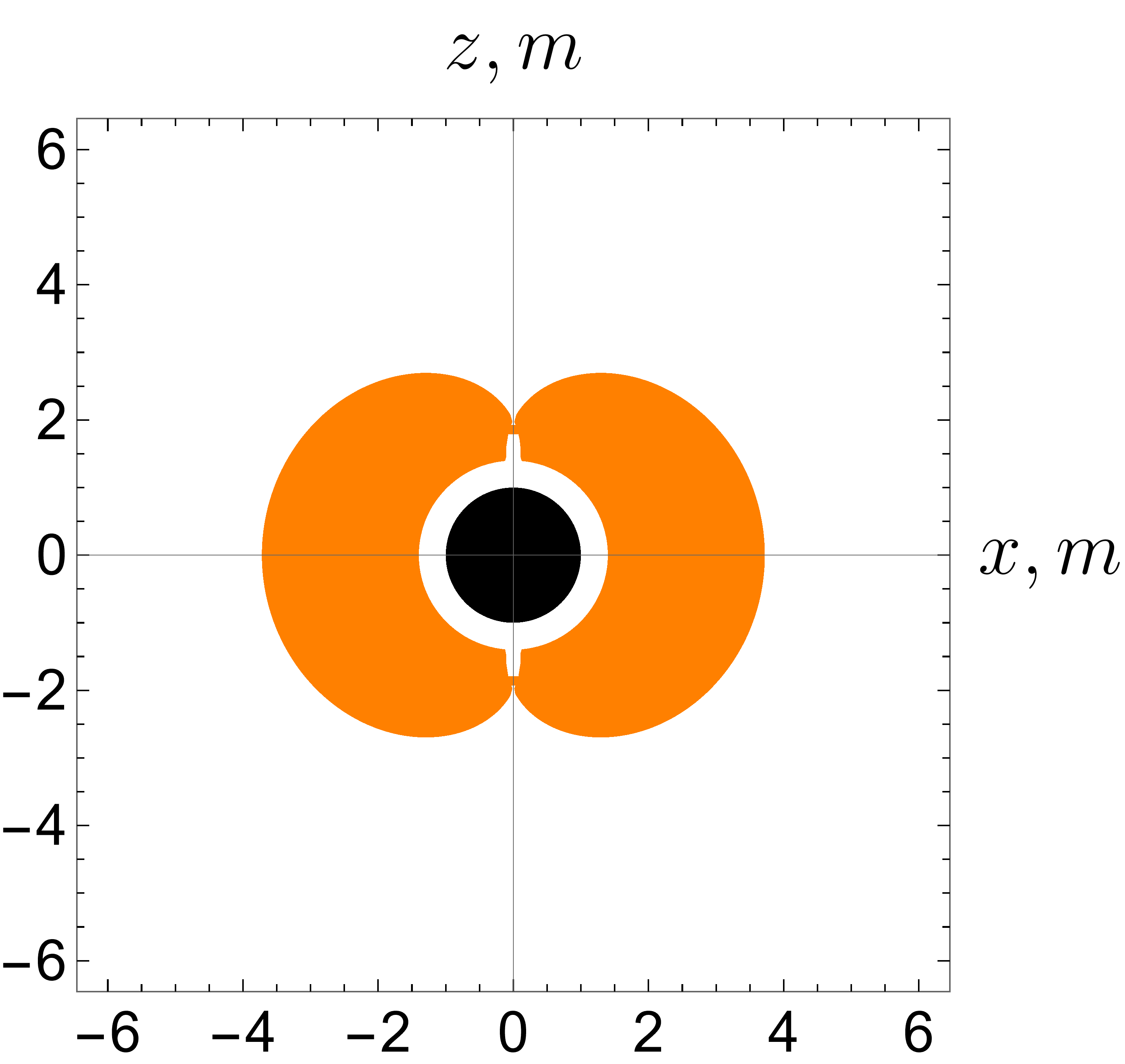}
    \includegraphics[width=4.8cm]{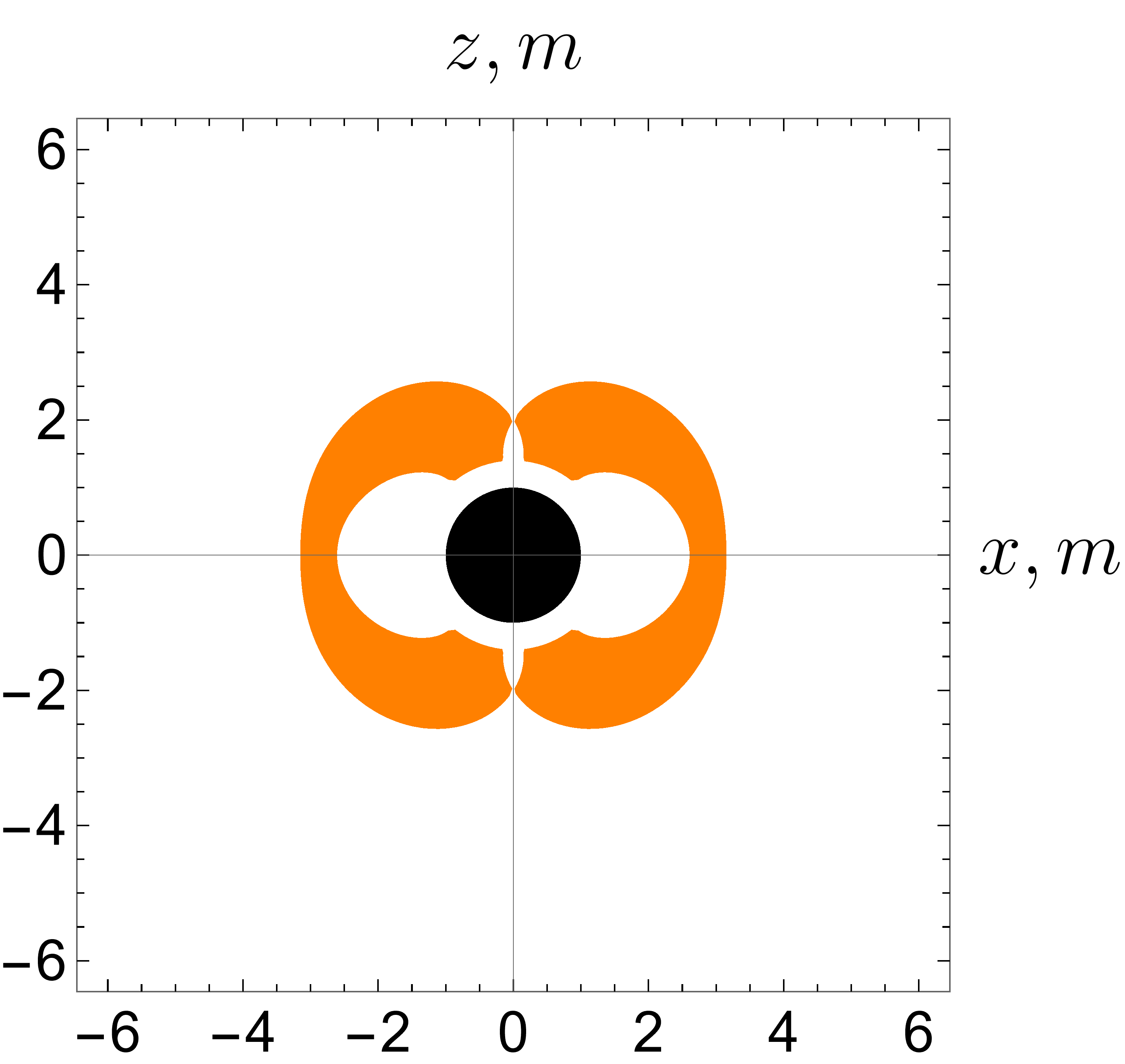} \\
     \includegraphics[width=4.8cm]{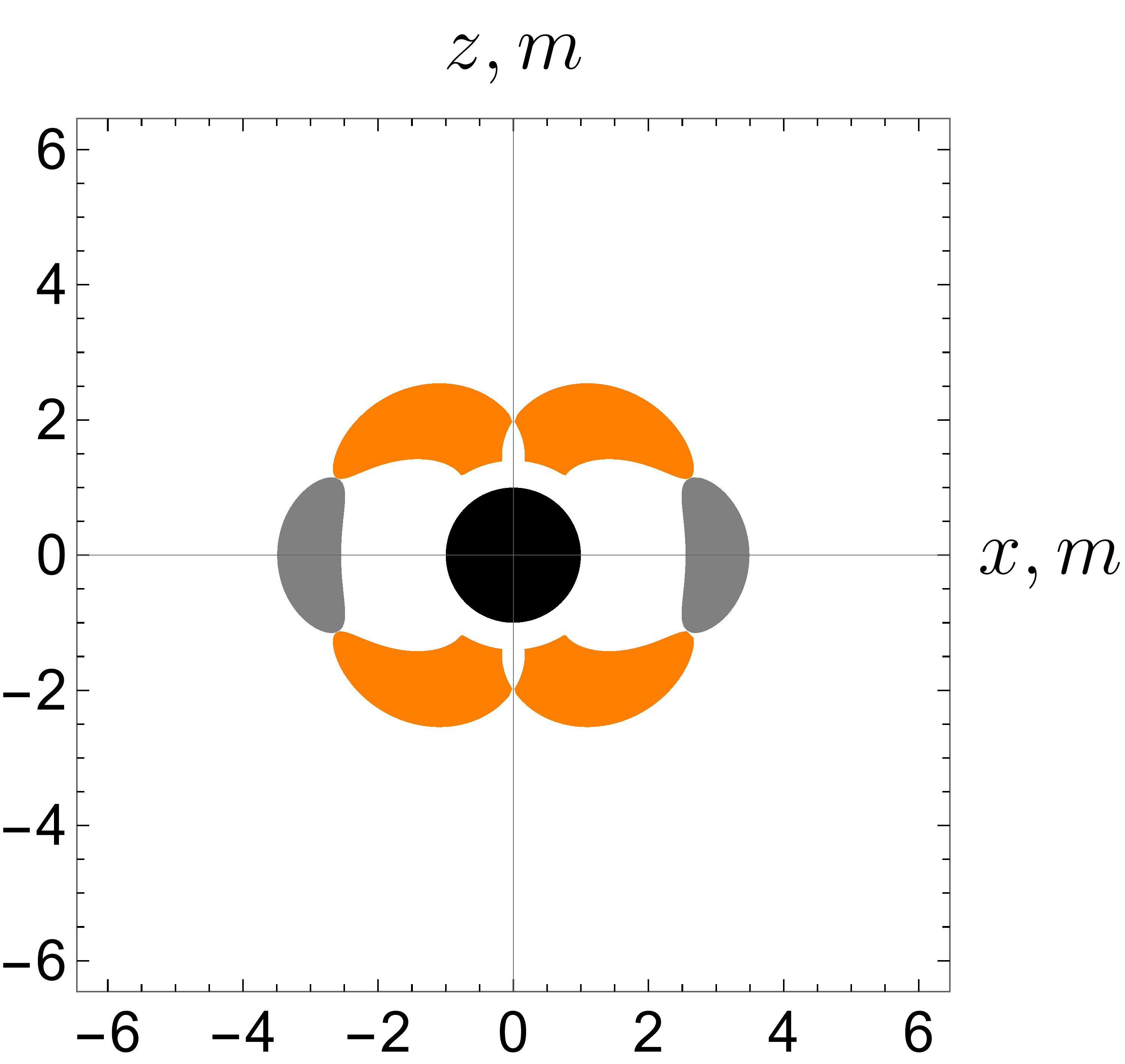} 
     \includegraphics[width=4.8cm]{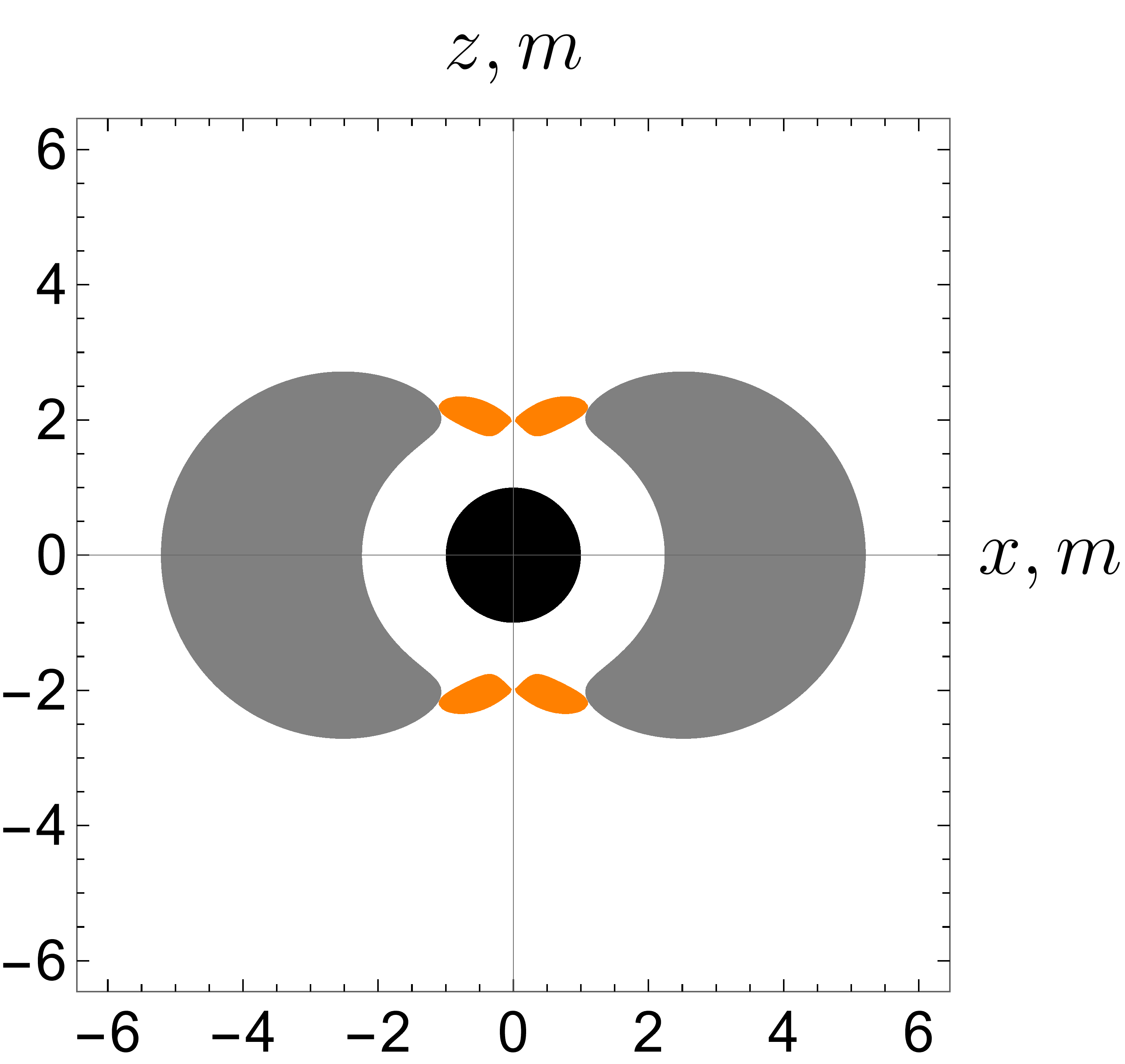}
     \end{tabular}
    \caption{WH2 with plasma distribution $f_\theta = \omega{_c}{^2}m{^2}(1+2\sin{^2}\theta)$, $f_r=0$ and spin parameter $a=0.999m$. The plasma frequencies for each panel are as follows: top$-$left $\omega{_c}/ \omega{_0} = 1$; top$-$right $\omega{_c}/ \omega{_0} = 2.8$; bottom$-$ left $\omega{_c}/ \omega{_0} = 3$; and bottom$-$right $\omega{_c}/ \omega{_0} = 3.8$. The forbidden region first forms along the equatorial plane and gradually expands toward the poles as the plasma frequency increases.}
\end{figure}

\begin{figure}[H]
\centering
    \begin{tabular}{ cc}
    \includegraphics[width=4.8cm]{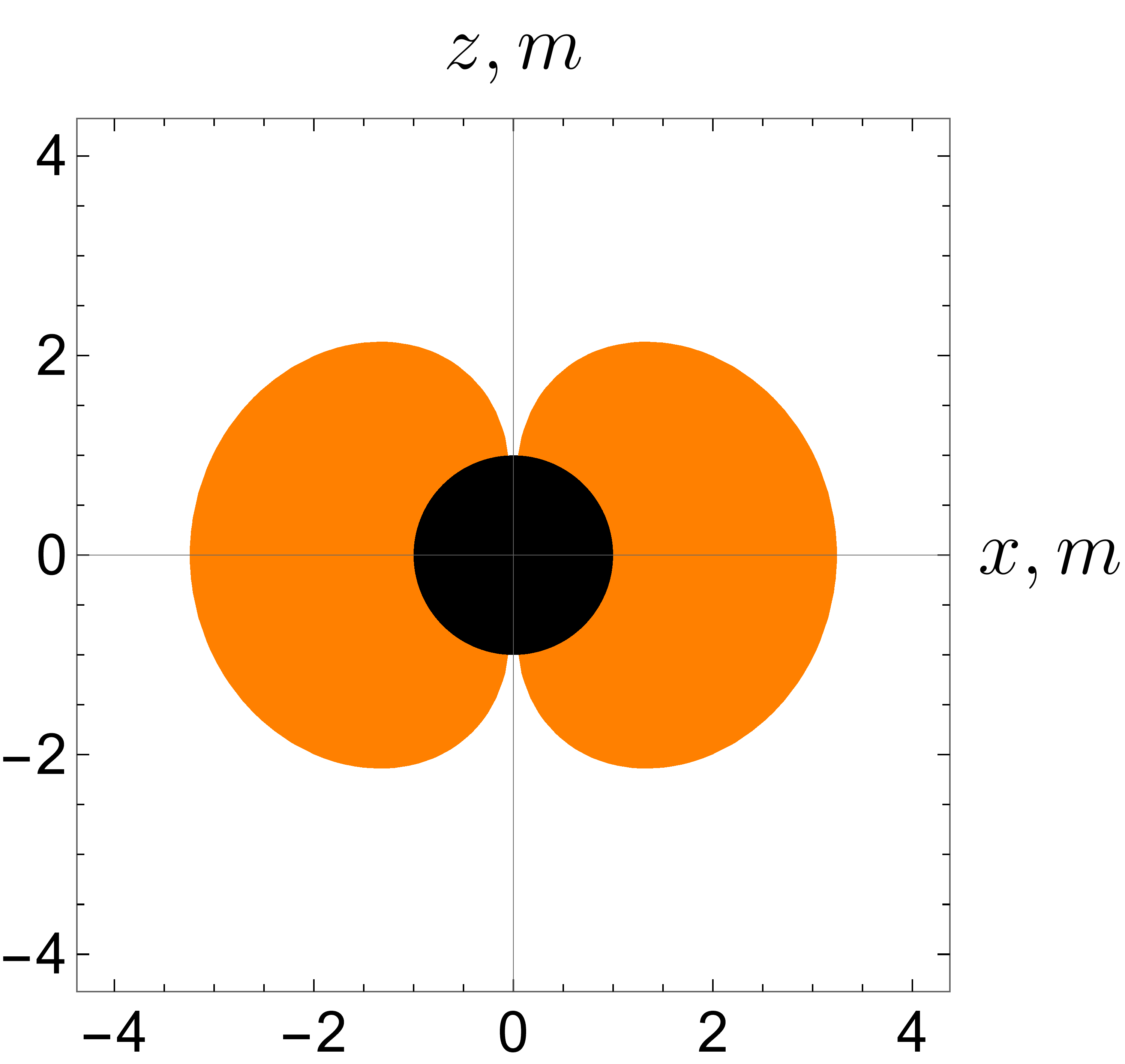}
    \includegraphics[width=4.8cm]{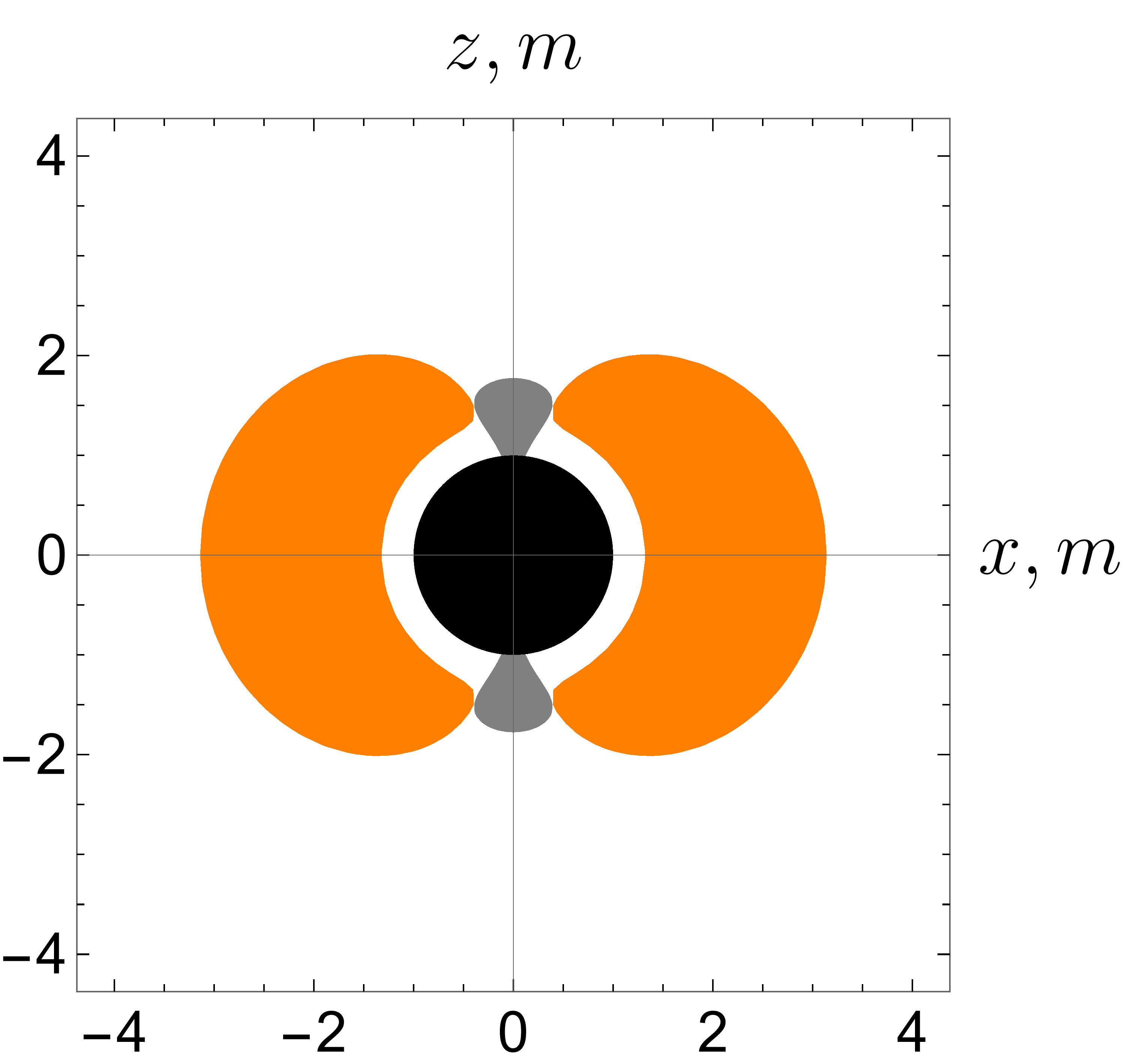} \\
    \includegraphics[width=4.8cm]{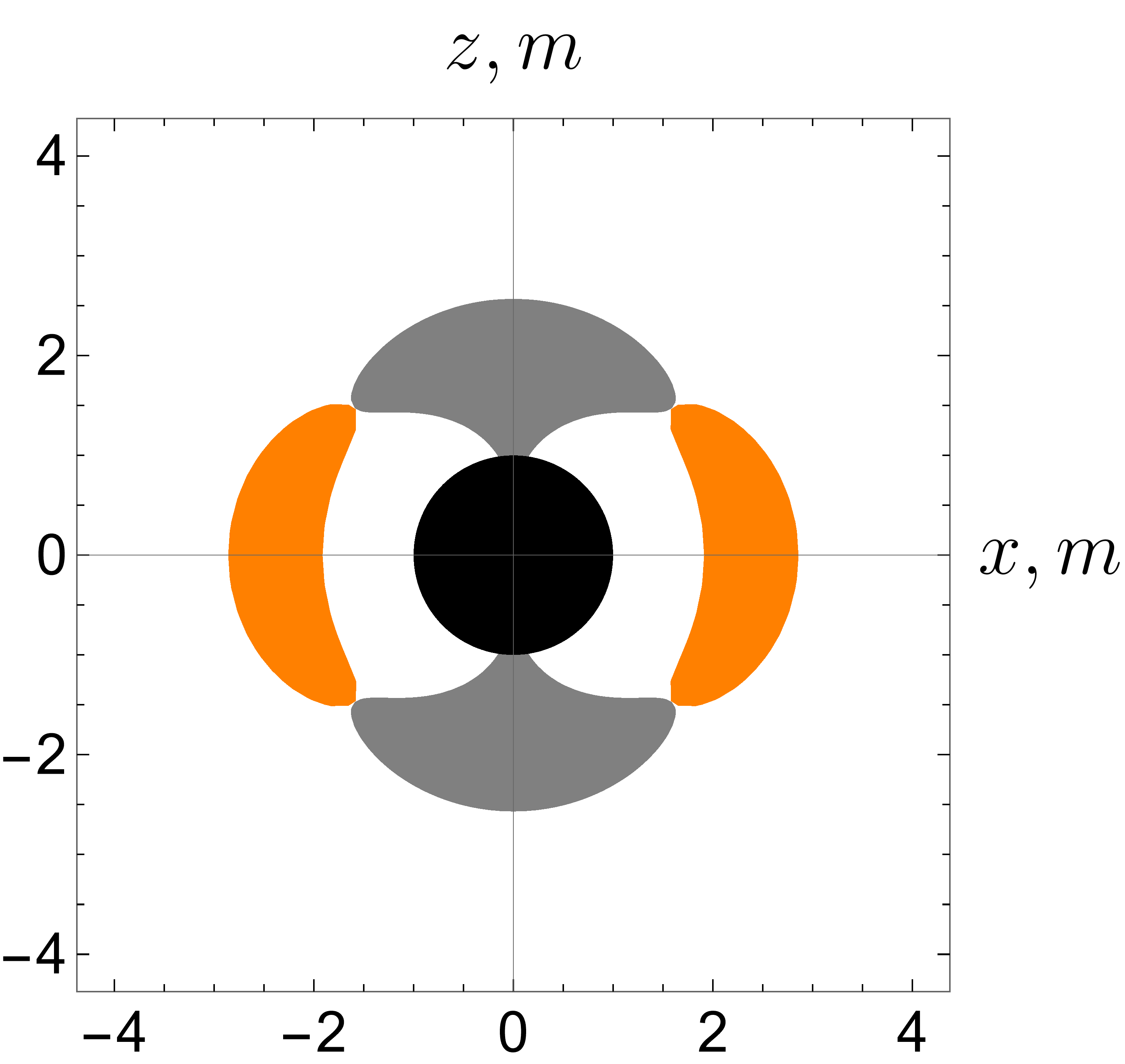}
    \includegraphics[width=4.8cm]{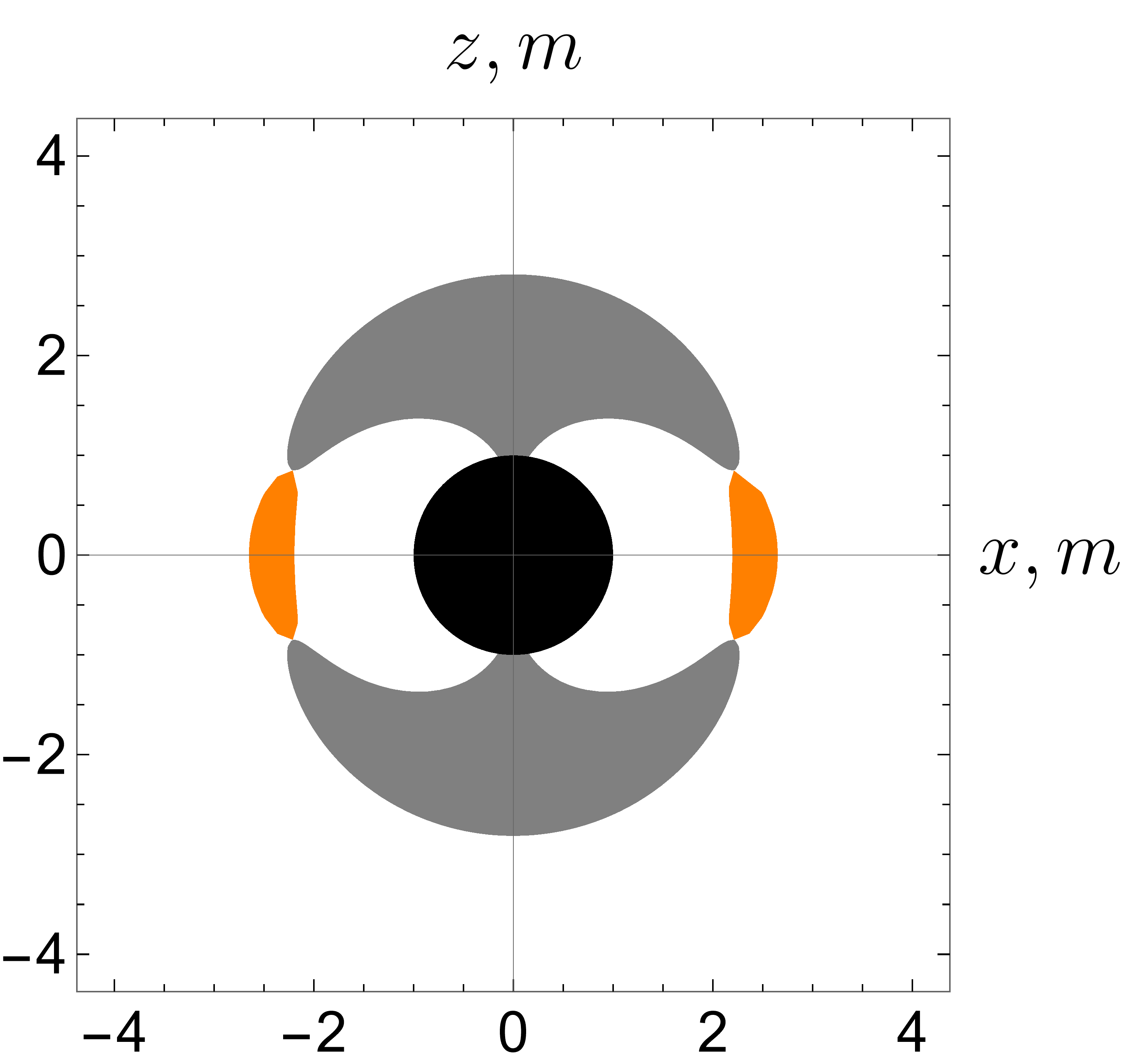}
\end{tabular}
    \caption{WH1 with plasma distribution $f_r = 4\omega{_c}{^2}m{^2}\sqrt{m/r}$, $f_\theta =0$ and spin parameter $a=0.999m$. The plasma frequencies for each panel are as follows: top$-$left $\omega{_c}/ \omega{_0} = 1.4$; top$-$right $\omega{_c}/ \omega{_0} = 1.8$; bottom$-$left $\omega{_c}/ \omega{_0} = 2.4$; and bottom$-$right $\omega{_c}/ \omega{_0} = 2.6$. The forbidden region originates near the poles and expands toward the equatorial plane with increasing plasma frequency.}
\end{figure}

\begin{figure}[t!]
\centering
\vspace{-3mm}
    \begin{tabular}{ cc}
    \includegraphics[width=4.8cm]{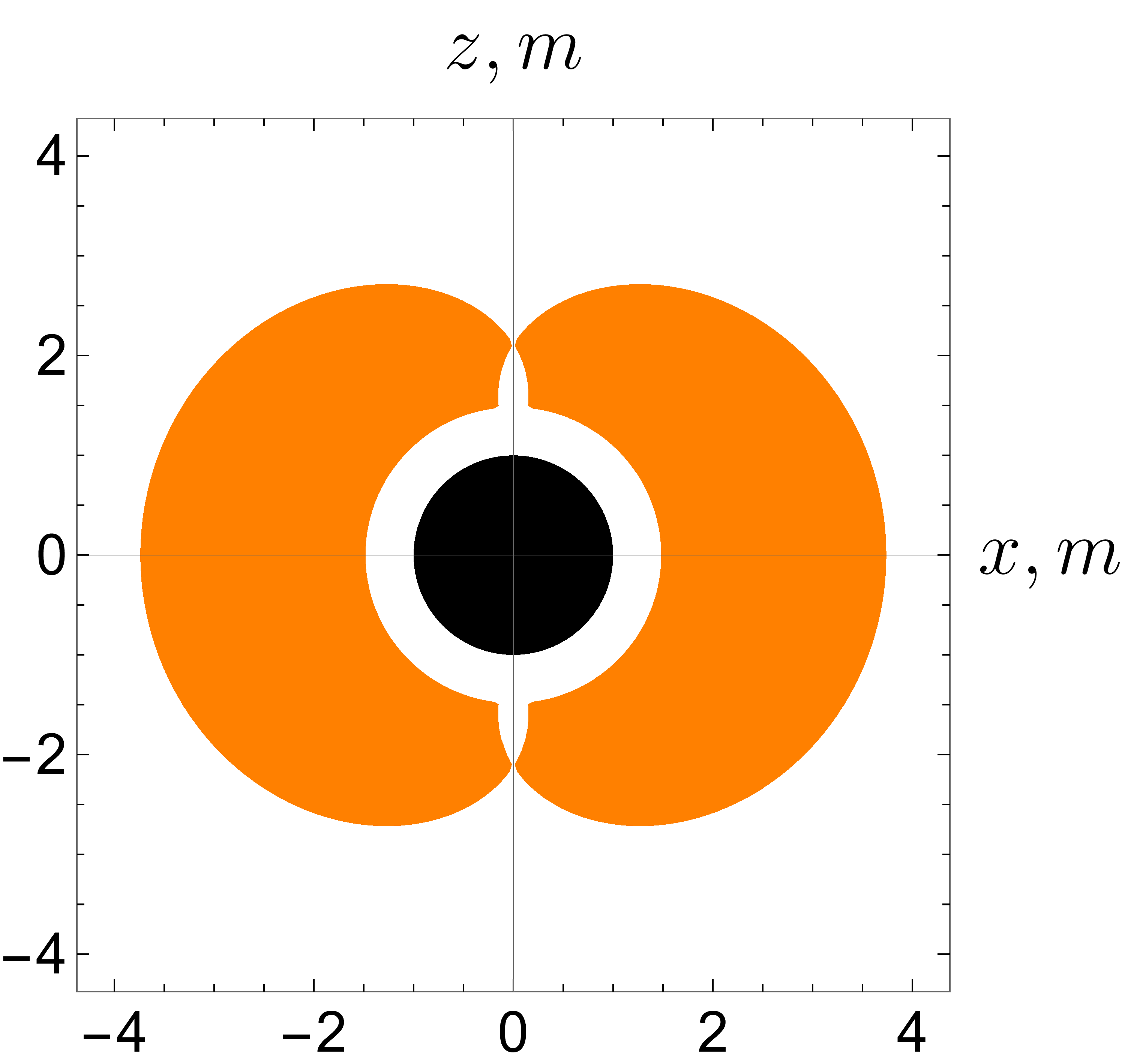}
    \includegraphics[width=4.8cm]{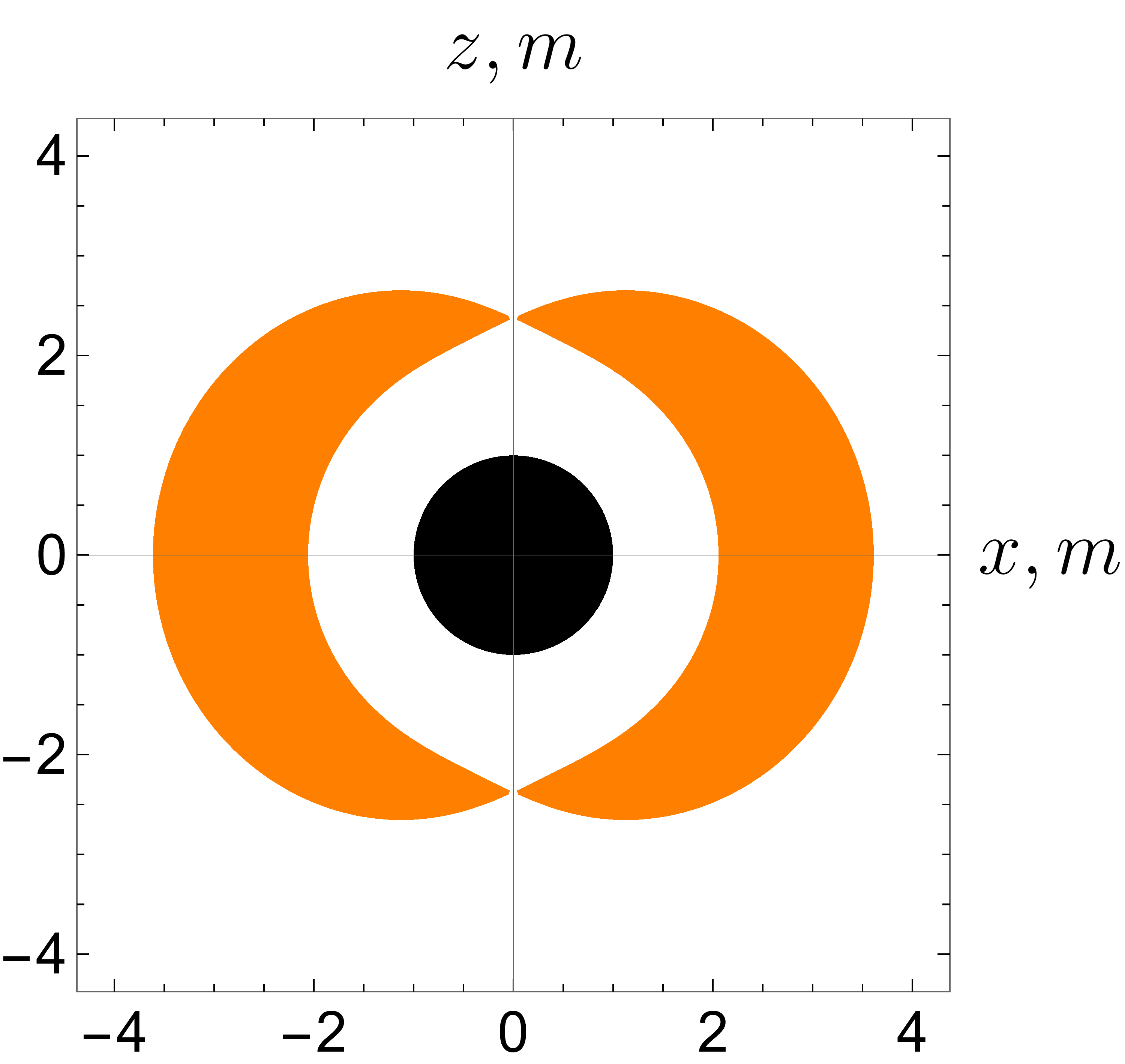} \\
    \includegraphics[width=4.8cm]{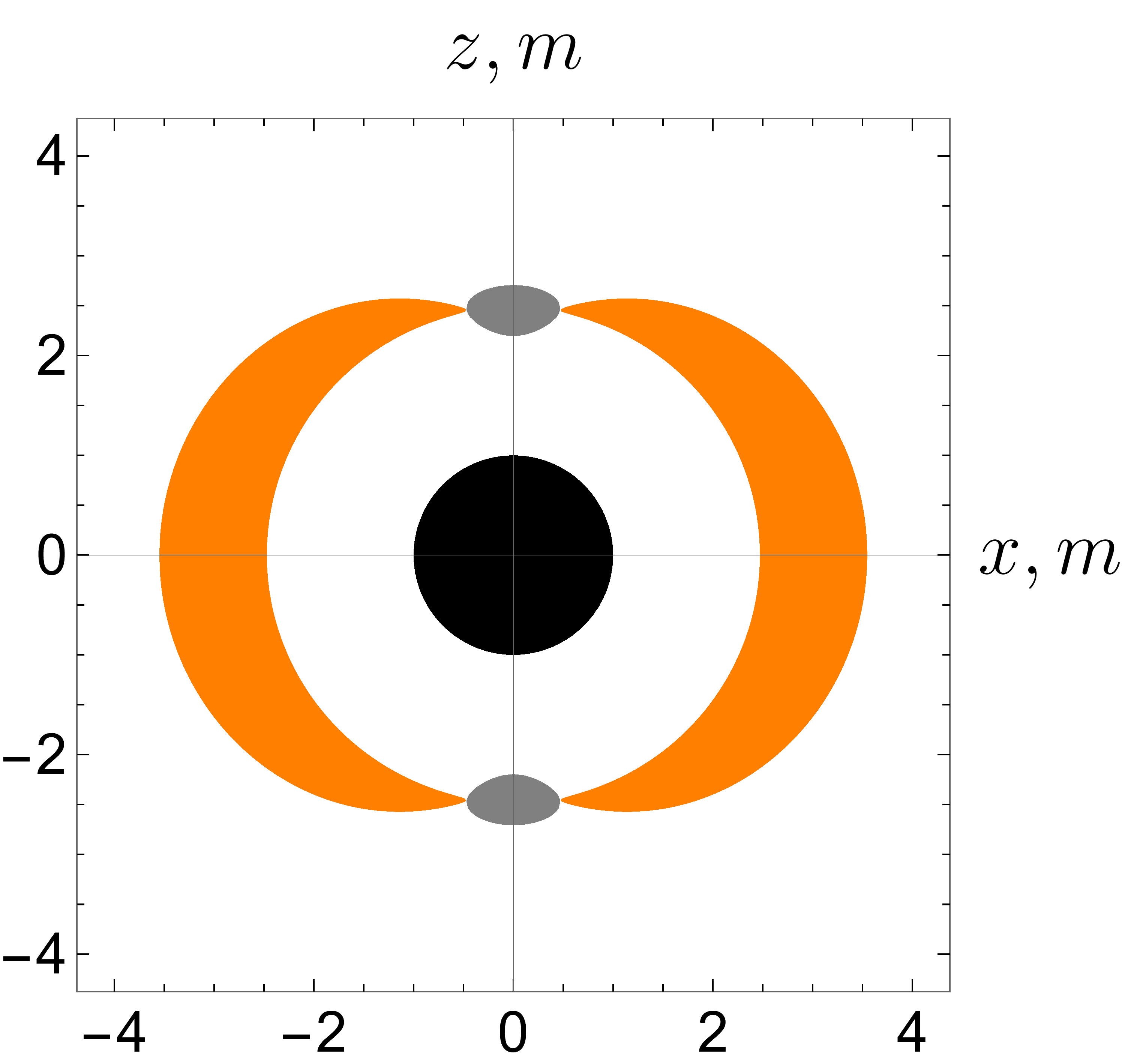}
    \includegraphics[width=4.8cm]{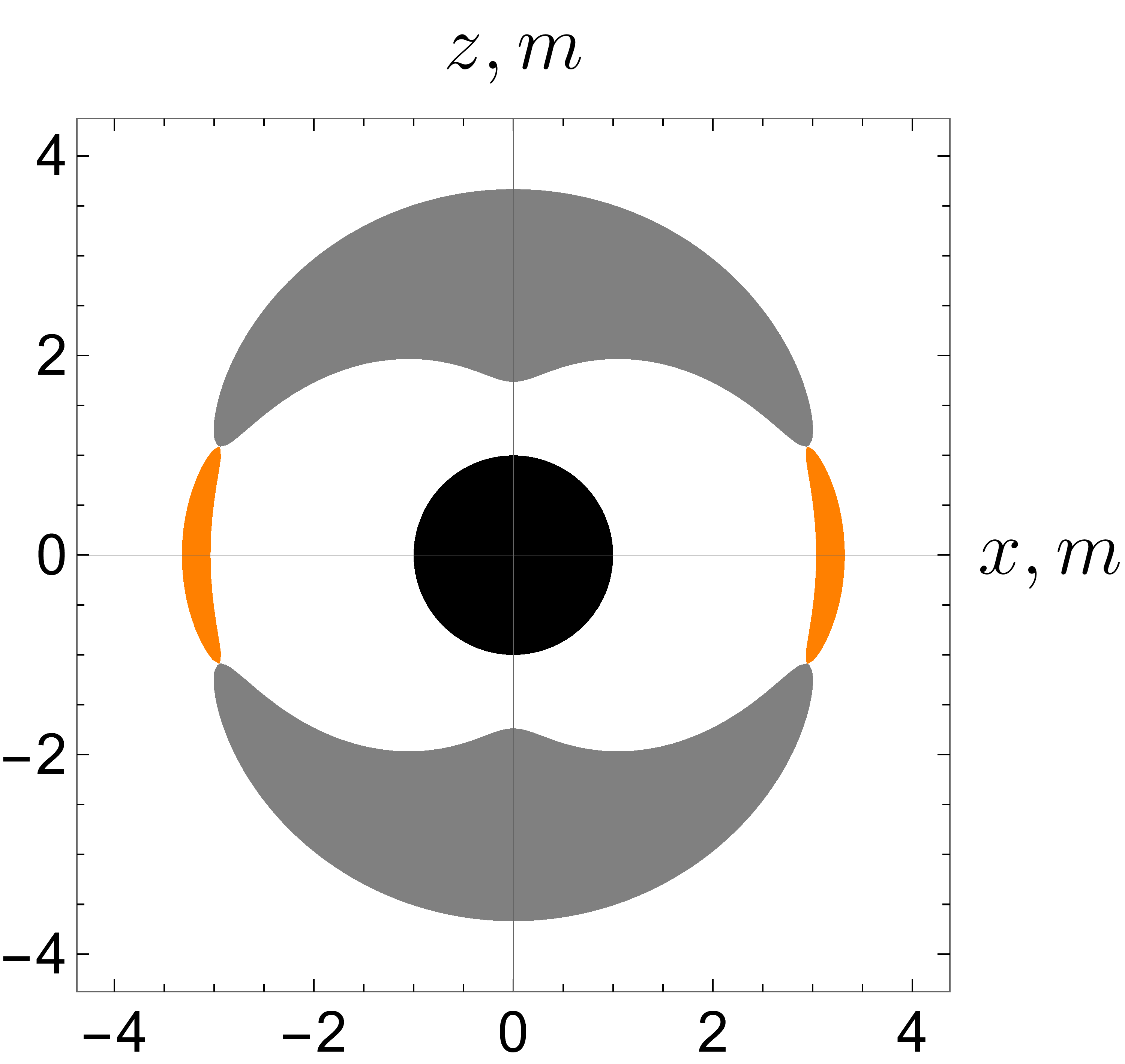} 
\end{tabular}
    \caption{WH2 with plasma distribution $f_r = \omega{_c}{^2} \sqrt{m{^3}r}$, $f_\theta =0$ and spin parameter $a=0.999m$. The plasma frequencies for each panel are as follows: top$-$left $\omega{_c}/ \omega{_0} = 2$; top$-$right $\omega{_c}/ \omega{_0} = 3.3$; bottom$-$left $\omega{_c}/ \omega{_0} = 3.5$; and bottom$-$right $\omega{_c}/ \omega{_0} = 3.75$. The forbidden region originates near the poles and expands toward the equatorial plane with increasing plasma frequency.}
\end{figure}
\newpage

\begin{figure}[H]
\centering
\vspace{-5mm}   
    \includegraphics[width=9.6cm]{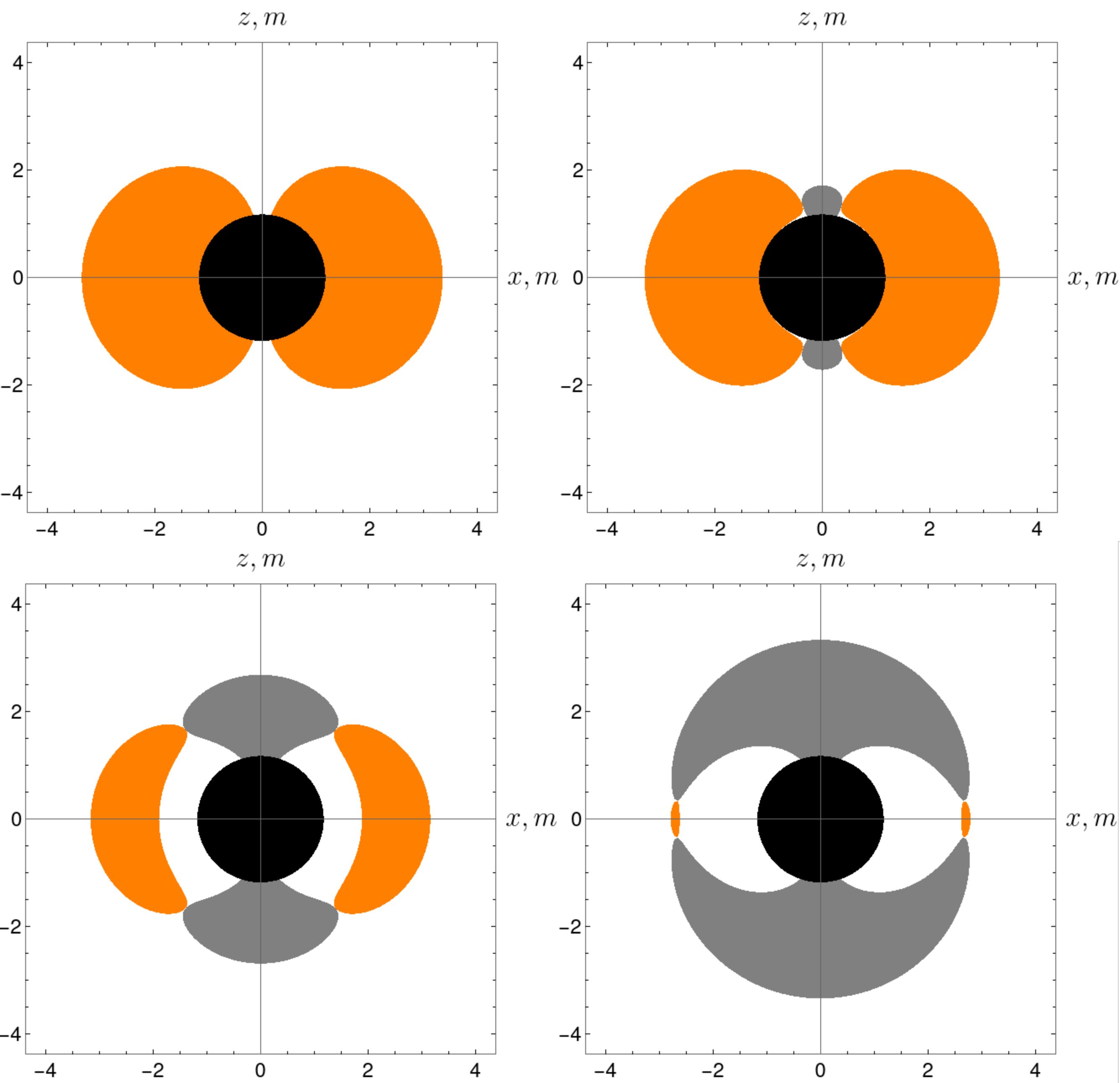}
    \caption{WH3 with plasma distribution $f_r = \omega{_c}{^2}\sqrt{m^3r}$, $f_\theta =0$ and spin parameter $a=0.999m$. The plasma frequencies for each panel are as follows: top$-$left $\omega{_c}/ \omega{_0} = 1$; top$-$right $\omega{_c}/ \omega{_0} = 1.5$; bottom$-$left $\omega{_c}/ \omega{_0} = 1.82$; and bottom$-$right $\omega{_c}/ \omega{_0} =1.83$. The forbidden region starts near the poles, while a second region forms at the equator, leading to the disappearance of the shadow for equatorial observers.}
   \end{figure}

\begin{figure}[t!]
\centering
\vspace{-5mm} 
    \includegraphics[width=9.6cm]{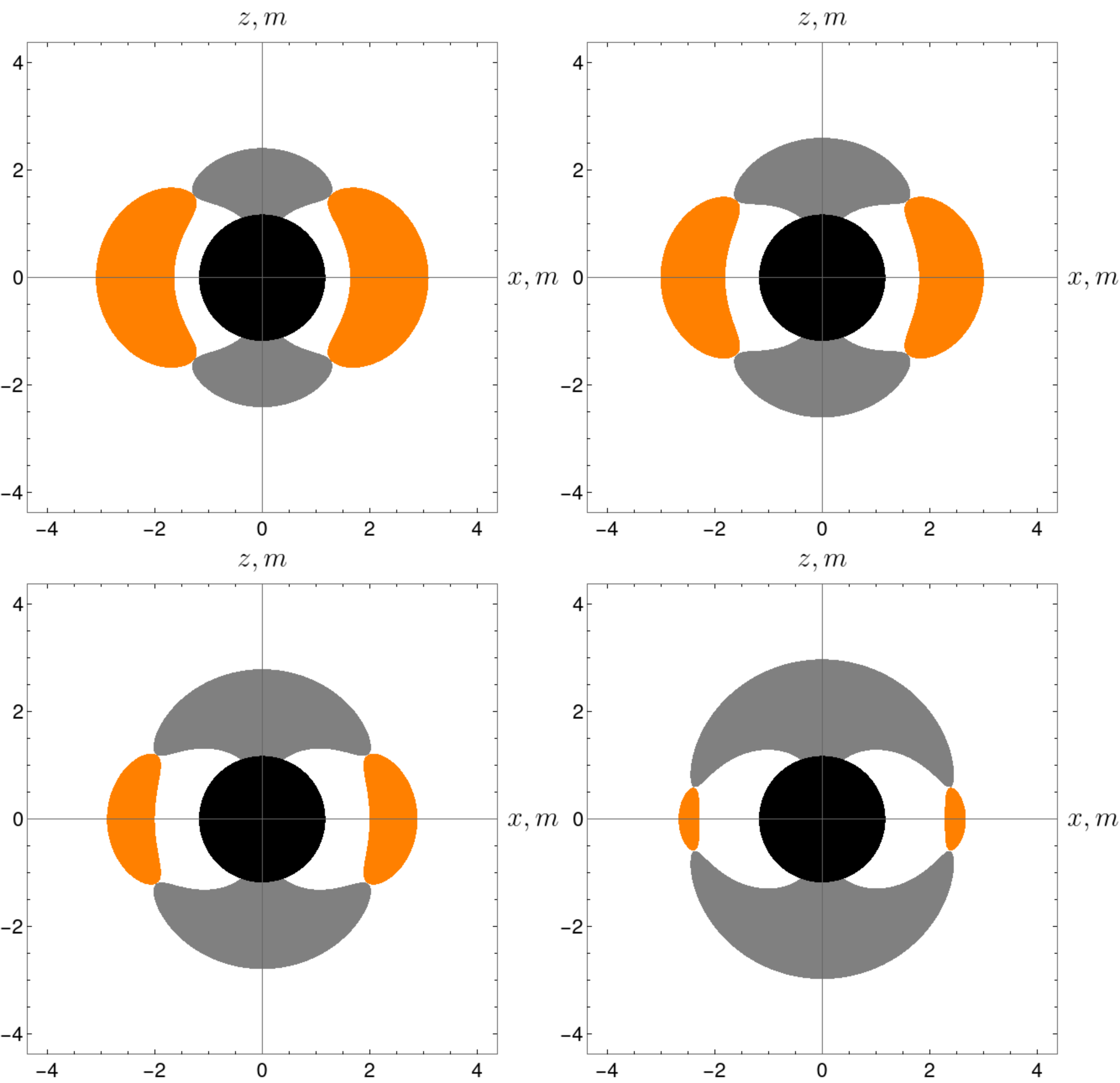}
    \caption{WH3 with plasma distribution with $f_r = 4\omega{_c}{^2}m{^2}\sqrt{m/r}$, $f_\theta=0$ and spin parameter $a=0.999m$. The plasma frequencies for each panel are as follows: top$-$left $\omega{_c}/ \omega{_0} = 1$; top$-$right $\omega{_c}/ \omega{_0} = 1.5$; bottom$-$left $\omega{_c}/ \omega{_0} = 1.82$; and bottom$-$right $\omega{_c}/ \omega{_0} =1.83$. The forbidden region starts near the poles, while a second region forms at the equator, leading to the disappearance of the shadow for equatorial observers.}
\end{figure}

\begin{figure}[p]
    \centering
    \includegraphics[width=15.5cm]{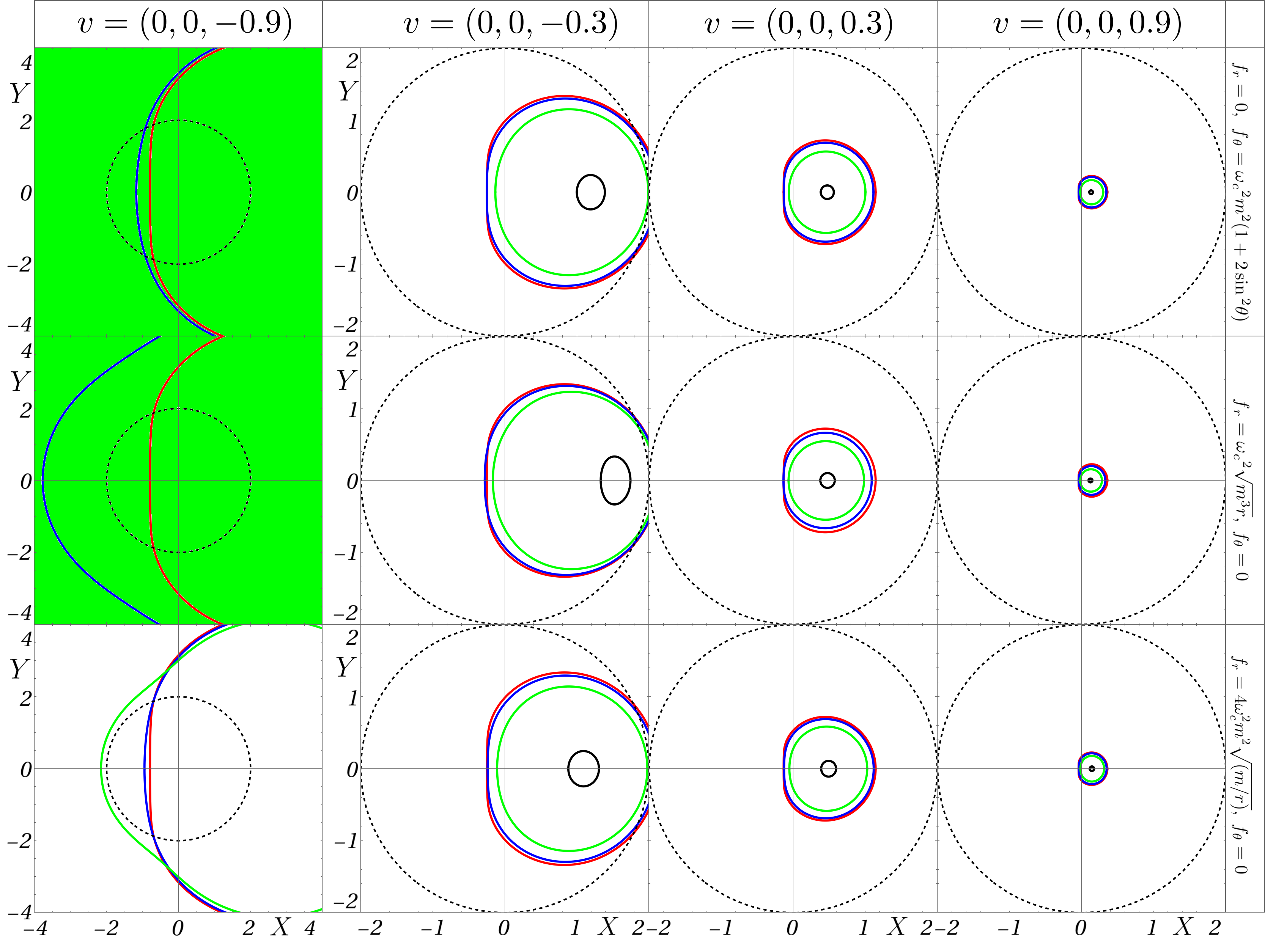}
    \caption{Shadows of the Kerr black hole for a moving observer located at $r_O=5m$, $\theta_O=\pi/2$, and spin parameter $a=0.999m$. Each column corresponds to a different observer velocity, while each row represents a distinct plasma distribution: the first row $f_r=0, f_\theta=\omega_c^2m^2(1+2\sin^2\theta)$; the second row $f_r=\omega_c^2\sqrt{m^3r}, f_\theta=0$; and the third row $f_r=4\omega_c^2m^2\sqrt{(m/r)},f_\theta=0$. The red contours correspond to the vacuum case. First row: $\omega_c/\omega_0=1$ (blue), $2$ (green), and $2.98$ (black). In the top-left panel, the green background indicates that the corresponding contour covers more than half of the observer’s sky, extending from $X=-10.1$ to $X=7.1$. Additionally, the black contour lies outside the field of view on the right, spanning $X=8.2$ to $X=9.3$. Second row: $\omega_c/\omega_0=2$ (blue), $3$ (green), and $3.88$ (black). In the leftmost panel, the green background indicates that the green contour covers more than half of the observer’s sky, extending from $X=-30.5$ to $X=8.1$. The black contour lies outside the field of view to the right, between $X=11.9$ to $X=15.4$. Third row: $\omega_c/\omega_0=1$ (blue), $2$ (green), and $3.35$ (black). The black contour lies outside the field of view to the right, between $X=6.0$ to $X=6.3$.}
    \label{fig:enter-label}
\end{figure}

\section*{Acknowledgments}
We gratefully acknowledge support by the Bulgarian NSF Grant KP-06-DV/8 under the funding program "VIHREN-2024".


\newpage

\begin{thebibliography}{tbds}



\bibitem{EHT1}
K. Akiyama et al. (Event Horizon Telescope), Astrophys. J. Lett. 875, L1-L6 (2019).

\bibitem{EHT2}
K. Akiyama et al. (Event Horizon Telescope), Astrophys. J. Lett. 910, L12-L13 (2021).

\bibitem{EHT3}
K. Akiyama et al. (Event Horizon Telescope), Astrophys. J. Lett. 930, L12-L17 (2022).

\bibitem{Moffat:2015}
J. W. Moffat,
``Modified Gravity Black Holes and their Observable Shadows'',
Eur. Phys. J. C75 (2015) 130.

\bibitem{Sakai}
N. Sakai, H. Saida, and T. Tamaki,
``Gravastar shadows'',
Phys. Rev. D90 (2014) 104013.

\bibitem{Cunha:2016a}
P. V.P. Cunha, C. Herdeiro, B. Kleihaus, J. Kunz, E. Radu
``Shadows of Einstein-dilaton-Gauss-Bonnet black holes'',
Phys.Lett. B768 (2017) 373.

\bibitem{Ahmedov:2014}
U. Papnoi, F. Atamurotov, S. Ghosh, B. Ahmedov,
``Shadow of five-dimensional rotating Myers-Perry black hole'',
Phys. Rev. D 90 (2014) 024073.

\bibitem{Herdeiro:2021}
H. C.D. Lima Junior, L. C.B. Crispino, P. V.P. Cunha, C. A.R. Herdeiro,
``Can different black holes cast the same shadow?'',
Phys. Rev. D103 (2021) 084040.

\bibitem{Shaikh:2019}
R. Shaikh, P. Kocherlakota, R. Narayan, P. Joshi,
``Shadows of spherically symmetric black holes and naked singularities'',
MNRAS 482 (2019) 52.

\bibitem{Nedkova:2019}
G. Gyulchev, P. Nedkova, T. Vetsov, S. Yazadjiev,
``Image of the Janis-Newman-Winicour naked singularity with a thin accretion disk'',
Phys. Rev. D100 (2019) 024055.

\bibitem{Nedkova:2020}
G. Gyulchev, J. Kunz, P. Nedkova, T. Vetsov, S. Yazadjiev,
``Observational signatures of strongly naked singularities: image of the thin accretion disk",
EPJC 80 (2020) 1017.

\bibitem{Nedkova:2021}
G. Gyulchev, P. Nedkova, T. Vetsov, S. Yazadjiev,
``Image of the thin accretion disk around compact objects in the Einstein–Gauss–Bonnet gravity'',
Eur. Phys. J. C 81 (2021) 885.

\bibitem{Nedkova:2013}
P. Nedkova, V. Tinchev, S. Yazadjiev,
``Shadow of a rotating traversable wormhole'',
Phys. Rev. D88 (2013) 124019.

\bibitem{Nedkova:2018}
G. Gyulchev, P. Nedkova, V. Tinchev, S. Yazadjiev,
``On the shadow of rotating traversable wormholes'',
EPJC 78 (2018) 544.


\bibitem{Kunz:2023}
H. Huang, J. Kunz, J. Yang, C. Zhang,
``Light ring behind wormhole throat: Geodesics, images and shadows,"
Phys.Rev.D 107 (2023) 104060.

\bibitem{Kunz:2016}
M. Zhou, A. Cardenas-Avendano, C. Bambi, B. Kleihaus, J. Kunz,
``Search for astrophysical rotating Ellis wormholes with x-ray reflection spectroscopy'',
Phys.Rev.D 94 (2016)  024036.

\bibitem{Shaikh:2019a}
S. Paul, R. Shaikh, P. Banerjee, T. Sarkar
``Observational signatures of wormholes with thin accretion disks'',
JCAP 03 (2020) 055.

\bibitem{Vincent:2020}
F.Vincent, M. Wielgus, M. Abramowicz, E. Gourgoulhon, J.-P. Lasota, T. Paumard, G. Perrin,
``Geometric modeling of M87* as a Kerr black hole or a non-Kerr compact object'',
A\&A 646 (2021)  A37.


\bibitem{Kocherlakota:2021}
P. Kocherlakota et al. (EHT Collaboration),
``Constraints on black-hole charges with the 2017 EHT observations of M87*",
Phys. Rev. D 103 (2021) 104047.


\bibitem{Eichhorn:2023a}
A. Eichhorn and A. Held,
``Quantum gravity lights up spinning black holes,"
JCAP 01 (2021) 032.

\bibitem{Eichhorn:2023}
A. Eichhorn, R. Gold, A. Held,
``Horizonless Spacetimes As Seen by Present and Next-generation Event Horizon Telescope Arrays,"
Astrophys.J. 950 (2023) 117.

\bibitem{Deliyski:2025}
V. Deliyski, G. Gyulchev, P. Nedkova, S. Yazadjiev,
``Observing naked singularities with the present and next-generation Event Horizon Telescope,"
Phys.Rev.D 111 (2025) 064068.


\bibitem{Flamm:1916}
L. Flamm
``Beiträge zur Einsteinschen Gravitationstheorie.''
Phys. Z., 17, 48 (1916)

\bibitem{EinsteinRosen}
A. Einstein and N. Rosen
``The Particle Problem in the General Theory of Relativity'',
Phys. Rev. 48, 73 (1935).

\bibitem{Morris:1988}
M. Morris, K. Thorne,
``Wormholes in spacetime and their use for interstellar travel: a tool for teaching general relativity'',
 	Am. J. Phys., 56 (5) (1988), pp. 395-412

\bibitem{Teo:1998}
E. Teo, ``Rotating traversable wormholes'',
Phys. Rev. D 58  (1998) 024014.

\bibitem{Deligianni:2021}
E. Deligianni, J. Kunz, P. Nedkova, S. Yazadjiev and R. Zheleva,
``Quasiperiodic oscillations around rotating traversable wormholes,"
Phys. Rev. D 104,024048 (2021).

\bibitem{Deligianni:2021a}
E. Deligianni, B. Kleihaus, J. Kunz, P. Nedkova and S. Yazadjiev,
``Quasiperiodic oscillations in rotating Ellis wormhole spacetimes,"
Phys. Rev. D 104, 064043 (2021).

\bibitem{Nedkova:2023}
V. Deliyski, G. Gyulchev, P. Nedkova, S. Yazadjiev,
``Polarized image of equatorial emission in horizonless spacetimes: traversable wormholes",
Phys.Rev.D 106 (2022) 104024.

\bibitem{Perlick:2022a}
V. Perlick, O. Yu. Tsupko
``Calculating black hole shadows: Review of analytical studies,"
Phys. Rept. 947 (2022) 1.

\bibitem{Rogers:2015}
A. Rogers,
``Frequency-dependent effects of gravitational lensing within
plasma,"
Mon. Not. Roy. Astron. Soc. 451, 4536 (2015).

\bibitem{Rogers:2017}
A. Rogers, 
``Escape and trapping of low-frequency gravitationally lensed rays by compact objects within plasma,"
Mon. Not. Roy. Astron. Soc. 465, 2151 (2017).

\bibitem{Rogers:2017a}
A. Rogers, 
``Gravitational Lensing of Rays through the Levitating Atmospheres of Compact Objects,"
Universe, 3, 3 (2017).

\bibitem{Mao:2014}
X. Er and S. Mao, 
``Effects of plasma on gravitational lensing,"
Mon. Not. Roy. Astron. Soc. 437, 2180 (2014).

\bibitem{Muhleman:1966}
D. O. Muhleman, I. D. Johnston, 
Phys. Rev. Lett. 17, 455 (1966).

\bibitem{Mihleman:1970}
D. O. Muhleman, R. D. Ekers, E. D. Fomalont, 
Phys. Rev. Lett. 24, 1377 (1970).

\bibitem{Bliokh:1989}
P. V. Bliokh and A. A. Minakov, 
“Gravitational Lenses”, 
Naukova Dumka, Kiev (1989) [in Russian].

\bibitem{Perlick:2000}
V. Perlick,
“Ray optics, Fermat’s principle and applications to general relativity”,
Springer, Heidelberg (2000).

\bibitem{Tsupko:2009}   
G. S. Bisnovatyi-Kogan and O. Yu. Tsupko,
``Gravitational Radiospectrometer,"
Gravitation and Cosmology, 15, 20 (2009).

\bibitem{Tsupko:2010}
G. S. Bisnovatyi-Kogan and O. Yu. Tsupko, 
``Gravitational lensing in a non-uniform plasma,"
Mon. Not. Roy. Astr. Soc. 404, 1790 (2010).

\bibitem{Tsupko:2015}
G. S. Bisnovatyi-Kogan and O. Yu. Tsupko, 
``Gravitational Lensing in Plasmic Medium,"
Plasma Physics Reports, 41, 562 (2015).

\bibitem{Morozova:2013}
V. Morozova, B. Ahmedov, and A. Tursunov, 
``Gravitational lensing by a rotating massive object in a plasma,"
Astrophys. Space Sci. 346, 513 (2013).

\bibitem{Tsupko:2013}
O. Yu. Tsupko and G. S. Bisnovatyi-Kogan, 
``Gravitational lensing in plasma: Relativistic images at homogeneous plasma,"
Phys. Rev. D 87, 124009 (2013).

\bibitem{Liu:2016}      
 C. Liu, C. Ding, J. Jing, 
``Effects of Homogeneous Plasma on Strong Gravitational Lensing of Kerr Black Holes,"
Chin. Phys. Lett. 34 (2017) 090401.

\bibitem{Perlick:2024}
V. Perlick, O. Yu. Tsupko
``Light propagation in a plasma on Kerr spacetime. II. Plasma imprint on photon orbits,"
Phys.Rev.D 109 (2024) 6, 064063.

\bibitem{Perlick:2025}
V. Perlick,
``Characterisation of circular light rays in a plasma,"
Class. Quant. Grav. 42 (2025) 17, 175013

\bibitem{Perlick:2015}
V. Perlick, O. Yu. Tsupko and G. S. Bisnovatyi-Kogan,
``Influence of a plasma on the shadow of a spherically symmetric black hole,"
Phys. Rev. D 92, 104031 (2015).


\bibitem{Perlick:2017}
V. Perlick, O. Tsupko,
`Light propagation in a plasma on Kerr spacetime: Separation of the Hamilton-Jacobi
equation and calculation of the shadow'',
Phys. Rev. D 95, 104003 (2017).

\bibitem{Perlick:2022}
V. Perlick,
``Light propagation in a plasma on an axially symmetric and stationary spacetime:
Separability of the Hamilton-Jacobi equation and shadow'',
J. Math. Phys. 63, 092501 (2022).


\bibitem{Badia:2021}
Javier Badia, Ernesto F. Eiroa,
``Shadow of axisymmetric, stationary and asymptotically flat black holes in the presence of plasma'',
Phys. Rev. D 104 (2021) 084055.

\bibitem{Badia:2023}
Javier Badia, Ernesto F. Eiroa,
``Shadows of rotating Einstein-Maxwell-dilaton black holes surrounded by a plasma'',
Phys. Rev. D 107 (2023) 124028 .

\bibitem {Briozzo:2023}
Briozzo, Gast\'on and Gallo, Emanuel and M\"adler, Thomas,
``Shadows of rotating black holes in plasma environments with aberration effects``,
Phys. Rev. D 107 (2023), 124004.

\bibitem{Grenzebach:2015}
A. Grenzebach, 
``Aberrational Effects for Shadows of Black Holes",
in D. Puetzfeld, C. Lämmerzahl, B. Schutz (eds)
{\it``Equations of Motion in Relativistic Gravity"}, Fundamental Theories of Physics, vol. 179, Springer, Cham (2015).     

 \bibitem{Penrose}
R. Penrose, 
``The Apparent Shape of a Relativistically Moving Sphere,"
Math. Proc. Camb. Phil. Soc. 55 (1959) 137.


\bibitem{Bardeen:1972}
J. Bardeen, W. Press,  S. Teukolsky,
``Rotating Black Holes: Locally Nonrotating Frames,
Energy Extraction, and Scalar Synchrotron Radiation'',
Astrophys. J. 178 (1972) 347.

\bibitem{Bardeen}
J. Bardeen,
``Timelike and null geodesies in the Kerr metric'',
in C. DeWitt and B. DeWitt, eds. Les Houches Summer School of Theoretical Physics: Black Holes,  Gordan and Breach, New
York, London and Paris (1973).







\end{thebibliography}
\end{document}